\documentclass{jfm}

\usepackage{graphicx}
\usepackage{newtxtext}
\usepackage{newtxmath}
\usepackage{natbib}
\usepackage{hyperref}
\hypersetup{
    colorlinks = true,
    urlcolor   = blue,
    citecolor  = black,
}

\newcommand{\RomanNumeralCaps}[1]

\usepackage{amssymb,amsmath}
\usepackage{array}
\usepackage{latexsym}
\usepackage{bezier}
\usepackage{psfrag}
\usepackage{bm}
\usepackage{float}
\usepackage{multirow}
\usepackage{mathrsfs}
\usepackage{color}
\usepackage{marginnote}
\usepackage{ulem}
\usepackage[pdf]{pstricks}

\newcommand{\rmi}{\mathrm{i}}
\newcommand{\rme}{\mathrm{e}}

\newcommand{\vu}{\mathbf{u}}

\newcommand{\vPsi}{\boldsymbol\Psi}
\newcommand{\vPhi}{\boldsymbol\Phi}

\newcommand{\bvec}[1]{\mbox{\bf #1}}
\newcommand{\Ri}{{\mathrm{R}_{\rm i}}}
\newcommand{\Ro}{{\mathrm{R}_{\rm o}}}
\newcommand\ci{\mathrm{i}}
\newcommand\ce{\mathrm{e}}
\newcommand\co{\mathrm{o}}
\newcommand{\bsy}[1]{{\boldsymbol{#1}}}                                         

\shortauthor{}

\title
  {Self-sustainment of coherent structures in counter-rotating Taylor-Couette flow}
  
\author
    []{B.\,Wang\aff{1},  R.\,Ayats\aff{1}, K.\,Deguchi\aff{2},
  F.\,Mellibovsky\aff{1} \corresp{\email{fernando.mellibovsky@upc.edu}} and A.\,Meseguer\aff{1}}

\affiliation{
  \aff{1} Departament de F{\'\i}sica, Universitat Polit\`ecnica de Catalunya, 08034, Barcelona, Spain
  \aff{2} School of Mathematics, Monash University, VIC 3800, Australia
}

\begin{document}
\maketitle

\begin{abstract}
We investigate the local self-sustained process underlying spiral turbulence in counter-rotating Taylor-Couette flow using a periodic annular domain, shaped as a parallelogram, two of whose sides are aligned with the cylindrical helix described by the spiral pattern. The primary focus of the study is placed on the emergence of drifting-rotating waves ({\sc drw}) that capture, in a relatively small domain, the main features of coherent structures typically observed in developed turbulence. The transitional dynamics of the subcritical region, far below the first instability of the laminar circular Couette flow, is determined by the upper and lower branches of {\sc drw} solutions originated at saddle-node bifurcations.  The mechanism whereby these solutions self-sustain, and the chaotic dynamics they induce, are conspicuously reminiscent of other subcritical shear flows.  Remarkably, the flow properties of {\sc   drw} persist even as the Reynolds number is increased beyond the linear stability threshold of the base flow. Simulations in a narrow parallelogram domain stretched in the azimuthal direction to revolve around the apparatus a full turn confirm that self-sustained vortices eventually concentrate into a localised pattern. The resulting statistical steady state satisfactorily reproduces qualitatively, and to a certain degree also quantitatively, the topology and properties of spiral turbulence as calculated in a large periodic domain of sufficient aspect ratio that is representative of the real system.
\end{abstract}

\begin{keywords}
Authors should not enter keywords on the manuscript, as these must be chosen by the author during the online submission process and will then be added during the typesetting process (see \href{https://www.cambridge.org/core/journals/journal-of-fluid-mechanics/information/list-of-keywords}{Keyword PDF} for the full list).  Other classifications will be added at the same time.
\end{keywords}

{\bf MSC Codes }  {\it(Optional)} Please enter your MSC Codes here

\section{Introduction}\label{secintro}
Nonlinear shear flow instabilities often lead to
\textit{intermittency}, a phenomenon that involves the coexistence of
laminar and turbulent flow regions.  The resulting turbulent-laminar
patterns take either the form of localised turbulent patches
surrounded by otherwise laminar flow or, surprisingly more orderly, of
coherent stripes that exhibit sharp interfaces tilted with respect to
the main direction of the flow. A paradigm for coherent intermittency
in rotating shear flows is the so-called \textit{Spiral Turbulence}
regime ({\sc spt}) that appears in Taylor-Couette flow, i.e. the fluid
flow between independently-rotating coaxial cylinders.  The {\sc spt}
regime consists of a turbulent helix that forms a coil within the
apparatus gap with a well-defined pitch and rotates at a fairly
constant angular speed.  This peculiar flow structure, first
discovered in the 1960s \citep{ColAtt67} and declared a puzzling
phenomenon by Feynman himself \citep{Fey64}, had not been reproduced
numerically until rather recently by means of very costly direct
numerical simulation ({\sc dns})
\citep{MeMeAvMa09_A,Dong2009,DoZhe2011}.

Similar oblique laminar-turbulent stripe patterns are common in other
shear flows featuring two extended space directions
\citep{PriGreChaDauSaa02,DuSchHen2010,Tucketal2014}. For a thorough
review on the rich variety of existing intermittent shear flow
phenomena, we refer the reader to the monograph by
\citet{TuChaBar2020} and references therein. Recent attempts at
elucidating the stripe formation mechanism have mostly been confined
to relatively simple parallel shear flows such as plane Couette or
plane Poiseuille flows. Both problems exhibit subcritical transition
to turbulence, namely transition in the absence of a linear
instability of the base flow, which is best tackled employing
dynamical systems theory. In this framework, simple solutions to the
Navier-Stokes equations, often called \textit{exact coherent
  structures} ({\sc ecs}), are shown to naturally play a central role
in organising the transitional and turbulent dynamics
\citep{Kerswell2005,EcScHoWe07,KaUhVe12,GraFlo21}.






The last three decades have witnessed overwhelming scientific activity
in the search for {\sc ecs} in many subcritical shear flows
\citep{Nag90,CleBu92,CleBu97,Wa03,FAIECK03,WEDKER04}, following the
discovery of a \textit{Self-Sustained Process} ({\sc ssp}) for
coherent structures in the absence of linear instability of the
laminar flow \citep{Wa97}. The process, which occurs at relatively
short length-scales and can therefore be observed in small periodic
domains, consists in a cyclic feedback mechanism whereby streamwise
vortices generate streaks through the lift-up mechanism, which in turn
become unstable to three-dimensional waves that feed energy back into
the streamwise vortices \citep{BOBRO88,HamKimWa95,Grossmann2000}.  The
instability of the streaks responds to an inviscid mechanism by which
three-dimensional waves are strongly amplified at a critical layer,
i.e. where the streak speed coincides with the wave speed
\citep{WangGiWa07,HaSh2010,DeguchiHall2015}.

For relatively small periodic boxes, the relationship between the
exact solution and the dynamics is understood to some extent.  Of the
{\sc ecs} emanated from saddle-node bifurcations, the lower (or
saddle) branch typically dictates the topology and amplitude of flow
perturbations that are capable of triggering transition, while the
upper (or nodal) branch generally regulates -or at least participates
in- the formation of the turbulent set.  In the subcritical transition
problem, the infinite-dimensional Navier-Stokes phase space typically
contains two stable (or metastable) invariant sets: the steady laminar
base flow and the chaotic/turbulent state. The basins of attraction of
these two sets meet along a codimension-1 manifold, usually known as
\textit{edge of chaos} \citep{ItTo01,SkYoEck06,SchGiLaDeLiEck08},
where the lower-branch solutions belong.
The upper branch {\sc ecs} are almost always unstable and their
participation in the generation of the chaotic set is not easily
dissected. Occasionally, however, upper-branch {\sc ecs} are linearly
stable, if only within sufficiently small domains or tightly
constrained symmetry conditions, in a neighbourhood of the saddle-node
bifurcation \citep{CleBu97,MeEck2011,MeEck2012}. In these cases,
further theoretical progress can be made as the path towards chaotic
dynamics admits a simpler analysis that can draw from a parallel with
low-dimensional dynamical systems
\citep{MeEck2012,KrEc2012,LuKaVeShiKo19}.

Transition and pattern formation in Taylor-Couette flow are more
involved than for parallel shear flows due to the interplay of shear
and rotation \citep{ALS86}. A notable difference with respect to
merely shear-driven flows is that {\sc spt} also persists in the
supercritcal regime of counter-rotating Taylor-Couette flow, beyond
the linear instability of the laminar circular Couette flow
\citep{PriGreChaDauSaa02,MeMeAvMa09_A}. Therefore, both the shear
and the centrifugal instabilities contribute their share to the
generation of streamwise vorticity, but this fact has gone largely
unnoticed in the literature, where the origin of the stripe may be solely
explained by the stability of both the basic flow and the autonomous
vortex emerged from the {\sc ssp}.

The aim of this paper is to investigate the dynamics induced by {\sc
  ecs} driven by the {\sc ssp} and to ascertain whether they may be held
responsible for the formation of the {\sc spt} regime observed in the
centrifugally-unstable region of parameter space.
The most natural place to look for nonlinear solutions is at the
linear critical point of the base flow, circular Couette flow in our
case. However, the nonlinear spiral solutions thus identified by
\citet{MeMeAvMa09_B} are only very mildly subcritical, which implies
that it is the centrifugal instability rather than the {\sc ssp} that
drives them. A wealth of solutions predicted by weakly nonlinear
theory \citep{ChoIoo94} followed the discovery of the subcritical
spirals \citet{DeAlt2013}, but none contributed to enlarge the known
region of subcriticality.  Shortly after, however, a highly
subcritical three-dimensional rotating-wave solution was found by
\citet{DeMeMe14}, although its dynamical relevance was not
investigated.

As noted earlier, the computation of laminar-turbulent banded patterns
is very costly, such that using narrow orthogonal domains suitably
tilted to align with the stripes has been decisive to the study of
this kind of laminar-turbulent patterns
\citep{BARTUC05,ShAvHo2013,ReKrSc19,PaDuHo20}.
The approach is easily undertaken for parallel shear flows, as a mere
change in the direction of the base flow suffices, but more
fundamental code modifications are necessary for cylindrical and
annular geometries. The required coordinate change was generalised by
\citet{DeAlt2013} to compute {\sc ecs} in parallelogram-shaped domains
wrapped within an annular geometry. Since their method of directly
solving the nonlinear algebraic equations was only applicable to
travelling-wave solutions, developing a {\sc dns} code in generalised
parallelogram-shaped periodic domains is requisite to efficiently
capture {\sc spt} dynamics, but has hitherto not been attempted to our
best knowledge.

The outline of the paper is as follows.  The problem formulation is
given in \S\ref{sec_formulation}, alongside a description of the
generalised parallelogram-shaped domain, its application to the
spectral space discretisation, and the numerical methods employed for
evolving the equations in time, and the coupling of the time-stepper
with a Poincar\'e-Newton-Krylov solver. Then \S\ref{sec_parameter}
briefly summarises the geometrical and physical parameters used for
the numerical calculations, and justifies the specific choice of the
domain shape. Since the rotating-wave solutions found by
\citet{DeMeMe14} were computed in a classical orthogonal small
periodic domain, the initial task in \S\ref{sec_subcritrow} is the
exploration of bifurcation scenarios leading to solutions of the same
family in small parallelogram domains.  In \S\ref{sec_stabdynflows},
the dynamical relevance of the solution is first analysed in the
subcritical regime to identify the {\sc ssp} and the onset of chaotic
dynamics. The possible relevance of these solutions to the
supercritical {\sc spt} regime is then discussed. Finally, the main
findings are summarised in \S\ref{sec_conclusions} along with
concluding remarks.

\section{Formulation of the problem}
\label{sec_formulation}
\begin{figure}
\centering
\includegraphics[width=.5\linewidth,clip=]{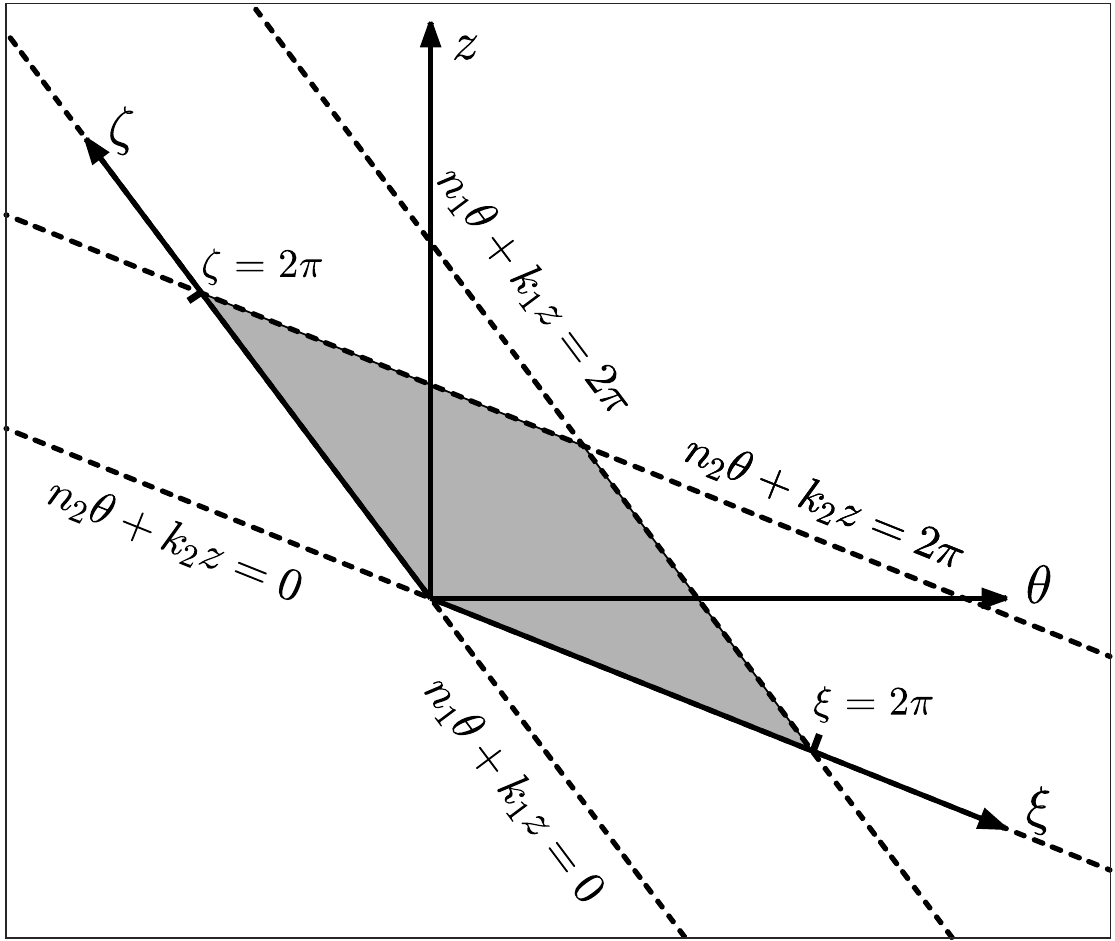}
\caption{Sketch of the parallelogram domain introducing the new
  variables $(\xi,\zeta)$ that replace the usual azimuthal and axial
  coordinates $(\theta,z)$.
}
\label{fig:parallel_scheme}
\end{figure}

Consider an incompressible fluid of dynamic viscosity $\mu$ and
density $\varrho$ (kinematic viscosity $\nu=\mu/\varrho$) completely
filling the gap between two concentric rotating cylinders whose inner
and outer radii and angular velocities are $r^*_{\rm{i}}$,
$r^*_{\rm{o}}$ and $\Omega_{\rm{i}}$, $\Omega_{\rm{o}}$,
respectively. A full set of independent dimensionless parameters
characterising the problem are the radius ratio
$\eta=r^*_{\rm{i}}/r^*_{\rm{o}}$, which fixes the geometry of the
annulus, and the circular Couette flow inner and outer Reynolds
numbers $\Ri=dr^*_{\rm{i}}\Omega_{\rm{i}}/\nu$ and
$\Ro=dr^*_{\rm{o}}\Omega_{\rm{o}}/\nu$, where
$d=r^*_{\rm{o}}-r^*_{\rm{i}}$ is the gap between the
cylinders. Henceforth, all variables will be rendered dimensionless
using $d$, $d^2/\nu$, and $\nu^2/d^2$ as units for space, time and the
reduced pressure ($p=p^*/\varrho$), respectively. The Navier--Stokes
equation, and the incompressibility and the zero axial net massflux
conditions become
\begin{eqnarray}\label{INSE}
  \partial_{t}\bvec{v} + (\bvec{v}\cdot\nabla)\bvec{v} & = & -\nabla p
  + \nabla^{2}\bvec{v} + f\,\hat{\bsy z},\label{INSE:NS}\\ \quad
  \nabla\cdot\bvec{v} & = & 0,\label{INSE:MASS}\\ Q(\bvec{v}) =
  \int_0^{2\pi}{\int_{r_{\rm{i}}}^{r_{\rm{o}}}{(\bvec{v}\cdot\hat{\bsy
        z})\,r\,dr\,dz}} & = & 0\label{INSE:ZERO},
\end{eqnarray}
where the axial forcing term $f=f(t)$ in (\ref{INSE:NS}) is
instantaneously adjusted to fulfil the constraint imposed by
(\ref{INSE:ZERO}), $\bvec{v}=(U,V,W)=U\,\hat{\bsy r} + V\,\hat{\bsy
  \theta} + W\,\hat{\bsy z}$ is the velocity of the fluid expressed in
cylindrical coordinates $(r,\theta,z)$, which satisfies no-slip
boundary conditions at the cylinder walls
\begin{equation}\label{NSBCS}
\bvec{v}|_{r=r_\ci}=(0,\Ri,0),\quad \bvec{v}|_{r=r_{\rm o}}=(0,\Ro,0), 
\end{equation}
with
\begin{equation}\label{defradii}
  r_\ci=\frac{r^*_\ci}{d} = \frac{\eta}{1-\eta},\quad
  r_\co=\frac{r^*_\co}{d} = \frac{1}{1-\eta},
\end{equation}
the non-dimensional inner and outer radii, respectively.  The basic,
laminar and steady \textit{circular Couette flow}, henceforth referred
to as {\sc ccf}, is
\begin{equation}\label{defccf}
\bvec{v}_{\rm b} = U_{\rm b}\,\hat{\bsy r} + V_{\rm b}\,\hat{\bsy
  \theta} + W_{\rm b}\,\hat{\bsy z} = \displaystyle \left(Ar+
\frac{B}{r}\right)\,\hat{\bsy \theta}, \;\; p_{\rm b}(r)=\int
\frac{V_{\rm b}^2}{r} \,{\rm d}r, \;\; f_{\rm b} = 0,
\end{equation}
with $ A=(\Ro-\eta\Ri)/(1+\eta)$ and
$B=\eta(\Ri-\eta\Ro)/	\left[(1-\eta)(1-\eta^2)	\right]$. In what follows we express
the velocity and pressure fields as
\begin{equation}\label{vbpluspert}
  \bvec{v} = \bvec{v}_{\rm b}(r) + \mathbf{u}(r,\theta,z;t),\quad
  p=p_{\rm b}(r) + q(r,\theta,z;t).
\end{equation}
The fields $q$ and $\mathbf{u}= u\,\hat{\bsy r} +
v\,\hat{\bsy \theta} + w\,\hat{\bsy z}$ are the deviations from the
equilibrium {\sc ccf} solution that, after formal substitution of
\eqref{vbpluspert} into \eqref{INSE}, satisfy
\begin{eqnarray}
\partial_t{\bf u} & = & -\nabla q + \nabla^2{\bf u}-({\bf v}_{\rm b} \cdot
\nabla){\bf u}-({\bf u} \cdot \nabla){\bf v}_{\rm b}-({\bf u} \cdot
\nabla){\bf u}, \label{hydro_nse} + f\,\hat{\bsy z}\\
\nabla \cdot {\bf u} & = & 0,\label{hydro_solen}\\
Q(\bvec{u}) & = & 0.\label{hydro_zero}
\end{eqnarray}
Although this solenoidal boundary value problem is naturally
formulated in cylindrical polar coordinates $(r,\theta,z)$, the
coherent flows addressed in this work are better captured and more
efficiently represented numerically on parallelogram domains such as
the one depicted in figure\,\ref{fig:parallel_scheme}. This type of
domains have been recently vindicated as minimal flow units to capture
mixed spiral modes and rotating-travelling waves with arbitrary
axial-azimuthal wavefront orientation in {\sc tcf}
\citep{DeAlt2013,DeguchiHall2015,AyDeMeMe20}. The parallelogram domain
is bounded by two consecutive ($2\pi$-shifted) wavefront loci and
might be naturally parametrised by introducing the new coordinates
\begin{equation}
  \xi  =  n_1\theta + k_1 z,\quad \zeta  = n_2\theta + k_2 z,
\label{xieta_thetar}
\end{equation}
or, conversely,
\begin{equation}
  \theta = \displaystyle \frac{k_2 \xi - k_1
    \zeta}{n_1k_2-n_2k_1},\quad z = \displaystyle \frac{-n_2 \xi
    + n_1 \zeta}{n_1k_2-n_2k_1}.
\label{thetar_xieta}
\end{equation}
Henceforth, we reformulate the boundary value problem
(\ref{hydro_nse}-\ref{hydro_solen}) within the parallelogram
assuming the pressure and velocity fields $q$ and $\vu$ are
$2\pi$-periodic in the two new coordinates $\xi$ and $\zeta$, thus satisfying
\begin{eqnarray}
  q(r,\xi+2\pi,\zeta;t) & = q(r,\xi,\zeta+2\pi;t) & = q(r,\xi,\zeta;t),\\
   \vu(r,\xi+2\pi,\zeta;t) & = \vu(r,\xi,\zeta+2\pi;t) & = \vu(r,\xi,\zeta;t).
\label{perbc_qv}
\end{eqnarray}
In what follows, we numerically discretise $q$ and $\vu$ within the
annular-parallelogram domain
\begin{equation}
(r,\xi,\zeta) \in \left[r_{\rm{i}},r_{\rm{o}}\right] 
\times[0,2\pi]\times[0,2\pi],
\label{ann_paral_domain}
\end{equation}
where the inner and outer radii of the cylinders are explicitly given in
(\ref{defradii}).
We characterise flows by their
associated \textit{normalised torque} at the inner and outer
cylinders, $\tau_{\rm{i}}$ and $\tau_{\rm{o}}$,
\begin{eqnarray}
  \tau_{\rm i, \rm o}=
\left. 1+\frac{\partial_r(r^{-1}\bar{v})}{\partial_r(r^{-1}v_{\rm
    b})}\right|_{r=r_{\rm i},r_{\rm o}},
\label{defnormtorque}
\end{eqnarray}
where $\bar{v}$ is the averaged azimuthal velocity in the angular and
axial directions. With this definition, $\tau_{\rm{i}}=\tau_{\rm{o}}=1$ for {\sc
  ccf}. Similarly, we will also characterise the flows by the
\textit{normalised kinetic energy} of the perturbation velocity field,
\begin{equation} E = \frac{1}{2E_{\rm b}}
\int_{0}^{2\pi}\mathrm{d}\zeta\int_{0}^{2\pi}\mathrm{d}\xi
\int_{r_{\rm{i}}}^{r_{\rm{o}}} \bvec{u}\cdot\bvec{u}\,
r\,\mathrm{d}r, \label{defenergyE}
\end{equation}
where $E_b$ is the kinetic energy of {\sc ccf}, namely
\begin{equation} E_{\rm b} = \frac{1}{2}
\int_{0}^{2\pi}\mathrm{d}\zeta\int_{0}^{2\pi}\mathrm{d}\xi
\int_{r_{\rm{i}}}^{r_{\rm{o}}}\bvec{v}_{\rm b}\cdot\bvec{v}_{\rm b} \,r\,\mathrm{d}r
=2\pi^2\int_{r_{\rm{i}}}^{r_{\rm{o}}}\left(Ar+\frac{B}{r}\right)^2r\,\mathrm{d}r.
\label{defenergyEb}
\end{equation}

\subsection{Direct numerical simulations in the annular-parallelogram domain}
The nonlinear boundary value problem
(\ref{hydro_nse}-\ref{hydro_solen}) is discretised using a solenoidal
Petrov-Galerkin scheme formerly formulated by \citet{EPJSTMAMM07}, and
suitably adapted to the annular-parallelogram domain
(\ref{ann_paral_domain}). In the transformed domain, the solenoidal
velocity perturbation is approximated by means of a Fourier $\times$
Fourier $\times$ Chebyshev spectral expansion ${\bf u}_{\rm s}$ of
order $N \times L \times M$ in $\xi \times \zeta \times r$,
respectively, of the form
\begin{equation}
    {\bf u}_{\rm s}(r,\xi,\zeta;t)=\sum_{\ell=-L}^{L} \sum_{n=-N}^{N} \sum_{m=0}^{M} 
    a_{\ell nm}^{(1)}(t) \vPhi_{\ell nm}^{(1)}(r,\xi,\zeta)+a_{\ell nm}^{(2)}(t) \vPhi_{\ell nm}^{(2)}(r,\xi,\zeta).
\label{specapx}
\end{equation}
Our aim here is to derive the dynamical system satisfied by the
coefficients $a_{\ell nm}^{(\imath)}(t)$, as symbolically represented by the
$2\times(M+1)\times(2L+1)\times(2N+1)$-dimensional state vector
$\mathbf{a}(t)$.
The binary superindex $\imath=\{1,2\}$ and the factor 2 in the count of unknowns follow from the two degrees of freedom per grid point that remain after taking into condiseration that the three velocity components are not independent, but linked by the solenoidal condition.
The vector fields $\vPhi_{\ell n m}^{(\imath)}$ constitute the elements of the
{\it trial} basis of solenoidal vector fields of the form
\begin{equation}
\vPhi_{\ell n m}^{(\imath)}(r,\xi,\zeta)=
\rme^{\rmi(n\xi+\ell\zeta)}
\vu_{\ell nm}^{(\imath)}(r) =
\rme^{\rmi(n\xi+\ell\zeta)}
\left(u_{\ell n m}^{(\imath)},v_{\ell n m}^{(\imath)},w_{\ell n m}^{(\imath)} \right),
\label{physbas}
\end{equation}
where $u_{\ell n m}^{(\imath)}$, $v_{\ell n m}^{(\imath)}$ and $w_{\ell n m}^{(\imath)}$ are the
radial, azimuthal and axial components of $\vu_{\ell nm}^{(\imath)}(r)$,
respectively. Each element of the trial basis satisfies the
divergence-free condition (\ref{hydro_solen}) that, in the
$(r,\xi,\zeta)$ variables, explicitly reads
\begin{equation} \left(\partial_r + \frac{1}{r}\right) u_{\ell nm}^{(\imath)} + 
  \frac{\rmi}{r}\left(n_1n + n_2\ell \right)v_{\ell n m}^{(\imath)}+
       {\rm i}\left(k_1n+k_2\ell \right) w_{\ell nm}^{(\imath)}=0.
\label{sol_cond_expl}
\end{equation}
Since $\vu$ represents the perturbation of the velocity field, it must
therefore vanish at the inner ($r=r_{\rm{i}}$) and outer ($r=r_{\rm{o}}$) walls of
the cylinders. Therefore, $\vPhi_{\ell n m}^{(\imath)}$ must also satisfy the
homogeneous boundary conditions
\begin{equation}
\vPhi_{\ell n m}^{(\imath)}(r_{\rm{i}},\xi,\zeta;t)=\vPhi_{\ell n m}^{(\imath)}(r_{\rm{o}},\xi,\zeta;t)=\mathbf{0}.
\label{hbc_phi}
\end{equation}
In what follows, we define the transformed radial coordinate
\begin{equation}\label{mapping}
x(r) = 2r-\frac{1+\eta}{1-\eta},
\end{equation}
that maps the radial domain $r\in[r_{\rm i},r_{\rm o}]$ to the
interval $x\in[-1,1]$. In addition, we define the radial functions
\begin{equation}\label{basicfunc}
 {\rm h}_m(r)=(1-x^2){\rm T}_m(x),\ 
 {\rm g}_m(r)=(1-x^2)^2{\rm T}_m(x),
\end{equation}
where ${\rm T}_m(x)$ is the Chebyshev polynomial of degree
$m$. Finally, we introduce the Chebyshev weight function ${\rm
  w}(x)=(1-x^2)^{-1/2}$, defined over the interval $(-1,1)$. The
functions introduced in (\ref{basicfunc}) satisfy
\begin{equation}\label{bcfg}
 {\rm h}_m(r_{\rm i})={\rm h}_m(r_{\rm o})={\rm g}_m(r_{\rm
 i})={\rm g}_m(r_{\rm o})={\rm D}{\rm g}_m(r_{\rm i})={\rm D}{\rm
 g}_m(r_{\rm o})=0,
\end{equation}
where ${\rm D}$ stands for the radial derivative ${\rm d}/{\rm
  d}r$. The solenoidal spectral method consists in devising complete
sets of vector fields (\textit{trial} functions) satisfying
(\ref{sol_cond_expl}) and (\ref{hbc_phi}). For $n n_1+\ell n_2 = 0$,
two such vector fields are
\begin{equation}\label{phybase_00}
\mathbf{u}_{\ell n m}^{(1)}(r)=(0\,,\,{\rm h}_m\,,\,0),\quad
\mathbf{u}_{\ell n m}^{(2)}(r)=( -\rmi(n k_1+\ell k_2) r {\rm g}_m\,,\,
0\,,\,{\rm D}[r {\rm g}_m]+{\rm g}_m),
\end{equation}
 with the third component of $\mathbf{u}_{\ell n m}^{(2)}$ replaced by
 $ {\rm h}_m$ whenever $n k_1+\ell k_2=0$.  Finally, for ${n n_1+\ell
   n_2 \neq 0}$, the solenoidal basis is
\begin{eqnarray}\label{phybase_kn}
\mathbf{u}_{\ell nm}^{(1)}(r) & =(-\rmi(n n_1+\ell n_2)  {\rm g}_m \, , \,
{\rm D}[r {\rm  g}_m] \, , \, 0),\\
\mathbf{u}_{\ell nm}^{(2)}(r) & =(0,-\rmi (n k_1+\ell k_2) r{\rm h}_m
\, , \, \rmi(n n_1+\ell n_2) {\rm h}_m),
\end{eqnarray}
except that the third component of $\mathbf{u}_{\ell nm}^{(2)}$ is
replaced by $ {\rm h}_m$ when $n k_1+\ell k_2=0$. The Petrov-Galerkin
solenoidal weak formulation is completed by introducing the
\textit{Hermitian product} of two arbitrary solenoidal \textit{trial}
and \textit{dual} fields $\vPhi$ and $\vPsi$, respectively, over the
annular-parallelogram domain (\ref{ann_paral_domain})

\begin{equation}
  (\vPsi,\vPhi)=\int_{r_{\rm{i}}}^{r_{\rm{o}}} r\,\rm{d}r
  \int_{0}^{2\pi} \rm{d}\xi
  \int_{0}^{2\pi} \rm{d}\zeta \;
  \vPsi^\dagger \cdot \vPhi.
\label{volint1}  
\end{equation}
Accordingly, we consider the dual basis for the projection space. In
particular, the basis for the case $n n_1+\ell n_2 = 0$ is
\begin{eqnarray}\label{projbase_k0}
\mathbf{\tilde{u}}_{\ell 0m}^{(1)}(r) & = &{\rm w}(0\,,\,r {\rm h}_m\,,\,0),\\
\mathbf{\tilde{u}}_{\ell 0m}^{(2)}(r) & = &r^{-2}{\rm w}
\left(\rmi(n k_1+\ell k_2) {\rm g}_m\,,\,0\,,\, {\rm D}_+{\rm g}_m
+2r^{-1}(1-x^2+rx){\rm h}_m\right),
\end{eqnarray}
where ${\rm D}_+={\rm D}+r^{-1}$, and the third component of
$\mathbf{\tilde{u}}_{\ell 0m}^{(2)}$ is replaced by $r {\rm h}_m$ if $n k_1+\ell k_2=0$.
Similarly, the basis for the case $n n_1+\ell n_2 \neq 0$ is
\begin{eqnarray}\label{projbase_kn}
\mathbf{\tilde{u}}_{\ell n m}^{(1)}(r) & = &{\rm w}
\left((n n_1+\ell n_2)\,r{\rm g}_m \,,\, r{\rm D}_+[r {\rm g}_m]+2xr^2{\rm h}_m \,,\,0\right),\\
\mathbf{\tilde{u}}_{\ell n m}^{(2)}(r) & = & {\rm w}
\left(0\,,\,\ci (n k_1+\ell k_2) r^2{\rm h}_m\,,\,-\ci \,(n n_1+\ell n_2)\,r{\rm h}_m\right).
\end{eqnarray}
These projection basis elements contain the Chebyshev weight function
${\rm w}(x)$ so that the resulting radial integration involved in
\eqref{volint1} can be computed exactly by suitable
quadrature formulas \citep{MoMoLe83,EPJSTMAMM07,CHQZ2007,CHQZ2010,MeBookFNM20}.

Formal substitution of the spectral expansion \eqref{specapx} into
\eqref{hydro_nse}, followed by Hermitian projection onto each one of
the dual basis elements leads to
\begin{equation}
  \left(\vPsi_{\ell n m}^{(\imath)} \;, \; \partial_t{\bf u}_{\rm S}\right) =
  \left(\vPsi_{\ell n m}^{(\imath)} \; , \nabla^2{\bf u}-({\bf v}_{\rm b} \cdot
  \nabla){\bf u}-({\bf u} \cdot \nabla){\bf v}_{\rm b}-({\bf u} \cdot
  \nabla){\bf u} + f\;\hat{\bsy z}\right),\label{MWR}
\end{equation}
for $\ell = -L,\ldots,L,\; n=-N,\ldots,N$, $m = 0,\ldots, M$ and $\imath=1,2$.  The
pressure deviation field drops upon projection, $\left(\vPsi_{\ell n
  m}^{(\imath)},\nabla q\right)=0$, by virtue of Stokes' theorem and has
therefore been omitted. Equation \eqref{MWR}, subject to the
constraint \eqref{INSE:ZERO}, constitutes a dynamical system
\begin{eqnarray}
\mathbb{A}_{\ell nm\imath}^{pqr\jmath} \;  \dot{a}_{pqr}^{(\jmath)} & = & \mathbb{B}_{\ell n m\imath}^{pqr\jmath} \; a_{pqr}^{(\jmath)} +
\mathbb{N}_{\ell nm}^{(\imath)}(\mathbf{a}) + f\,\mathbb{F}_{\ell nm}^{(\imath)} \label{dynsys}\\
Q_m^r a_{00r}^{(2)} & = & 0\label{constr}
\end{eqnarray}
for the axial forcing $f(t)$ and amplitudes $a_{\ell n m}^{(\imath)}(t)$, where
repeated indices must be interpreted following the index summation
convention. The zero-net-massflux constraint \label{r} reduces to a
mere linear equation for the $n=l=0$, $\imath=2$ coefficients upon substitution of
\eqref{specapx}.  The quadratic form $\mathbb{N}_{\ell
  nm}^{(\imath)}(\mathbf{a})$ appearing in (\ref{dynsys}) corresponds to the
projection of the nonlinear convective term, $ \left(\vPsi_{\ell n m}^{(\imath)}
\;, ({\bf u} \cdot \nabla){\bf u}\right)$, that is computed
pseudospectrally using Orszag's $3/2$-dealiasing rule
\citep{CHQZ2007,CHQZ2010}.  Overall, the resulting stiff system of
      {\sc ode} is integrated in time by means of a fourth order
      linearly-implicit Backwards Differentiation scheme with explicit
      polynomial extrapolation of the nonlinear terms, conveniently
      started with a fourth order Runge-Kutta method.

\subsection{Computation and stability analysis of invariant solutions}
In the $(\xi,\zeta)$ coordinate system, a travelling wave is
represented by the spectral expansion
\begin{equation}
  \begin{array}{l}
    {\bf u}_{\rm s}(r,\xi,\zeta;t)=\\
    \displaystyle{=\sum_{\ell=-L}^{L} \sum_{n=-N}^{N} \sum_{m=0}^{M} 
    \left[\check{a}_{\ell nm}^{(1)}
      \vPhi_{\ell nm}^{(1)}(r,\xi,\zeta)+
      \check{a}_{\ell nm}^{(2)}
      \vPhi_{\ell nm}^{(2)}(r,\xi,\zeta)\right]\rme^{\displaystyle \rmi n(\xi-c_\xi
      t)}\rme^{\displaystyle\rmi \ell(\zeta-c_\zeta t)}},
  \end{array}
  \label{specapx_tw}
\end{equation}
where $c_\xi$ and $c_\zeta$ are the unknown wave speed components in the $\xi$
and $\zeta$ directions, respectively, so that the Fourier-Chebyshev
spectral coefficients of this particular type of solution read
\begin{equation}
a_{\ell nm}^{(\imath)}(t) = \check{a}_{\ell nm}^{(\imath)} \rme^{\displaystyle -\rmi n c_\xi
  t}\rme^{\displaystyle - \rmi \ell c_\zeta t},
\label{a_lmn_tw}
\end{equation}
with the complex constant $\check{a}_{\ell nm}^{(\imath)}$ representing the wave
shape unambiguously except for arbitrary rotations and shifts. In this
case, formal introduction of expansion (\ref{specapx_tw}) in
\eqref{hydro_nse} followed by Hermitian projection leads to the system
of nonlinear algebraic equations for the travelling wave coefficients
$\check{a}_{\ell nm}^{(\imath)}$ ($\check{\mathbf{a}}$ for brevity):
\begin{eqnarray}
 \left[\mathbb{B}_{\ell n m \imath}^{pqr\jmath} + \rmi (n c_\xi + \ell c_\zeta)
 \mathbb{A}_{\ell n m \imath}^{pqr\jmath}\right] \check{a}_{pqr}^{(\imath)} +
 \mathbb{N}_{\ell n m}^{(\imath)}(\check{\mathbf{a}}) + \check{f}\,\mathbb{F}_{\ell n m}^{(\imath)} & = & 0,
 \label{nonlina_lmn_tw}\\
 Q_m^r \check{a}_{00r}^{(2)} & = & 0,
\end{eqnarray}
where $\check{\mathbf{a}}$ appearing in the nonlinear term is the
state vector representing the Fourier-Chebyshev coefficients $
\check{a}_{\ell nm}^{(\imath)}$ of the travelling wave, and $\check{f}$ is the axial pressure gradient required to enforce the
zero-net-flux condition. The azimuthal and axial degeneracy of
solutions, associated to drift speeds $c_\xi$ and $c_\zeta$, is
removed using two additional phase constraints in the same way as is
done for the computation of rotating-travelling waves in pipe flow
\citep{MeEck2011}. The system is solved using a \textit{matrix-free}
Newton-Krylov method \citep{Kel03}, implicitly using {\sc gmres} as
matrix-free solver \citep{Trebooknla}. The converged nonlinear
solutions are tracked in parameter space using pseudo-arclength
continuation schemes \citep{Kuz04}.

The linear stability of a travelling wave solution of
(\ref{nonlina_lmn_tw}) is formulated by adding disturbances of very
small amplitude $|\varepsilon_{\ell n m}^{(\imath)}|\ll|\check{a}_{\ell nm}^{(\imath)}|$  to its Fourier-Chebyshev coefficients
following
\begin{eqnarray}
a_{\ell nm}^{(\imath)}(t) & = & \left(\check{a}_{\ell nm}^{(\imath)}+\varepsilon_{\ell n m}^{(\imath)}
\ce^{\sigma t} \right)\rme^{\displaystyle -\rmi n c_\xi
  t}\rme^{\displaystyle - \rmi \ell c_\zeta t},\label{linstab_lmn_tw}\\
f(t) & = & \check{f} + \phi\ce^{\sigma t},
\end{eqnarray}
and the forcing perturbation $|\phi|\ll|\check{f}|$ is such that
coefficient perturbations comply also with the zero-massflux condition
$Q_m^r \varepsilon_{00r}^{(2)}=0$.  Formal substitution of the perturbed
solution (\ref{linstab_lmn_tw}) in (\ref{dynsys}), and subsequent
neglect of
quadratic perturbation terms, leads to the
constrained generalised eigenvalue problem
\begin{eqnarray}
  \left( \sigma -\rmi n c_\xi - \rmi \ell c_\zeta \right)\mathbb{A}_{\ell n m\imath}^{p
    q r\jmath}\varepsilon_{\ell n m}^{(\imath)} & = & \mathbb{B}_{\ell n m\imath}^{p q
    r\jmath}\varepsilon_{\ell n m}^{(\imath)} +
  \mathbf{D}_{\mathbf{a}}\mathbb{N}_{\ell n m\imath}^{p q r\jmath}(\check{\mathbf{a}})\varepsilon_{\ell n m}^{(\imath)} +
  \phi\,\mathbb{F}_{\ell n m}^{(\imath)},\label{linstab_lmn_tw_eigprob}\\
  Q_m^r \varepsilon_{00r}^{(2)} & = & 0
\end{eqnarray}
where $\mathbf{D}_{\mathbf{a}}\mathbb{N}_{\ell n m\imath}^{p q r\jmath}(\check{\mathbf{a}})$ is the linear action of the Jacobian of $\mathbb{N}$ evaluated
at the travelling wave solution $\check{\mathbf{a}}$, and $(p,q,r,\jmath) \in
[0,L]\times[-N,N]\times[0,M]\times\{1,2\}$. This linear action therefore allows to
formulate the generalised eigenvalue problem
(\ref{linstab_lmn_tw_eigprob}) in a matrix-free form, so that Arnoldi
methods (\cite{Trebooknla}) can be applied to compute the leading
eigenvalues $\sigma_j$ and their associated eigenvectors
$\mathbf{\varepsilon}_j=\{\varepsilon_{\ell m n}^{(\imath)}\}_j$.

To compute relative periodic orbits (i.e., modulated travelling waves)
beyond their region of linear stability, a Poincar\'e-Newton-Krylov
method is devised. The method is essentially an adaptation of the one
used for the computation of modulated travelling waves in plane
Poiseuille flow \citep{MeMe2015}. In this case, the method solves the
nonlinear system of equations resulting from root finding for the map
defined by consecutive crossings of a Poincar\'e section $\mathbb{P}$:
\begin{equation}
  \mathbf{a} \rightarrow \widetilde{\mathbf{a}} =
  \mathbb{P}(\mathbf{a})= \mathbb{T}(\Delta \xi,\Delta \zeta)
  \bm{\varphi}(\mathbf{a};\mathrm{T}),
\label{pnk_1}
\end{equation}
where $\bm{\varphi}(\cdot;t)$ is the action of the uniparametric group
or \textit{flow} generated by (\ref{dynsys}), $\mathrm{T}$ is the
modulation period of the relative periodic orbit, and
\begin{equation}
\left[\mathbb{T}(\Delta \xi,\Delta \zeta)\mathbf{a}\right]_{\ell n m}^{(\imath)}
= \rme^{\displaystyle -\rmi n \Delta \xi }\rme^{\displaystyle -\rmi
  \ell \Delta \zeta}a_{\ell n m}^{(\imath)}
\label{pnk_Tshift}
\end{equation}
is a double-shift operator, removing the drift of the relative periodic orbit in the
two homogeneous parallelogram coordinates $\xi$ and $\zeta$.  Finally,
the above described time-stepping formulation, as well as the
travelling wave and relative periodic orbit solvers, are enforced to
satisfy zero net flux condition for the perturbation field.

\section{Choice of geometrical and physical parameters}
\label{sec_parameter}

The present study is done for the same cylinder radius ratio of
$\eta=0.883$ as employed by \cite{ALS86} in their Taylor-Couette
apparatus. The outer Reynolds number is fixed to $\Ro=-1200$, at which
\cite{MeMeAvMa09_A} found numerically that {\sc spt} is sustained
within sufficiently large domains for values of the inner cylinder
Reynolds number ranging in $\Ri \in [540, 640]$ (see line $\Gamma_2$
in figure\,3 of that paper).
Below the lower threshold for {\sc spt}, spot-like intermittency is
observed for an interval, bordering on the other side with a regime of
interpenetrating-spirals that prevails in the range $\Ri \in [450,
  480]$.

The critical point above which circular Couette flow becomes linearly
unstable with respect to non-axisymmetric perturbations occurs at
$\Ri=447.35$. The bifurcating nonlinear spiral-wave solution branches
found by \citet{MeMeAvMa09_B} can only be continued down to
$\Ri=445.65$, which makes them qualify as only mildly
subcritical. While none of the mixed-mode solutions detected by
\citet{DeAlt2013} reached this value of $\Ri$, the subcritical rotating waves
computed by \citet{DeMeMe14} do indeed exist in a much wider region of the linearly stable regime, extending to as low as $\Ri>377.3$.

\begin{figure}           
\begin{center}
\begin{tabular}{cc}
  (a) & (c) \\
   \raisebox{-1.8cm}{\includegraphics[width=.475\linewidth]{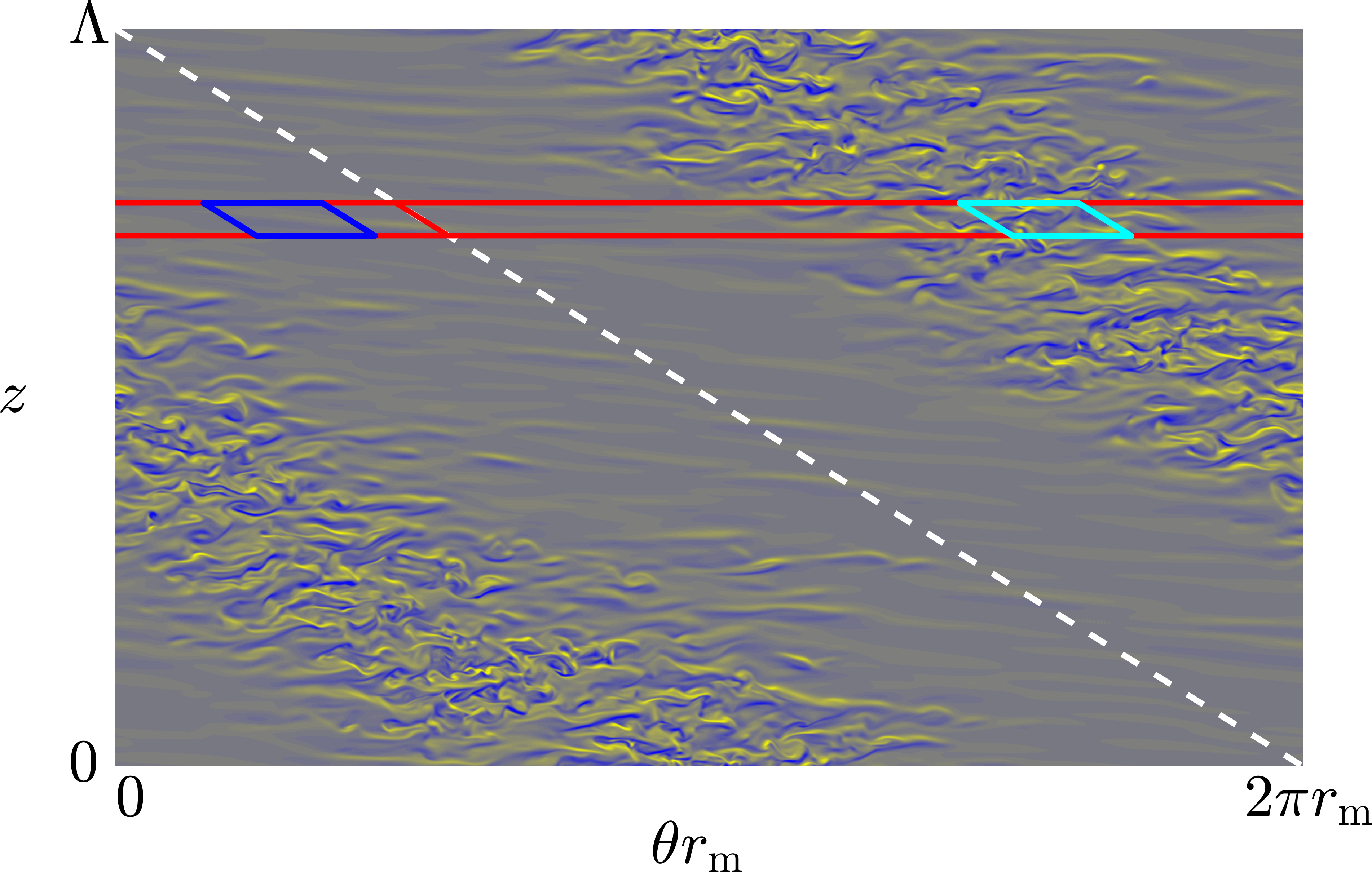}}&
  \begin{tabular}{l}
    \raisebox{1cm}{\includegraphics[width=.4\linewidth]{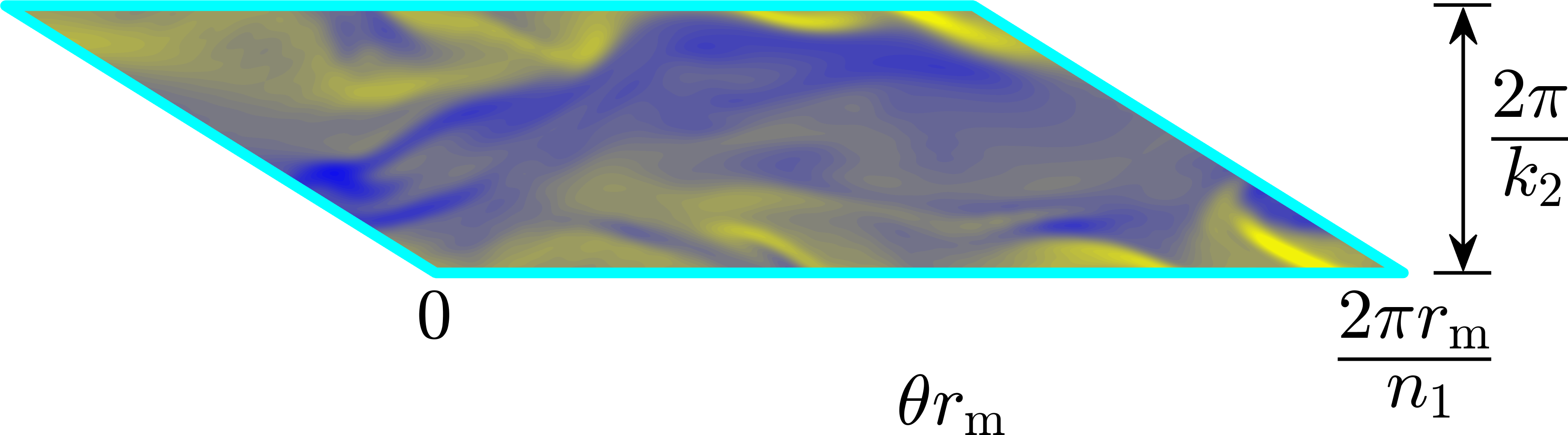}}  \\[-2em]
    \raisebox{0cm}{\includegraphics[width=.4\linewidth]{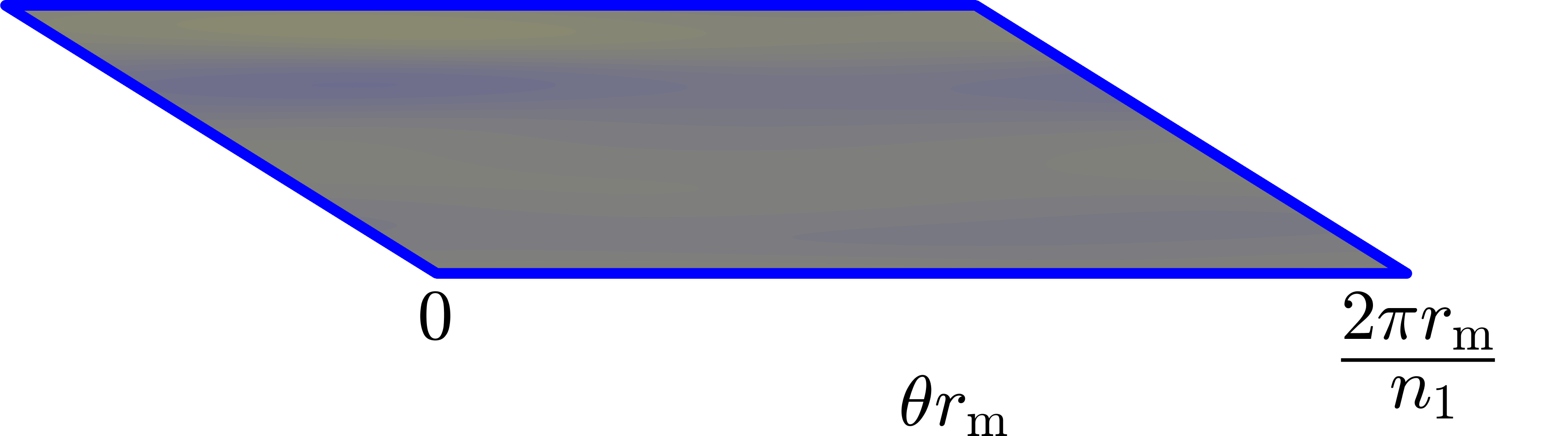}}     \end{tabular}
\end{tabular}\\
(b) \\[0.5em]
\includegraphics[width=0.9\linewidth]{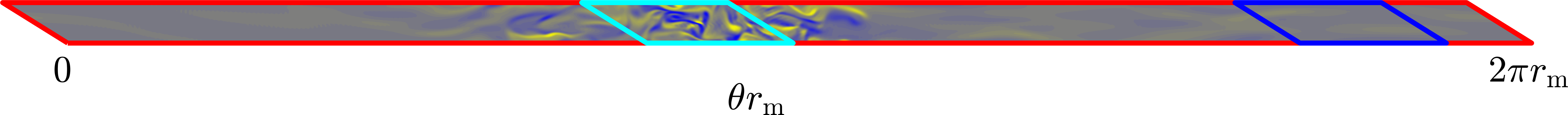}
\end{center}
\caption{Instantaneous fields of statistically steady states obtained
  with {\sc dns} for $(\eta,\Ro,\Ri)=(0.883,-1200,600)$. All pannels
  show radial vorticity $\omega_r\in[-4000,4000]$ colourmaps at the
  intermediate radius $r_{\rm m} = (r_{\rm{i}}+r_{\rm{o}})/2 \approx
  8.05$.  (a) {\sc spt} in the full orthogonal domain of periodic
  aspect ratio $\Lambda=31.4$, with spectral resolution $(L,N,M) =
  (322,322,42)$ (See online movie 1). The white dashed line indicates the tilt of the
  stripe and of the minimal narrow domain that can capture {\sc spt}
  (solid red enclosure).  (b) {\sc spt} in a narrow long parallelogram
  domain $(n_1,k_1,n_2,k_2)=(1,0.2,0,4.5)$ with resolution $(L,N,M) =
  (18,322,42)$ (See online movie 2). The solid cyan parallelogram delimits the minimal
  domain required for self-sustained turbulent dynamics and used
  throughout the paper. (c) Turbulent (top) and laminar (bottom)
  snapshots at different time instants along the same simulation
  within a narrow and short parallelogram domain
  $(n_1,k_1,n_2,k_2)=(10,2,0,4.5)$, using $(L,N,M) = (18,32,42)$
  spectral modes (See online movie 3).}
  \label{figdomain}
\end{figure}

Figure\,\ref{figdomain}a shows radial vorticity, ${\omega_r = (\nabla
  \times \mathbf{v})_r}$, colour maps for a snapshot of the
statistically steady {\sc spt} obtained at $\rm{R}_{\rm{i}}=600$ in a
sufficiently large computational domain following
\citet{MeMeAvMa09_A}.
The unwrapped domain is a rectangle in the $\theta$-$z$ plane as
usually employed in {\sc spt} calculations, and its height-to-gap
aspect ratio is $\Lambda \approx 31.4$ (corresponding to an axial
wavenumber $k = 2\pi/\Lambda = 0.2$).
Within the turbulent stripe, wavy vortices of relatively short axial
and azimuthal wavelength emerge. These small-scale flow structures
seem to decorrelate rather fast along the spiral slope direction.
This statistical property of {\sc spt} suggests that a narrow
periodic domain winding horizontally around the full apparatus gap but
sheared to preserve the tilt of the spiral (dashed white line in panel
a) would in principle suffice to capture minimally its main
topological and dynamical features.

In fact, a localised turbulent stripe arises when {\sc dns} is
performed in an azimuthally long but axially narrow
parallelogram domain such as shown in figure\,\ref{figdomain}b, whose
comparative size is indicated in figure\,\ref{figdomain}a as the
enclosure bound by a red line. The corresponding generalised
wavenumbers for this domain (see figure \ref{fig:parallel_scheme})
were chosen as follows. First we note that the pitch of {\sc spt} is
enforced by the dimensions of the full domain through the ratio
$k/n=2\pi/\Lambda$. A periodic parallelogram domain that can sustain
spirals of the same pitch must have the slope of two of its sides
prescribed by azimuthal and axial wavenumbers keeping the same ratio
$k_1/n_1=k/n=0.2$.  Picking $n_1=1$, and therefore
$(n_1,k_1)=(1,0.2)$, the parallelogram domain winds exactly once
around the apparatus circumference, with its two opposing sides that
are aligned with the slope of the spiral falling on the same helical
surface. The other pair of sides, defined by the values $(n_2,k_2)$,
can be chosen arbitrarily at this stage. For simplicity we take
$n_2=0$, thus forcing the parallelogram to extend horizontally and
align with the azimuthal coordinate $\xi\equiv\theta$.  Notice that
the axial height of the domain is then simply specified as
$2\pi/k_2$. Inspecting the typical size of flow structures in
figure\,\ref{figdomain}a, shows that $k_2=4.5$ is an adequate choice
for the narrow domain size to fit exactly one single streak most of
the time.
  
We note here that in the full domain, the symmetry of the system in
the axial direction allows for a turbulent spiral band structure with
an opposite tilt. However, in the narrow but long parallelogram
domain, the symmetry is broken. The choice of the wavenumber ratio
$k_1/n_1=0.2$ makes the realisation of the opposite spiral impossible,
which could be computed using instead the wavenumber ratio $k_1/n_1=-0.2$.
  
Finally, if the focus is to be placed exclusively on the small-scale
flow structures that exist within the core of {\sc spt} rather than on
the large-scale interactions that result in intermittency and
laminar-turbulent interfaces, but want to keep at the same time the
potential influence of the characteristic tilt of the spiral pattern,
the long domain  of figure\,\ref{figdomain}b can be further reduced by
shortening it along the azimuthal direction into the small parallelogram
enclosed by the solid cyan/blue line.
The two snapshots of figure\,\ref{figdomain}c illustrate the dynamics in the small parallelogram domain. The flow is temporally chaotic and alternates rather quiescent, spatially-coherent and smoothly evolving {\it laminar} phases (bottom), with sudden bursts of violent, spatially-disordered and rapidly fluctuating (i.e. spatiotemporally chaotic)  {\it turbulent} transients (top). The former periods are seemingly representative of the laminar regions of {\sc spt}, while the latter display characterisitic features typical of the vortical flow structures found in the core of turbulent spirals.
%
Here we have chosen to shorten the coiling length of the long domain
by setting an azimuthal wavenumber $n_1=10$ following
\citet{DeMeMe14}, such that the first pair of wave numbers is replaced
by $(n_1,k_1)=(10,2.0)$ and a periodical tiling of the parallelogram
along the coil fits exactly 10 times.

It is precisely this minimal domain of figure\,\ref{figdomain}c, with
$k_2=4.5$, that we employ in studying the {\sc ecs} in
\S\ref{sec_subcritrow}, although some continuation in the $k_2$
parameter has also been performed to ellucidate the complex
bifurcation scenario of {\sc drw}. The {\sc dns} results shown in
figure\,\ref{figdomain}b,c will be discussed in
\S\ref{sec_stabdynflows}.

\section{Bifurcation scenarios of (relative) equilibria}
\label{sec_subcritrow}

\begin{figure}                                                                 
 \begin{center}
  \begin{tabular}{c}
   \includegraphics[width=.65\linewidth]{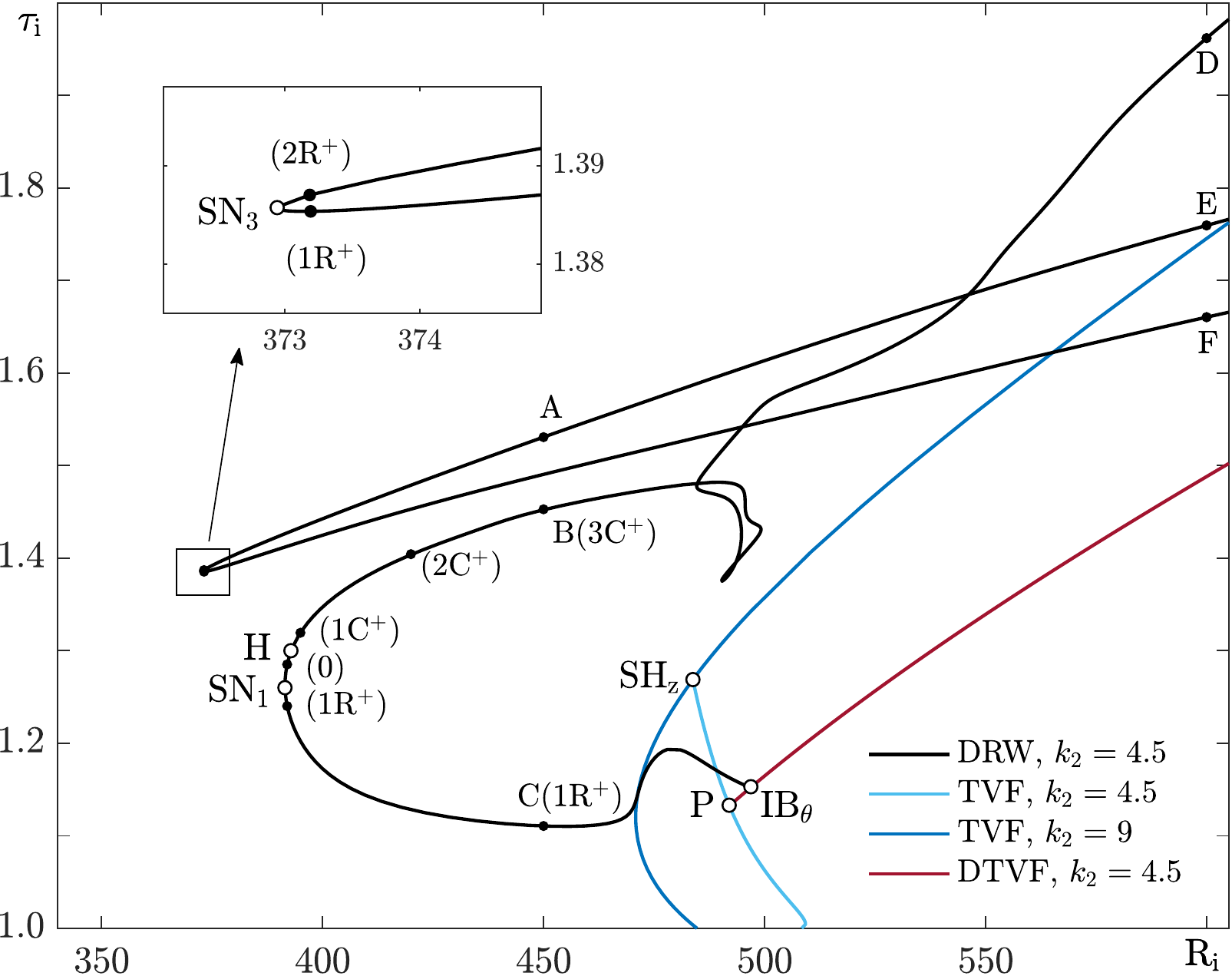} 
   \end{tabular}
  \end{center}                                                                 
\caption{Bifurcation diagram of {\sc ecs} in the small box of
  figure~\ref{figdomain}c. Shown are solution branches of stationary
  ({\sc tvf}, blue) and drifting ({\sc dtvf}, red) Taylor vortex flow,
  as well as of drifting-rotating waves ({\sc drw}, black).  White
  bullets denote saddle-node ({\sc sn}), pitchfork ({\sc p}), Hopf
  ({\sc h}), azimuthal invariance breaking ({\sc ib}$_{\rm \theta}$)
  and axial-modulational subharmonic ({\sc sh}$_{\rm z}$) bifurcation
  points.  The number of unstable real eignenvalues (${\rm R}^{+}$)
  and complex conjugate pairs (${\rm C}^{+}$) of some solutions along
  the branches are given in parentheses. }
    \label{stab_analysis_subrw4p5}   
\end{figure}

Figure \ref{stab_analysis_subrw4p5} summarises the collection of {\sc
  ecs} that we have managed to capture in the 
small parallelogram domain described above (see figure~\ref{figdomain}c), including
stationary, drifting and rotating-drifting wave solutions.
Solution branches have been tracked via Newton-Krylov arclength
continuation. The spectral resolutions used lie within the range
$(L,N,M) \in [8,16]\times[8,16]\times[24,50]$, always ensuring that no
further increase resulted in qualitative or significant quantitative
variation as to flow properties or bifurcation scenarios.

Of particular interest in figure\,\ref{stab_analysis_subrw4p5} is a
family of three-dimensional travelling-wave solutions that have
non-zero phase speeds in both the azimuthal and axial directions and
will henceforth be referred to as \textit{Drifting Rotating Waves}
({\sc drw}, black). Continuation reveals that {\sc drw} branches are
remarkably subcritical, with the corresponding saddle-node points
({\sc sn}$_1$ and {\sc sn}$_3$) located at inner Reynolds numbers
$\Ri^{\rm SN_1}=391.5$ and $\Ri^{\rm SN_3}=372.9$ preceding by far the
linear instability of the base flow ({\sc ccf}) at $\Ri=447.35$.  We
will see later that, by tuning the geometrical parameters that define
the domain size and shape, the {\sc drw} solution can be shown to
essentially bifurcate from Taylor Vortex Flow ({\sc tvf}), which in
turn bifurcates from {\sc ccf}, a multi-stage bifurcation scenario
that was already reported by \cite{DeMeMe14}. In our particular choice
of domain, however, the bifurcation details are more involved.

The {\sc tvf} solution branch (dark blue curve) bifurcates
subcritically from {\sc ccf} for $k_2=9$ following an axisymmetric
centrifugal instability at $\Ri=484.7$. The branch evolves for a short
$\Ri$-range and turns in a saddle-node at $\Ri=470.8$ before
undergoing an axial-modulational subharmonic bifurcation ({\sc
  sh}$_{\rm z}$) at $\Ri=483.8$, whence another branch of {\sc tvf}
(light blue), of fundamental wave number $k_2=4.5$, is
issued. Systematical Arnoldi linear stability analysis along the {\sc
  tvf} branches shows that the $k_2=9$ solution stabilises briefly
across the saddle-node, and that this stability is transfered to the
$k_2=4.5$ solution at the supercritical {\sc sh}$_{\rm z}$
point. The $k_2=4.5$ {\sc tvf} solution, which bifurcates at
$\Ri=508.4$ from {\sc ccf} at the other end, exhibits a pichfork
bifurcation ({\sc p}) half way along the branch ($\Ri^{\rm P}=492.1$)
that breaks the Z$_2$ axial reflection symmetry and produces a pair of
mutually-symmetric branches of axially-Drifting Taylor Vortex Flow
solutions ({\sc dtvf}, red). The {\sc dtvf} branch soon undergoes an
azimuthal invariance breaking bifurcation ({\sc ib}$_{\rm \theta}$) that
generates the highly subcritical {\sc drw} (black). The {\sc ib}$_{\rm
  \theta}$ bifurcation is formally a Hopf bifurcation, as a complex
conjugate pair of eigenvalues crosses into the positive-real-half of
the complex plane, but it introduces no dynamics other than the trivial drift along the group orbit corresponding to axial shifts.

\begin{figure}                                                                 
 \begin{center}
   \begin{tabular}{cc}
     \parbox{1cm}{\raggedleft {\sc tvf}\\at {\sc sh}$_{\rm z}$}\raisebox{-1.5cm}{\includegraphics[width=.3\linewidth]{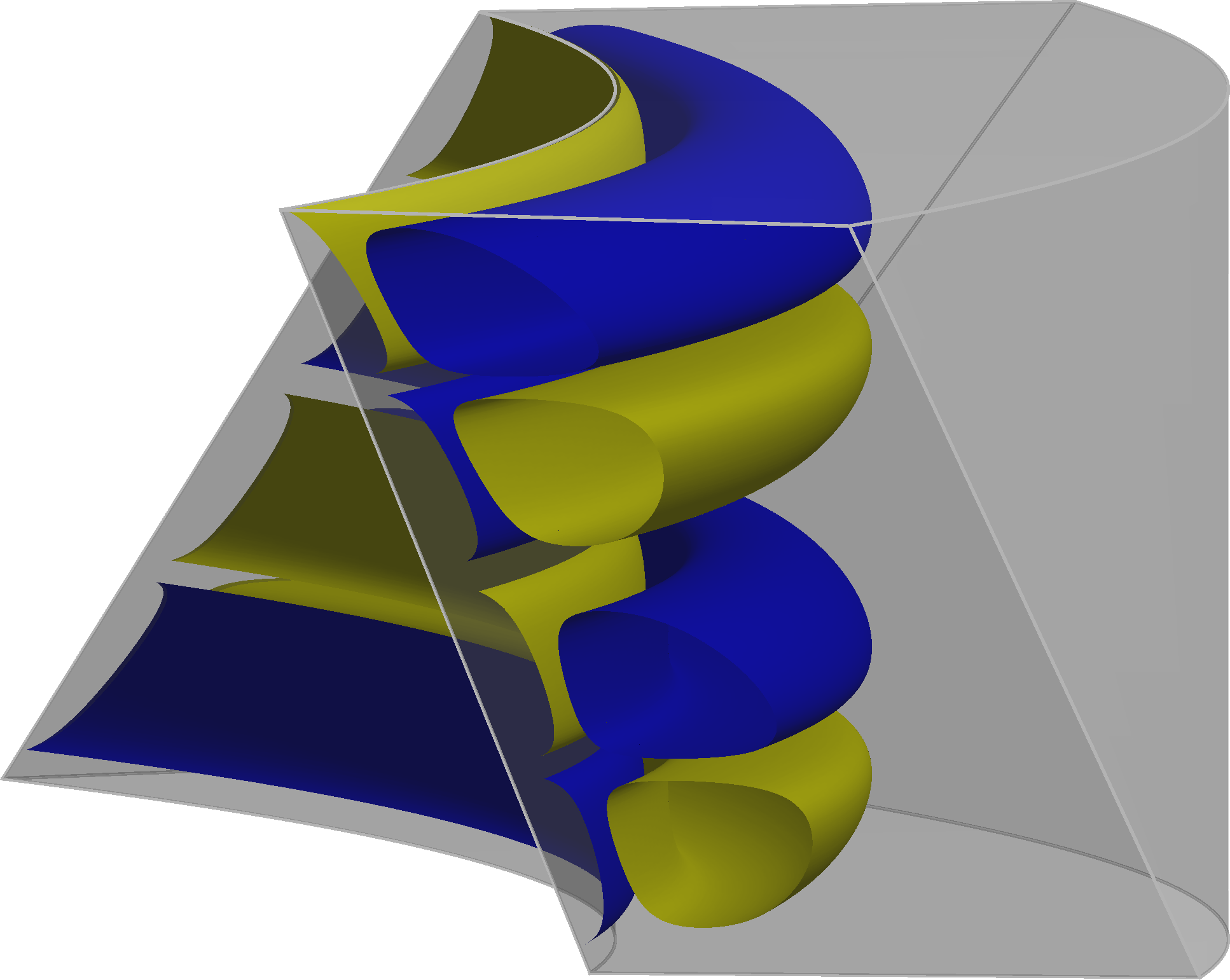}}&
     \parbox{1cm}{\raggedleft {\sc tvf}\\at {\sc p}}\raisebox{-1.5cm}{\includegraphics[width=.3\linewidth]{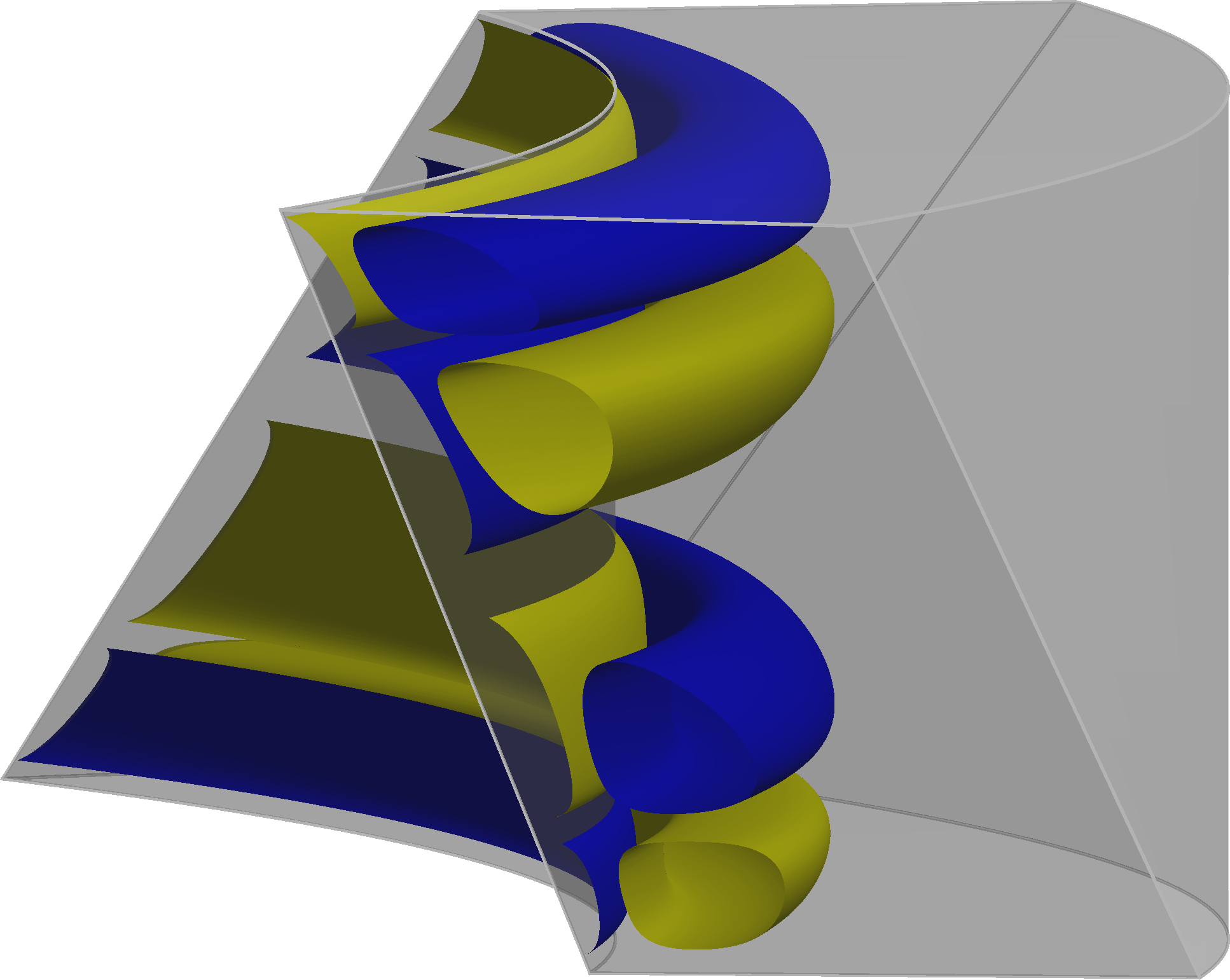}}\\
     \parbox{1cm}{\raggedleft {\sc dtvf}\\at {\sc ib}$\rm _\theta$}\raisebox{-1.5cm}{\includegraphics[width=.3\linewidth]{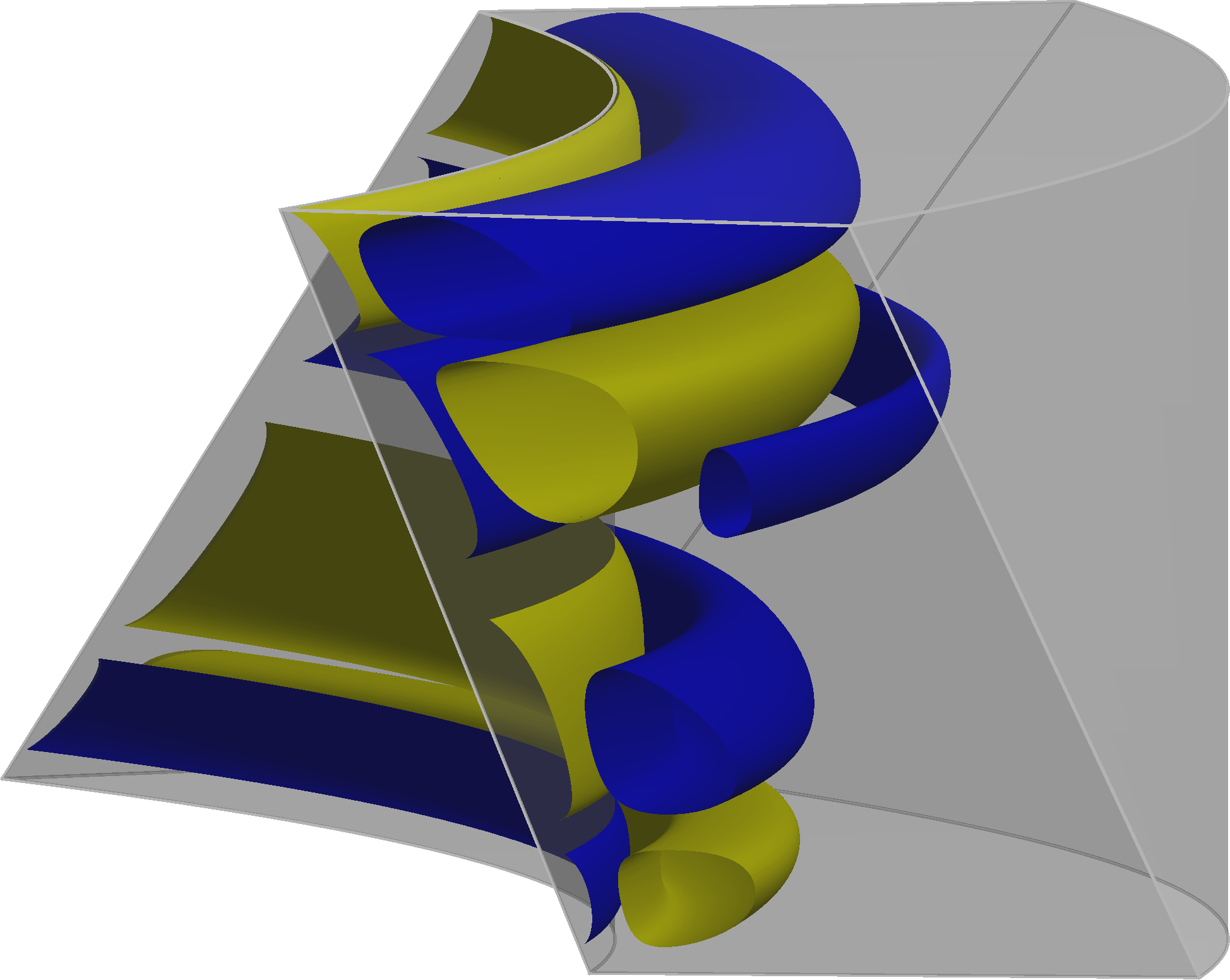}}&
     \parbox{1cm}{\raggedleft {\sc drw}\\at {\sc sn}$_1$}\raisebox{-1.5cm}{\includegraphics[width=.3\linewidth]{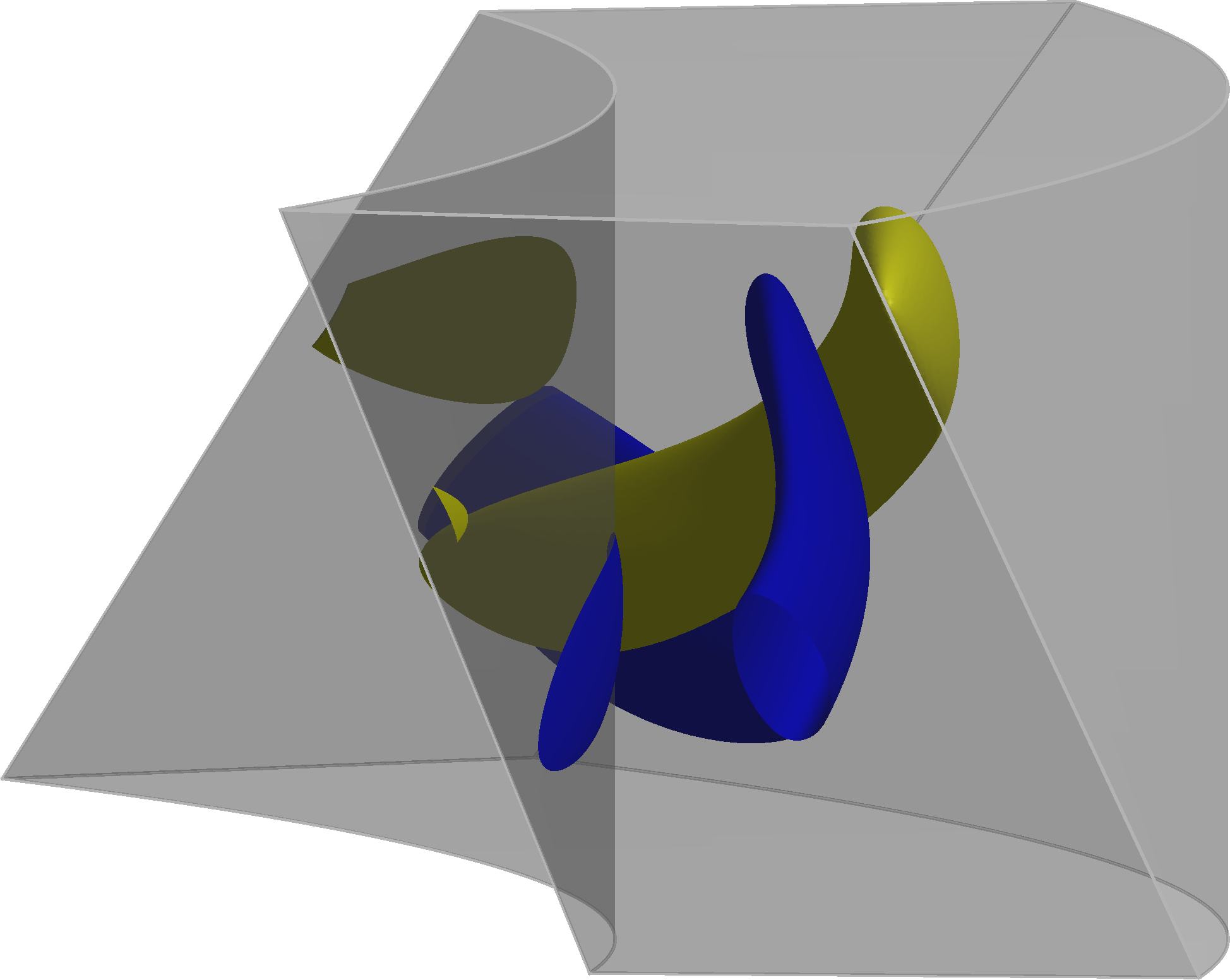}}\\
   \end{tabular}
 \end{center}                                                                 
\caption{Distinct types of {\sc ecs} at specific points labelled in
  figure\,\ref{stab_analysis_subrw4p5}. Shown are positive (yellow)
  and negative (blue) isosurfaces of azimuthal vorticity at
  $\omega_\theta=\pm100$ for {\sc tvf} at {\sc sh}$\rm_z$, {\sc tvf} at {\sc
    p} and {\sc dtvf} at {\sc ib}$\rm_\theta$, and at
  $\omega_\theta=\pm600$ for {\sc drw} at {\sc sn}$_1$.}
  \label{fig_3dplots3}   
\end{figure}

The flow topology of the $k_2=9$ and $k_2=4.5$ {\sc tvf} solutions is
illustrated in figure\,\ref{fig_3dplots3} at points {\sc sh}$_{\rm z}$ and {\sc
  p}, respectively, through a couple of azimuthal vorticity
($\omega_\theta$) iso-surfaces.
They both possess a reflectional symmetry in the axial direction that
blocks all possibility of axial drift, which, in combination with
azimuthal invariance, keeps them stationary. Although the $k_2=4.5$
{\sc tvf} preserves the vortical arrangement of the $k_2=9$ solution,
contiguous vortex pairs are no longer identical as a consequence of the two-fold axial modulation enacted by the {\sc sh}$_{\rm z}$ bifurcation. The reflectional symmetry is nonetheless preserved and the solutions remain stationary.

The symmetry is finally broken at the pitchfork point
{\sc p}, such that the resulting solution drifts axially, as expected
from bifurcation theory in the presence of symmetries
\citep{ChoIoo94}. Azimuthal invariance is preserved and the flow
structure is still very much alike that of Taylor vortices, which
accounts for its being dubbed \textit{Drifting}-{\sc tvf} ({\sc
  dtvf}). The final breaking of the mirror symmetry is clear from the
$\omega_\theta$ isosurfaces of figure~\ref{fig_3dplots3} for {\sc
  dtvf}.

All these axisymmetric flows are characterised by the presence of
vortical structures in the vicinity of the inner cylinder, the reason
being that the centrifugal instability is confined to $r<r_{\rm n}=\sqrt{-B/A}$, according to Rayleigh's
stability condition. 
Here $r_{\rm n}$ is the nodal radius, where $A$ and $B$ are the constants in the base flow (\ref{defccf}).
On the other hand, the three-dimensional
structures of the {\sc drw}, also depicted in
figure~\ref{fig_3dplots3} at {\sc sn}$_1$, exhibit remarkably
different properties that will be discussed in the next section.

\begin{figure}                                               
  \begin{center}
    \begin{tabular}{llll}
      \raisebox{.3\linewidth}{(a)} &
      \multicolumn{3}{c}{\hspace{-1.8em}\includegraphics[width=.86\linewidth]{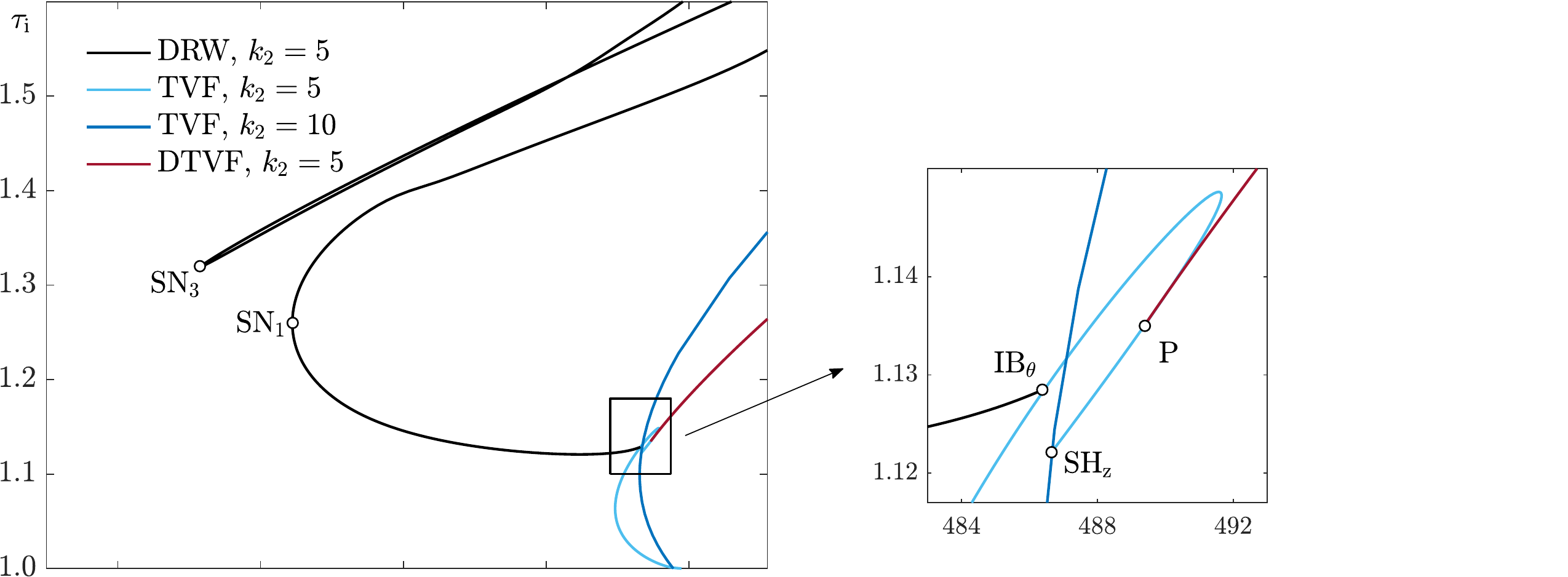}} \\
      \raisebox{.3\linewidth}{(b)} &
      \includegraphics[width=.42\linewidth]{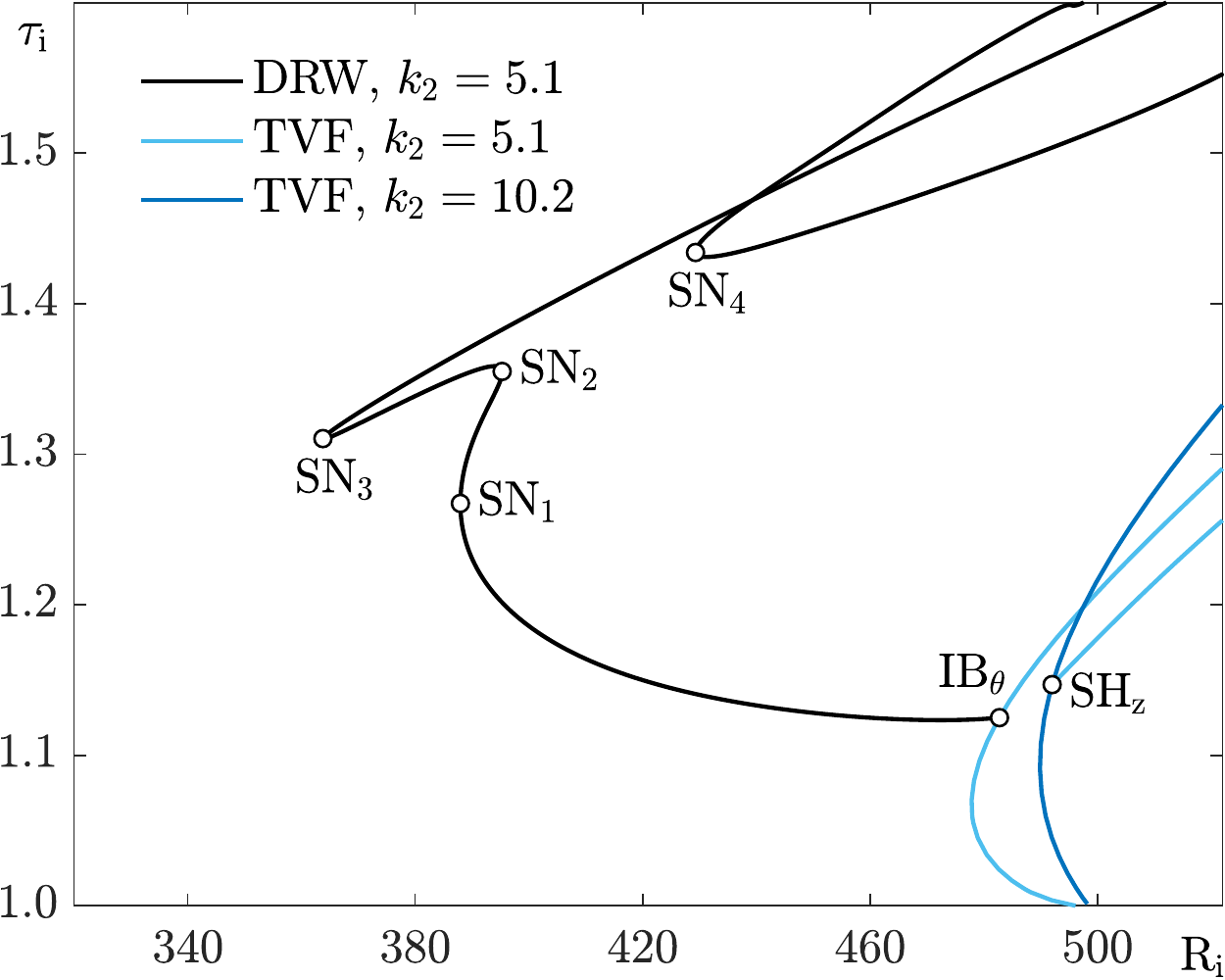} &
      \raisebox{.3\linewidth}{(c)} & 
      \includegraphics[width=.42\linewidth]{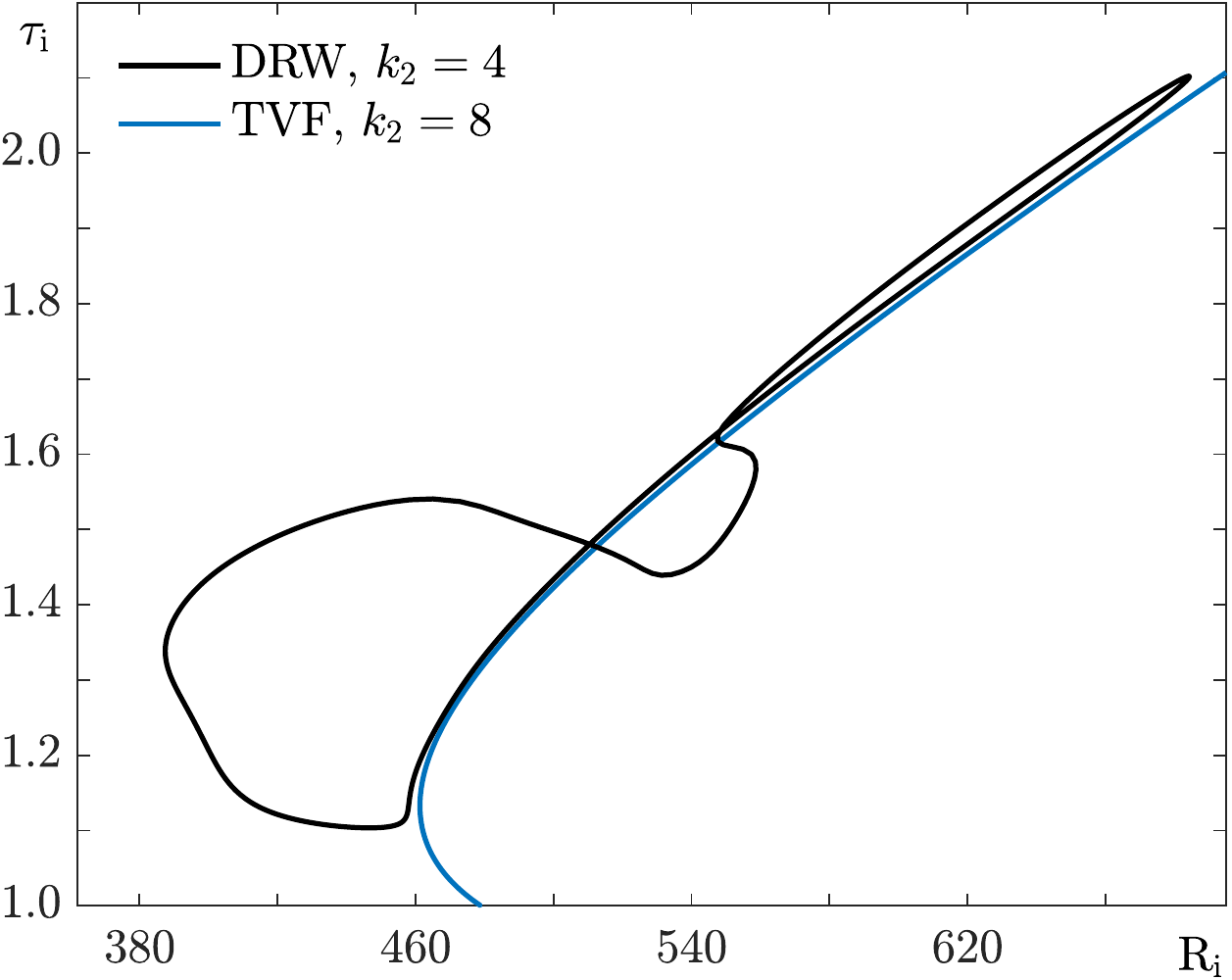}
  \end{tabular}
\end{center}
\caption{Bifurcation diagram of {\sc ecs} for (a) $k_2=5$, (b)
  $k_2=5.1$, and (c) $k_2=4$. Colour coding and labels as for figure
  \ref{stab_analysis_subrw4p5}.  }
  \label{fig_continuations}
\end{figure}

A second, seemingly unrelated branch of {\sc drw} is laid out in
figure\,\ref{stab_analysis_subrw4p5} that bifurcates in a saddle-node
{\sc sn}$_3$ at $\Ri^{\rm SN_3}=372.9$. The two apparently
disconnected families of {\sc drw} are, in point of fact, one and the
same, as evinced by continuation in the $k_2$ parameter in
figure~\ref{fig_continuations}.
Increasing the axial wavelength to $k_2=5$
(figure~\ref{fig_continuations}a) the lower-torque branch of the {\sc
  sn}$_3$-related wave approaches the higher-torque branch of the {\sc
  sn}$_1$-related solution. At slightly higher $k_2$ the two branches
collide in a straightforward codimension-2 double-zero bifurcation
point, where each splits in two separate segments that are spliced in
a new arrangement. As a result, two mutually-facing saddle-node points
arise, {\sc sn}$_2$ and {\sc sn}$_4$, which are illustrated for
$k_2=5.1$ in figure~\ref{fig_continuations}b. This type of structural
instability of pairs of saddle-nodes with respect to small changes in
the size of the domain has also been identified in other shear flows
\citep{DeNa2011,MeMe2015,AyMeMe20}. The new branch containing {\sc
  sn}$_4$ becomes disconnected, while the other one exhibits a triplet
of saddle-nodes and connects directly to $k_2=5.1$ {\sc tvf} at the
azimuthal invariance breaking point {\sc ib}$_{\rm \theta}$. The
transfer of {\sc ib}$_{\rm \theta}$ from the {\sc dtvf} to the {\sc
  tvf} branch, which follows a codimension-2 Hopf-pitchfork
bifurcation, has in fact already been effected at $k_2=5$, as is clear
from the inset of figure~\ref{fig_continuations}a. In fact, the
subharmonic {\sc tvf} branch (light blue) has outdone, in terms of
subcriticality, the {\sc tvf} branch from which it bifurcates (dark
blue). And to complicate things further, the branch has also bent to
include a second saddle-node that moves fast towards higher $\Ri$ as
$k_2$ is increased.



Another interesting phenomenon also occurs when $k_2$ of {\sc drw} is
reduced from 4.5.  The branch associated with {\sc sn}$_1$ disconnects
from {\sc dtvf} and becomes an isola as illustrated in
figure\,\ref{fig_continuations}c for $k_2=4$ (black line). In
addition, the Reynolds number at which {\sc sn}$_3$ occurs increases,
making it disappear from the range of the diagram.



We conclude from the above parametric study that although the topological
arrangement of solution branches is rather sensitive to small changes
in $k_2$, their amply subcritical nature is a robust feature of
{\sc drw} solutions.

\section{Structure and stability of DRW, and their role in the dynamics}
\label{sec_stabdynflows}


We shall see presently some evidence that {\sc drw} solutions appear to play a central role in organising the flow dynamics both in the subcritical and supercritical
regimes of counter-rotating Taylor-Couette flow. 
In \S 5.1 we discuss the particular flow structure of {\sc drw} and relate it
 to the {\sc ssp}. We then analyse in \S 5.2 the stability of the waves and their contribution to
the formation of a chaotic set in the subcritical regime. 
Finally, we inspect in \S 5.3 the early supercritical regime in order to gauge the part {\sc
  drw} solutions play in driving supercritical turbulent dynamics.

\subsection{Flow structure of {\sc drw}}

\begin{figure}                                                                 
 \begin{center}
   \begin{tabular}{ccc}
     & (a) & (b)\\
     \parbox{1cm}{\raggedleft {\sc drw}\\at {\sc a}} &
     \raisebox{-2em}{\includegraphics[width=.42\linewidth]{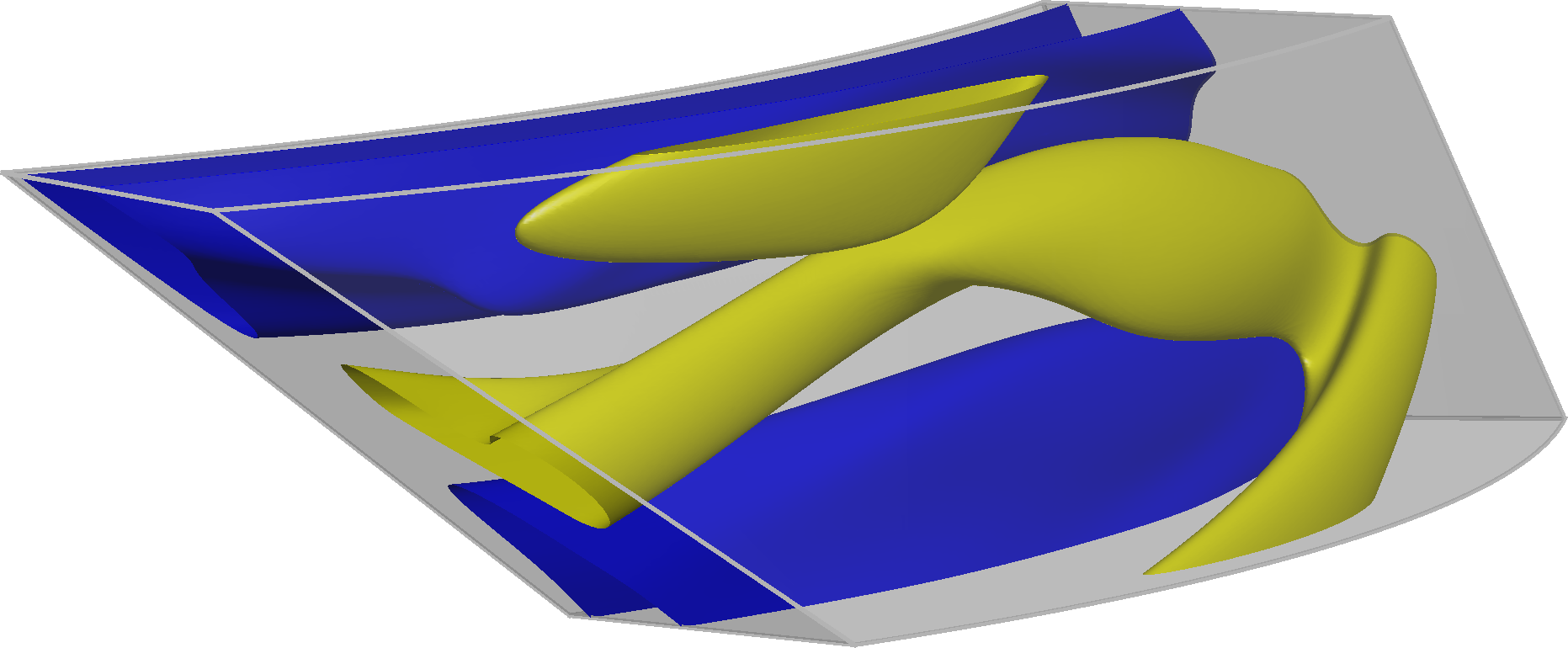}} &
     \raisebox{-2em}{\includegraphics[width=.42\linewidth]{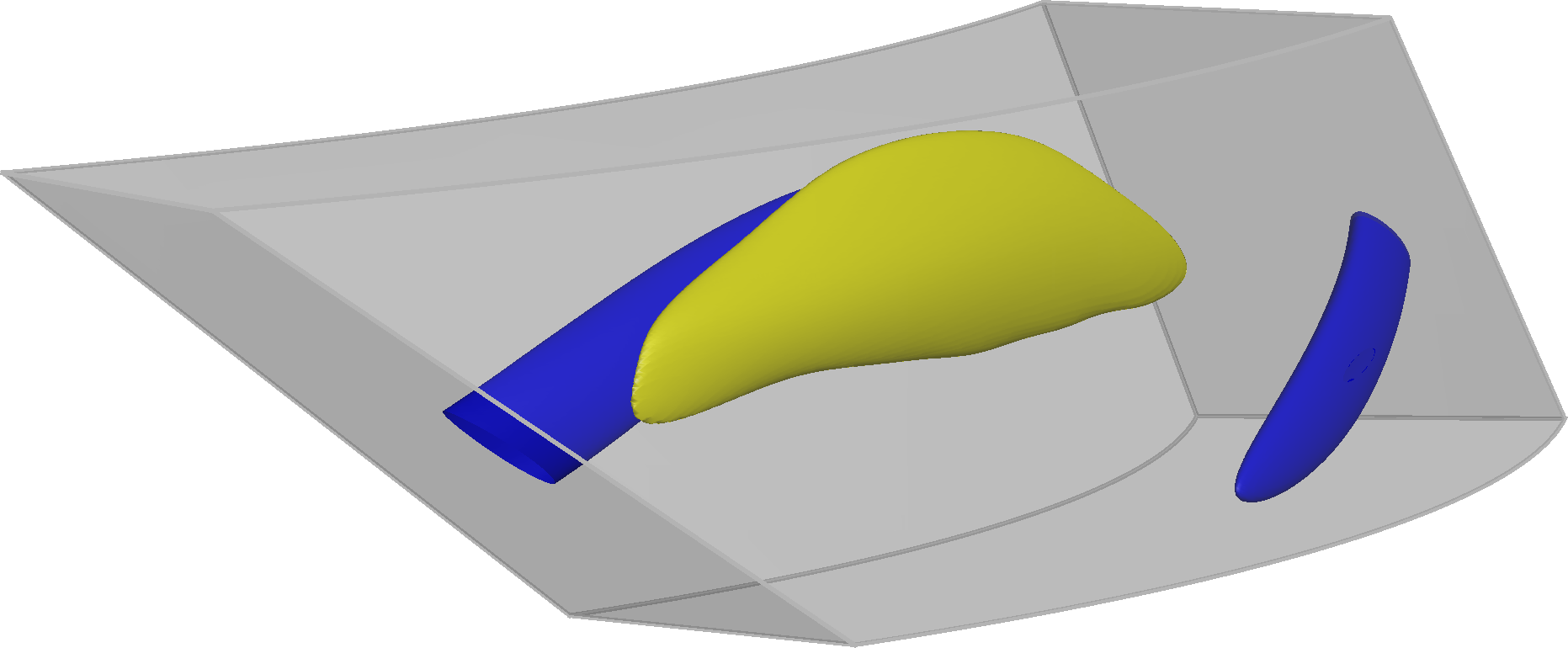}} \\
     \parbox{1cm}{\raggedleft {\sc drw}\\at {\sc b}} &
     \raisebox{-2em}{\includegraphics[width=.42\linewidth]{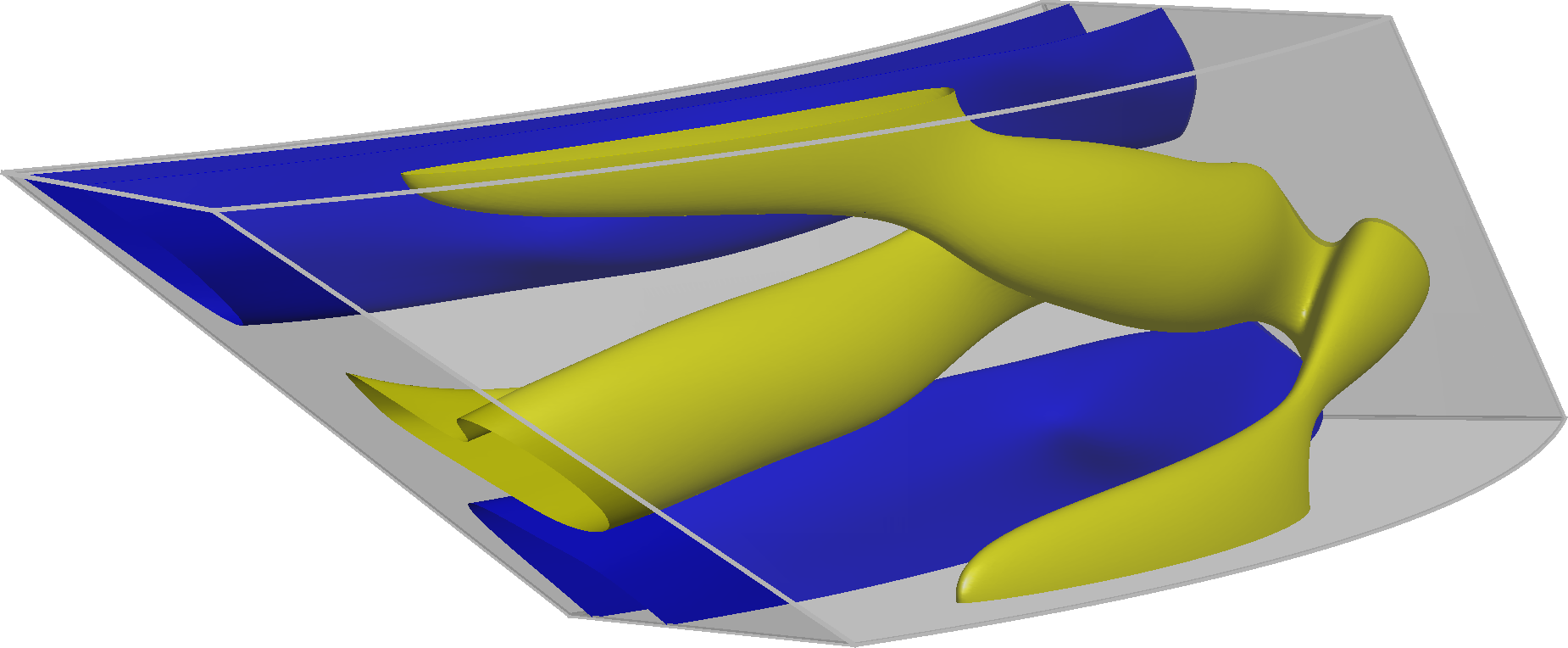}} &
     \raisebox{-2em}{\includegraphics[width=.42\linewidth]{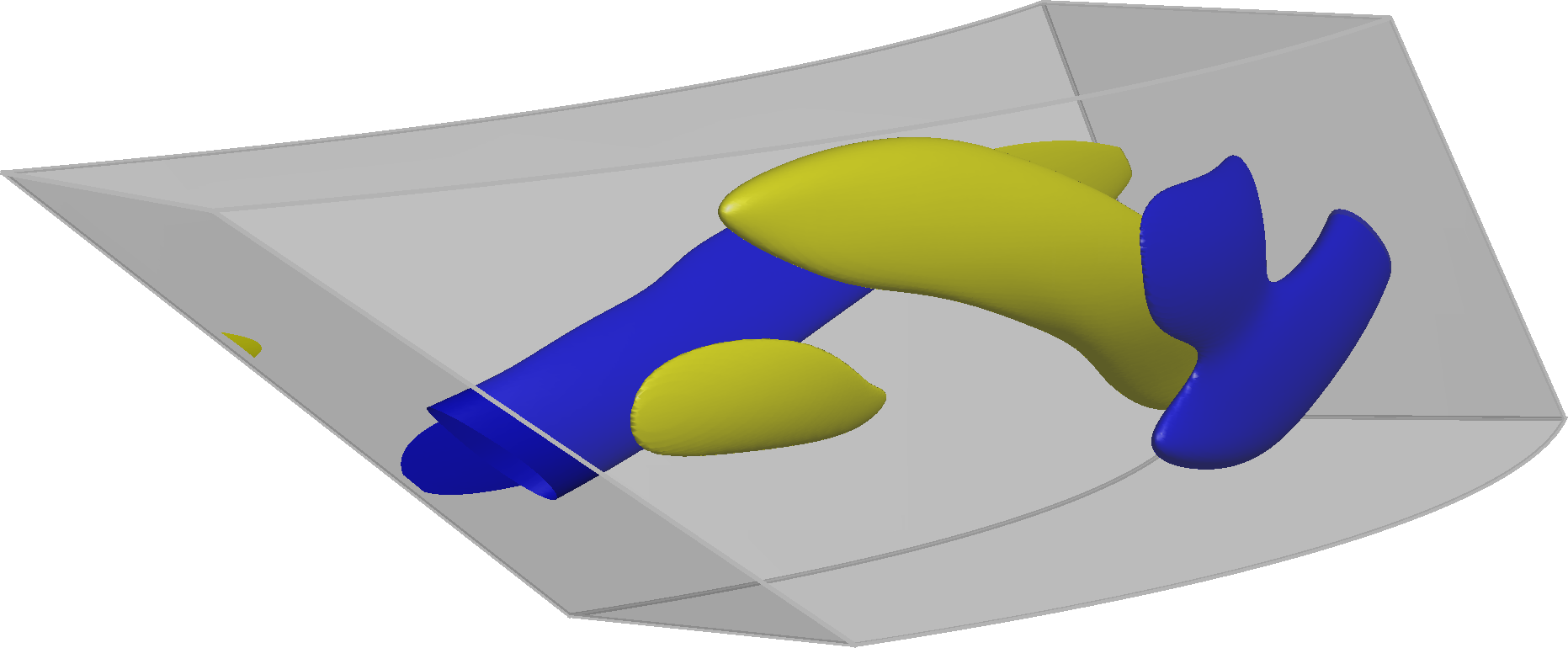}}\\
     \parbox{1cm}{\raggedleft {\sc drw}\\at {\sc c}} &
     \raisebox{-2em}{\includegraphics[width=.42\linewidth]{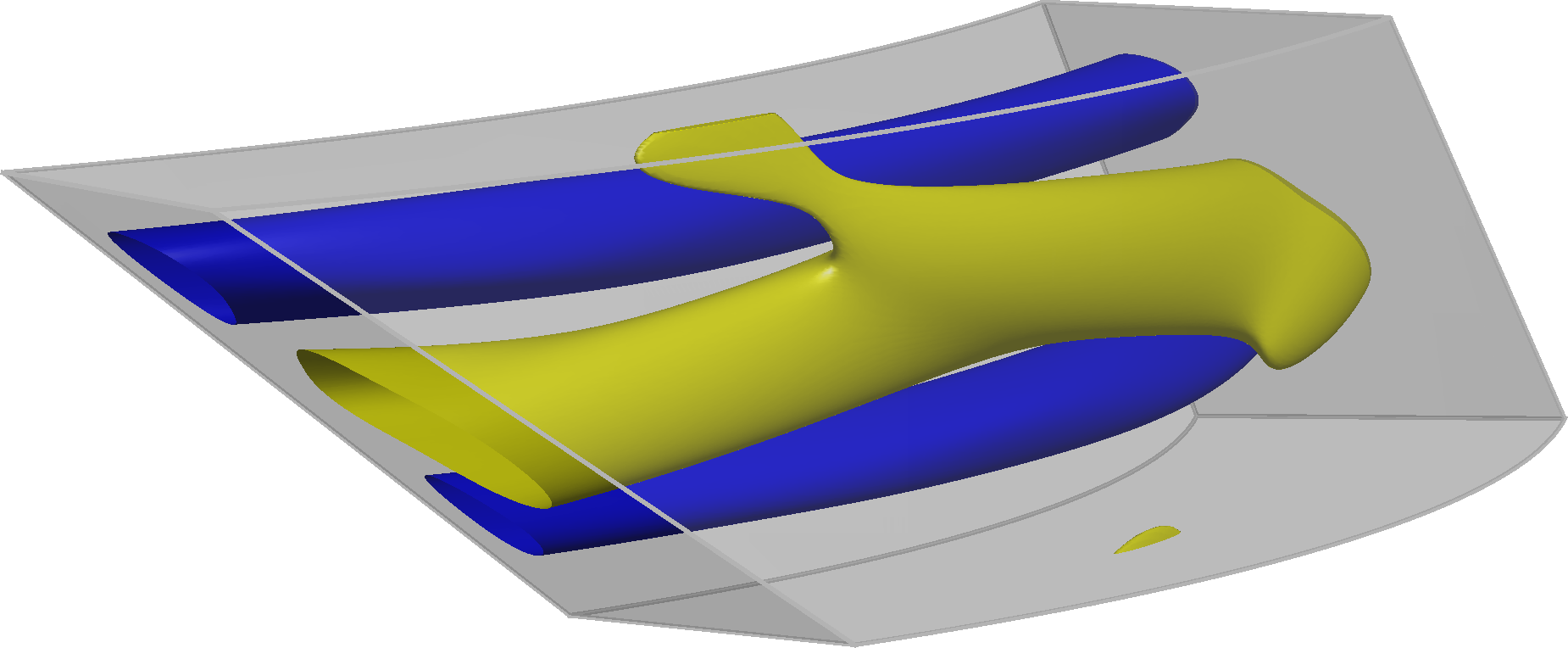}} &
     \raisebox{-2em}{\includegraphics[width=.42\linewidth]{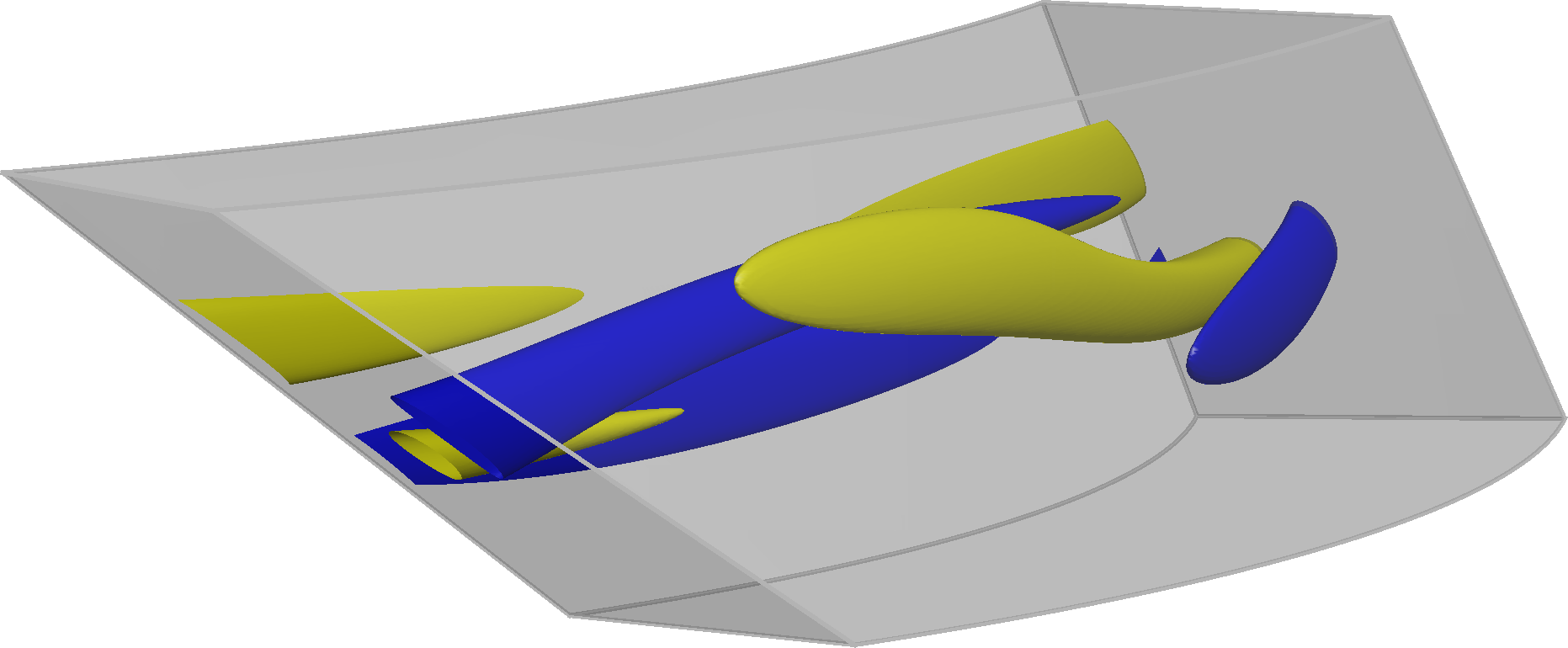}}
   \end{tabular}
 \end{center}                                                                 
 \caption{Three-dimensional flow structure of {\sc drw} solutions at
   subcritical $\Ri=450$, corresponding to points {\sc a}, {\sc b} and
   {\sc c} of figure\,\ref{stab_analysis_subrw4p5}. Shown are positive
   (yellow) and negative (blue) isosurfaces of (a) perturbation
   azimuthal velocity $v$, and (b) azimuthal vorticity
   $\omega_\theta$.  The corresponding isosurface levels are
   $v=\{-100,250\}$ and $\omega_\theta=\pm1000$ (solution A),
   $v=\{-100,250\}$ and $\omega_\theta=\pm600$ (B), and $v=\{-40,90\}$
   and $\omega_\theta=\pm300$ (C).}
  \label{fig_3dplots2}   
\end{figure}

Let us begin with a detailed analysis of the flow structure and
properties of the several coexisting {\sc drw} solutions in the
$k_2=4.5$ domain at subcritical $\Ri=450$.  Figure \ref{fig_3dplots2}
provides the three-dimensional characterisation of the flow structure
of {\sc drw} at the representative points labeled {\sc a}, {\sc b},
and {\sc c} in figure \ref{stab_analysis_subrw4p5}.
Of the solutions ensuing saddle node {\sc sn}$_1$, point {\sc c} is
chosen to represent the lower-torque branch. Two isosurfaces of the
azimuthal perturbation velocity ($v$) shown in the top row of figure
\ref{fig_3dplots2}a reveal the presence of a low-speed streak close to the
inner cylinder. For the higher-torque branch (point {\sc b}), this
slow streak induces a strong velocity distortion that reaches the
middle of the gap, an effect that is all the more pronounced for the
solutions originated at saddle-node {\sc sn}$_3$.  For the solutions
related to {\sc sn}$_3$, only point {\sc a} on the higher-torque
branch is shown, as the corresponding lower-torque solution is very
similar.
Another characteristic feature of the flow field is the presence of a vortex sheet,
which is visualised in figure\,\ref{fig_3dplots2}b through a couple of
azimuthal vorticity ($\omega_{\theta}$) isosurfaces. The position of
the sheet does not change much in the azimuthal direction, but the
sign of the vorticity fluctuates in a sinuous fashion. 
It is precisely the interaction of these two fields, the fluctuating
vortex layer and the quasi-azimuthally-invariant streak, that might be
held responsible for the self-sustaiment of {\sc drw} in the absence
of a linear instability of {\sc ccf}. The physical reasons for the
requirement of a feedback mechanism from the wave to the streak field can be
ascertained by examining the centrifugally unstable region adjacent to the inner cylinder, bounded by
$r_{\rm i}<r<r_{\rm n}$.  Although the
Rayleigh criterion predicts centrifugal inviscid instability of {\sc
  ccf} in this region, the laminar base flow remains stable because
the effective Taylor number is still slightly short of the critical
threshold \citep{EsGr96,Deguchi_JFM17}. The streak field of {\sc drw}
must therefore be supported by the Reynolds stresses induced by the
wave-like vortex layer, rather than centrifugal effects.

\begin{figure}                                                                 
 \begin{center}
   \begin{tabular}{cccc}
      & (a) & (b) & (c) \\[0.5em]
           \parbox{1cm}{\raggedleft {\sc drw}\\at {\sc a}} &
           \raisebox{-5em}{\includegraphics[width=.22\linewidth]{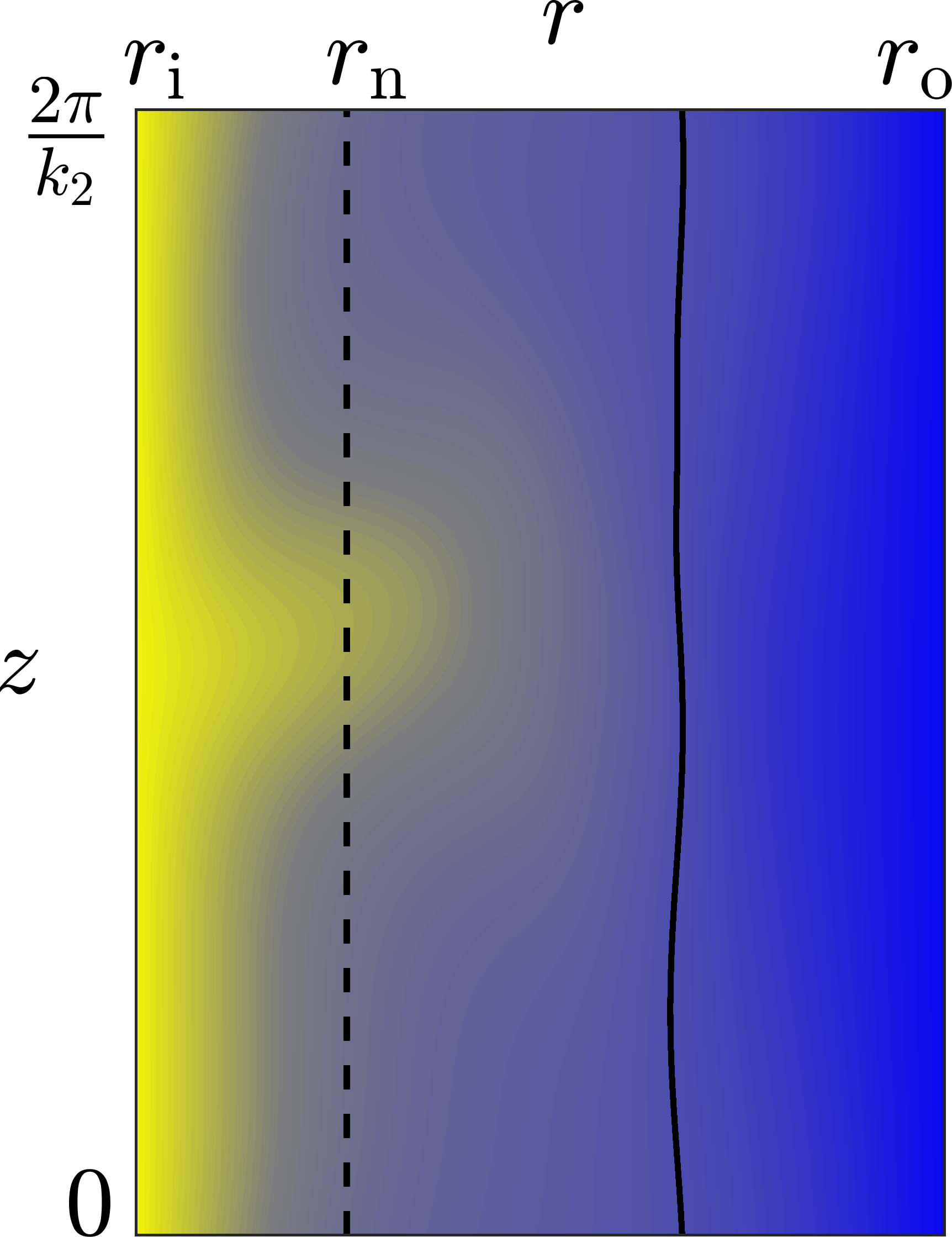}} & 
           \raisebox{-5em}{\includegraphics[width=.188\linewidth]{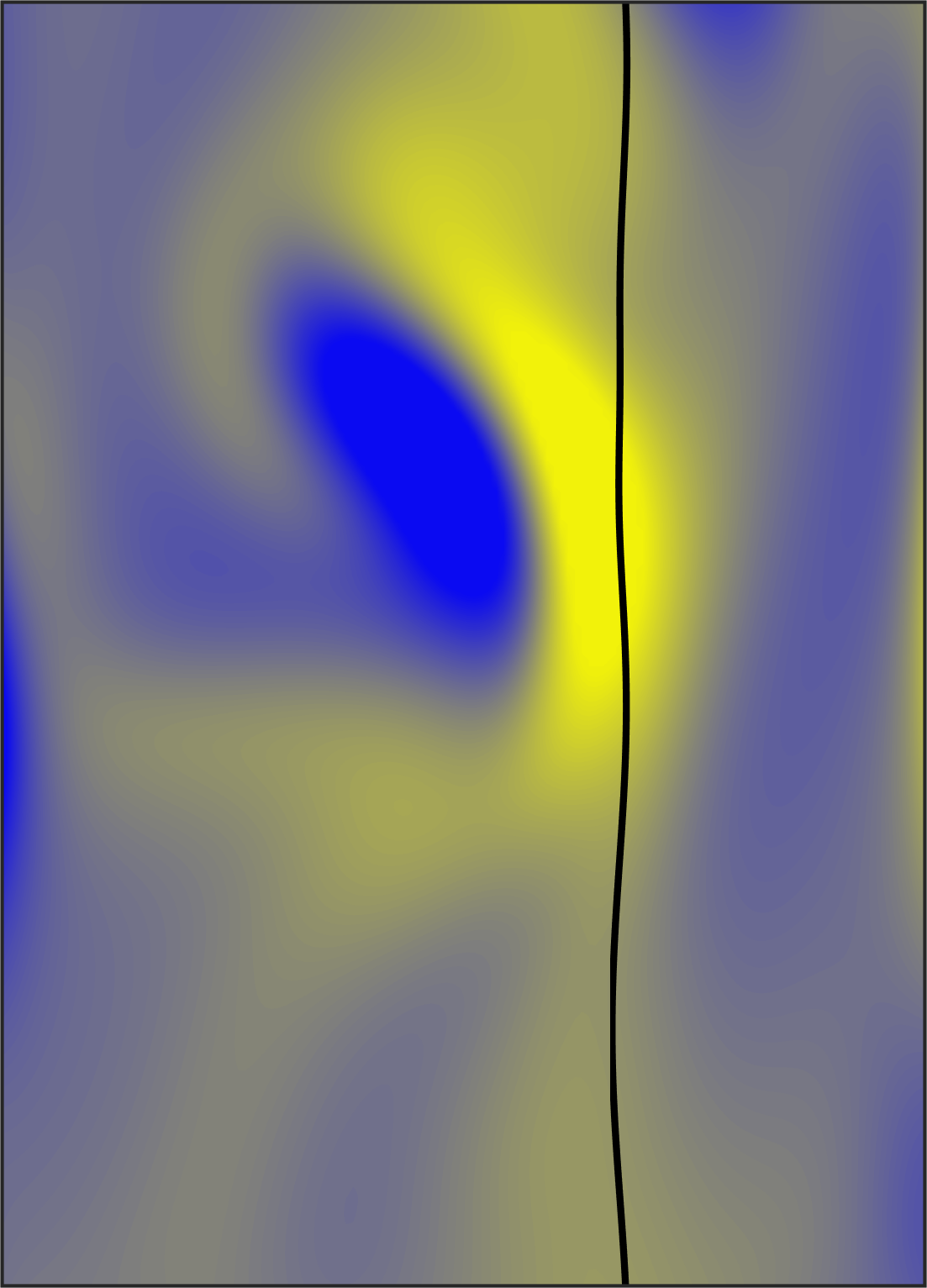}} & 
           \raisebox{-5em}{\includegraphics[width=.188\linewidth]{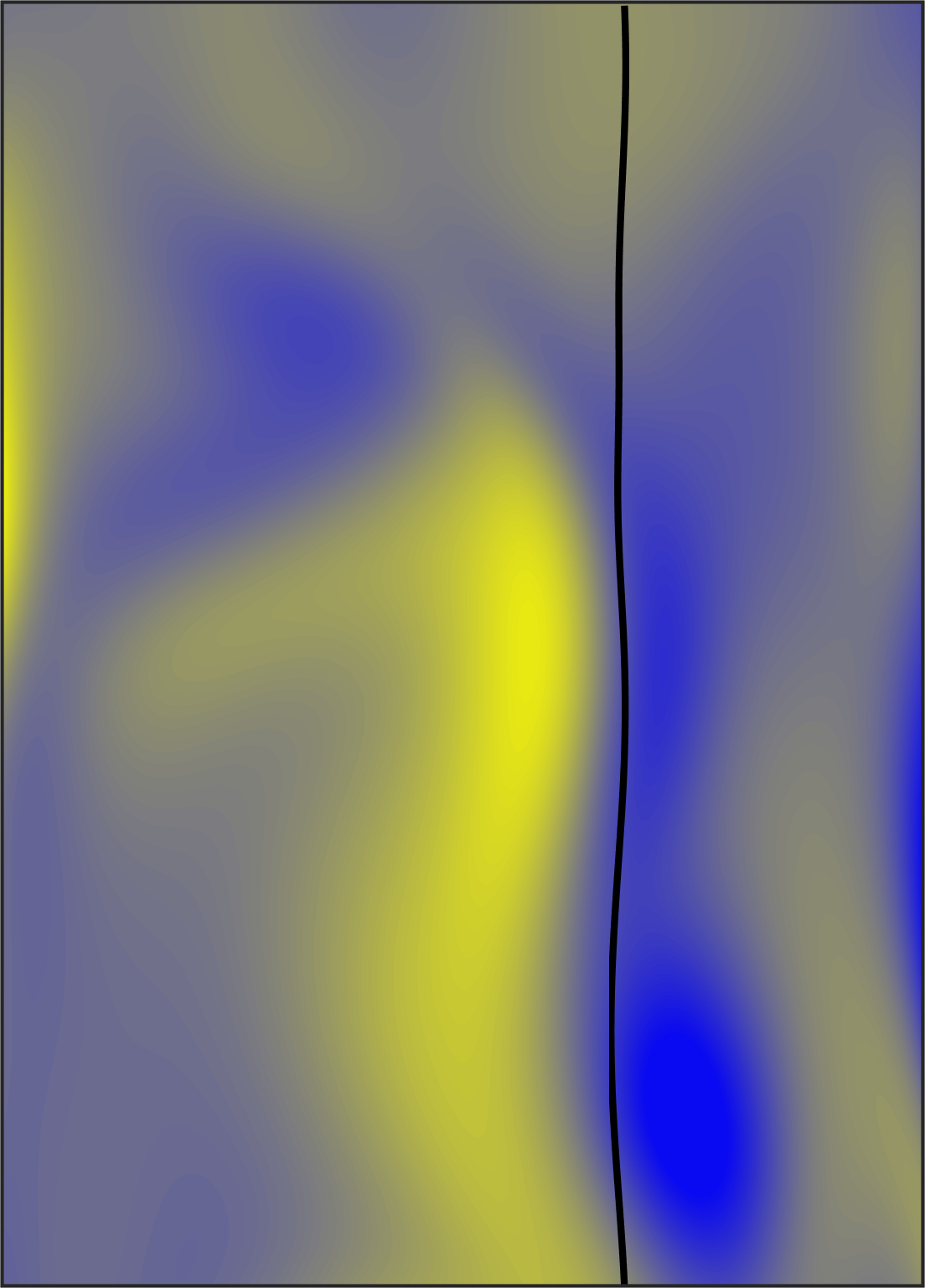}} \\[5em]
           \parbox{1cm}{\raggedleft {\sc drw}\\at {\sc b}} &
           \raisebox{-5em}{\includegraphics[width=.22\linewidth]{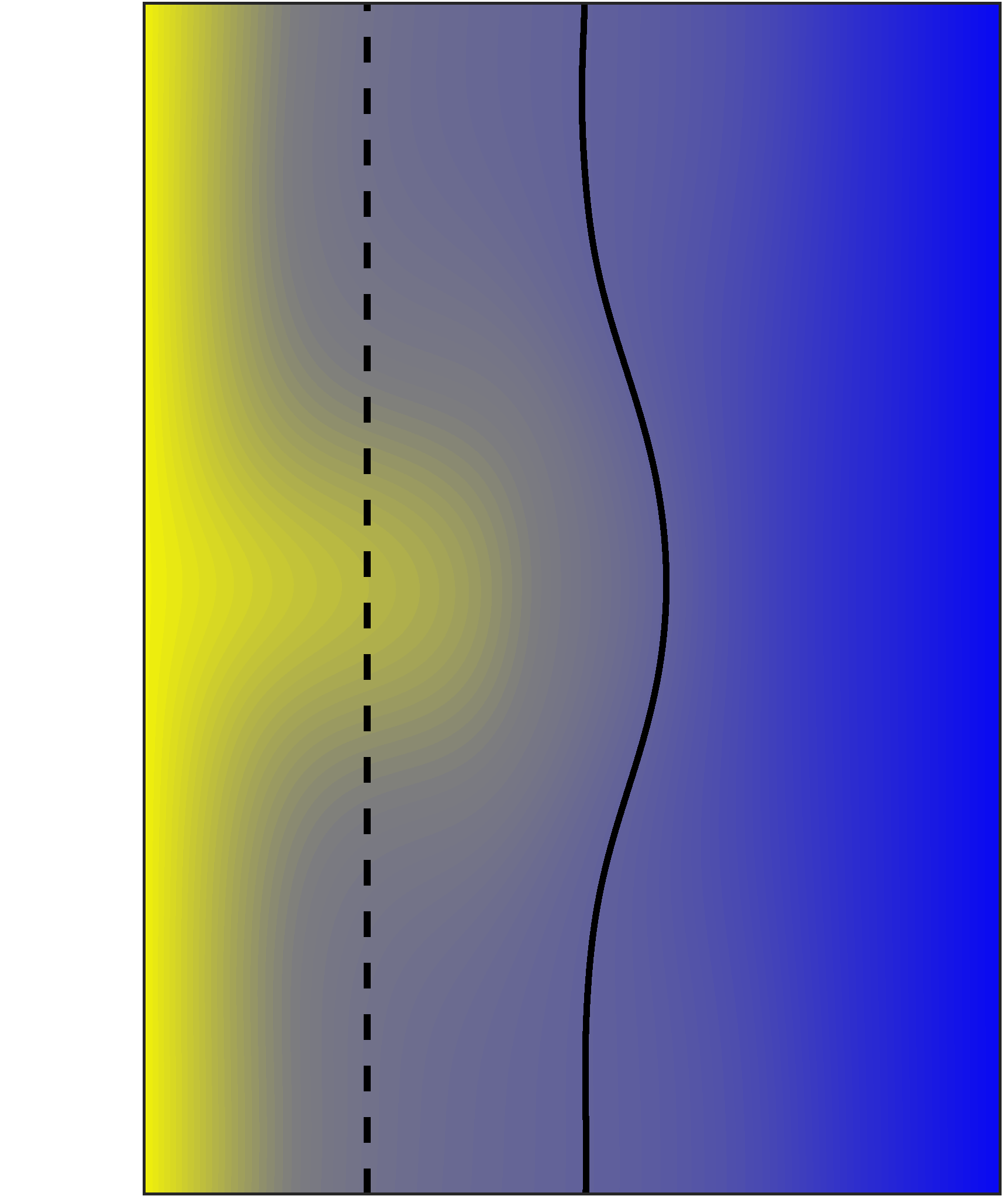}}& 
           \raisebox{-5em}{\includegraphics[width=.188\linewidth]{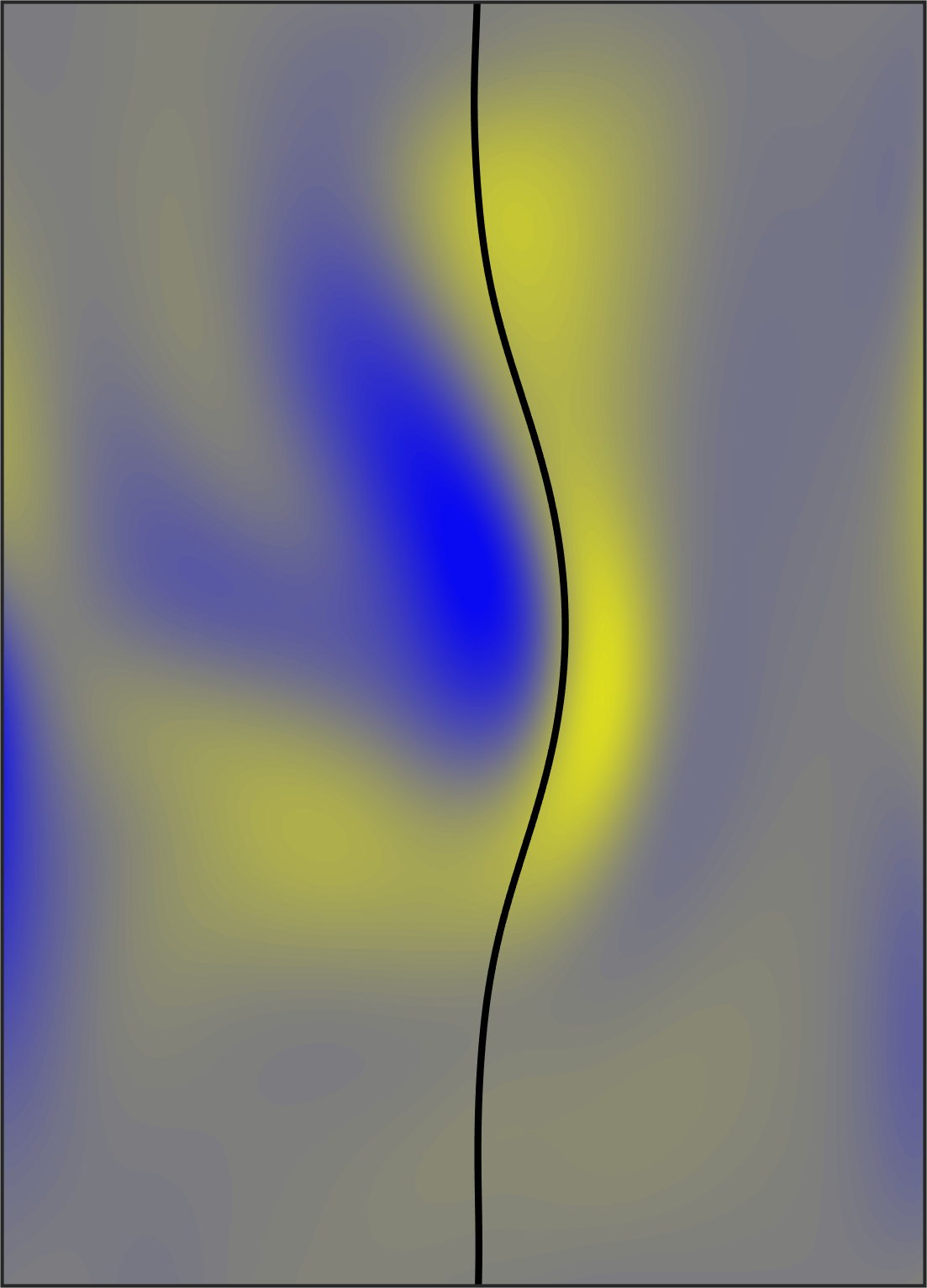}} & 
           \raisebox{-5em}{\includegraphics[width=.188\linewidth]{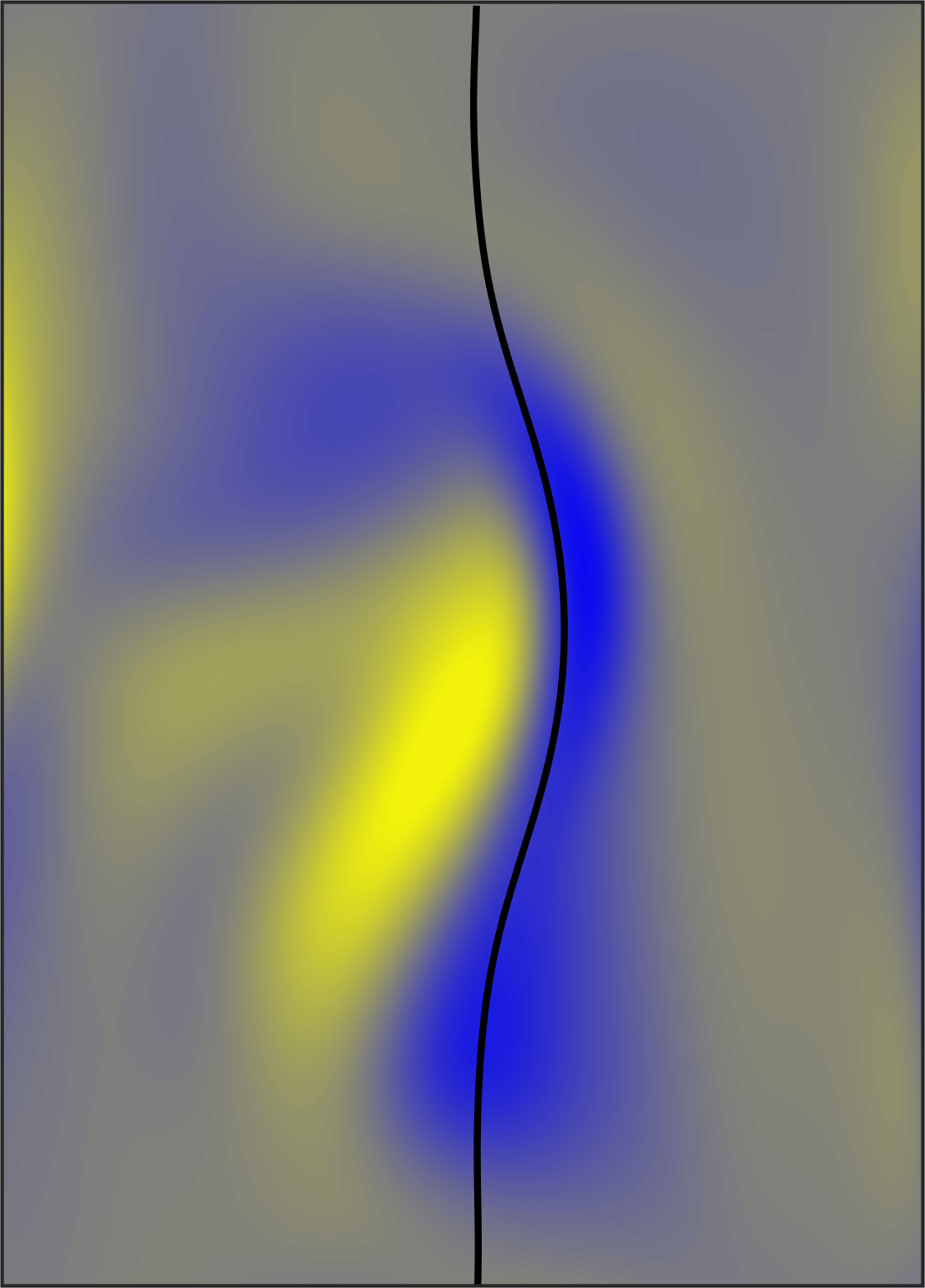}} \\[5em]
           \parbox{1cm}{\raggedleft {\sc drw}\\at {\sc c}} &
           \raisebox{-5em}{\includegraphics[width=.22\linewidth]{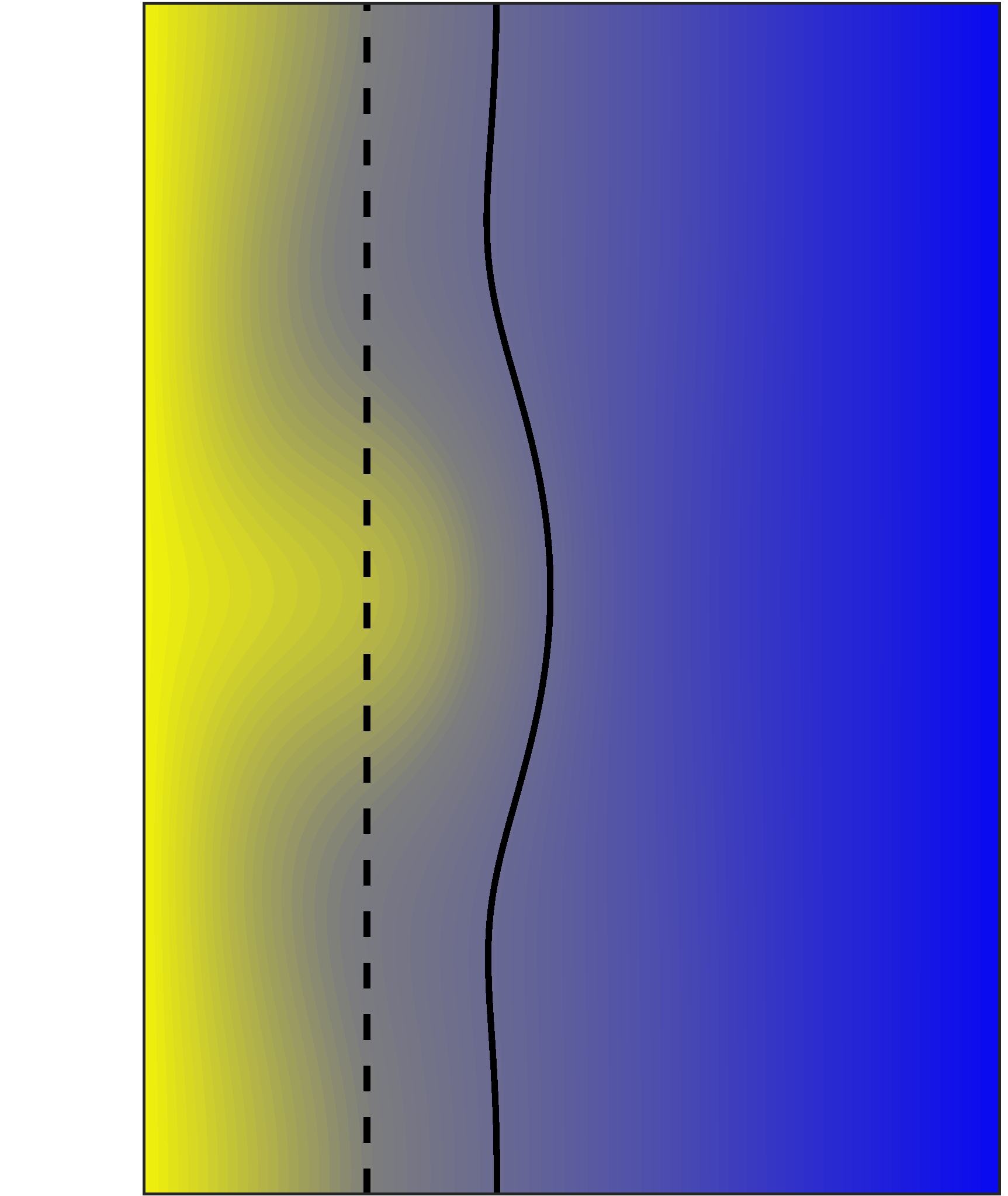}} &
           \raisebox{-5em}{\includegraphics[width=.188\linewidth]{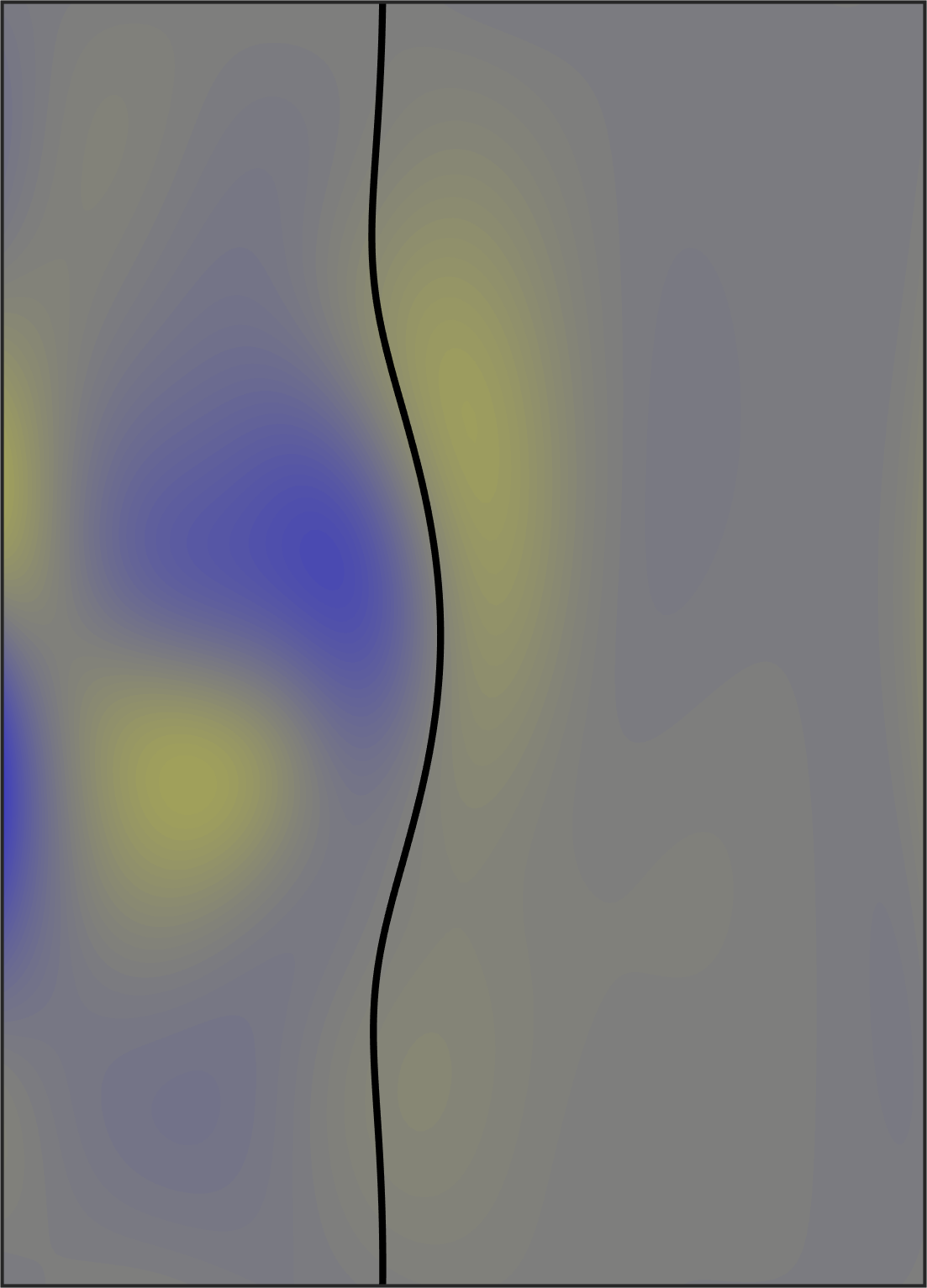}} & 
           \raisebox{-5em}{\includegraphics[width=.188\linewidth]{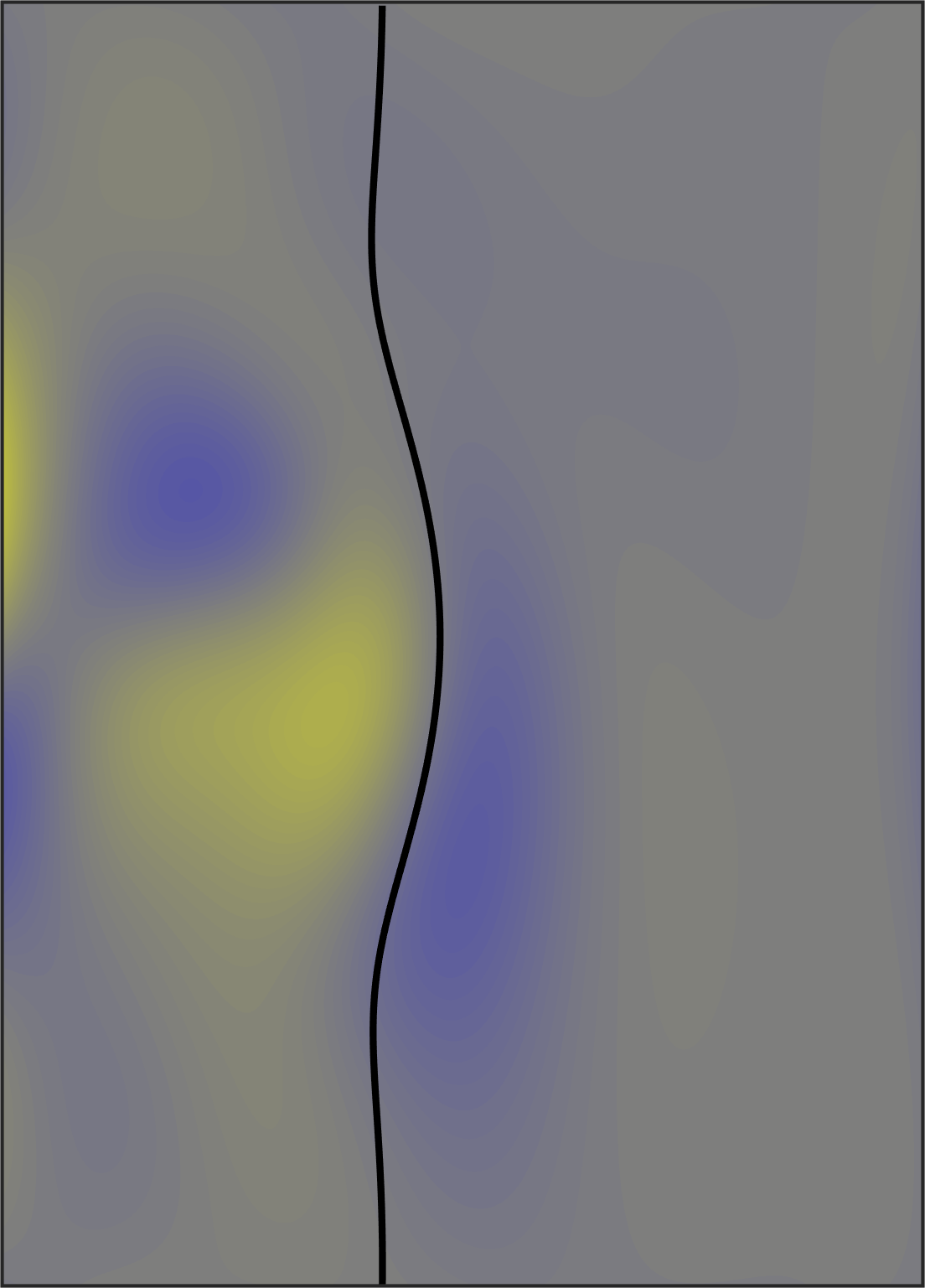}}
   \end{tabular}
 \end{center}                                                                 
 \caption{Comparison of (a) the streak field as signified by azimuthally-averaged
   distribution of total azimuthal velocity $\langle V
   \rangle_{\theta}/r\in[-1200,450]$, and the wave field, represented by azimuthal
   vorticity $\omega_{\theta}\in[-1000,1000]$, at (b) $\theta=0$ and
   (c) $\theta=\pi/n_1=0.314$, for {\sc drw} solutions labeled {\sc
     a}, {\sc b} and {\sc c} in
   figure\,\ref{stab_analysis_subrw4p5}. Lines indicate the location
   of the critical layer (solid) and the nodal radius $r_{\rm n}$
   (dashed).}
 \label{fig_vor_vel_field}   
\end{figure}

The comparison of both fields, shown together in
figure\,\ref{fig_vor_vel_field}, clearly discloses the similarity of
their interaction with the {\sc ssp} that is typical of parallel shear
flows.
Here we conjecture that the generation of the fluctuating field might
likely be explained by the inviscid linear instability of the streak
field, where the critical layer holds the key to the validation of
this hypothesis. The critical layer is formally a singularity of the
inviscid problem and asymptotic theory shows that the amplitude of the
wave must surge around it.  In parallel flows, the nature of this wave
amplitude growth in {\sc ecs} is well known
\citep{WangGiWa07,HaSh2010,DeguchiHall2015}.  The critical layer is
the place at which the streak velocity coincides with the phase speed
of the wave. There the advection term becomes small and tends to
vanish, such that the inviscid approximation breaks
\citep{Lin55,DraRe81}. In defining streaks and critical layers in the
Taylor-Couette problem, we note that subtle differences arise from the
parallel flow cases.  The velocity field of {\sc drw} solutions can be
rewritten as a function of $r$ and two phase variables
$\theta-c_{\theta} t$ and $z-c_z t$, where the constants $c_{\theta}$
and $c_z$ can be computed easily from $c_{\xi}$ and $c_{\zeta}$.  The
axial phase velocity $c_z$ is small and hence the critical layer is
approximately determined by comparing the azimuthal phase velocity
$c_{\theta}$ alone with the streak field. As it happens, the spanwise
phase velocity of drifting waves can be shown to vanish in the
asymptotic limit of increasing Reynolds number for parallel shear
flows, and the same may presumably be expected when the geometry is
curved. Since $c_{\theta}$ is an angular velocity, the streak field
must be defined as $\langle V\rangle_{\theta}/r$, using the azimuthal
average $\langle \bullet \rangle_{\theta}$ of the total (base plus
perturbation) velocity $V=v_b+v$.
The black line in figure\,\ref{fig_vor_vel_field}a indicates the critical
layer thus computed on top of the streak field. The
azimuthal cross-sections of $\omega_{\theta}$ colour maps at $\theta=\{0,\pi/n_1\}$ (panels b and c) unequivocally show that the
vortex sheet amplitude is strongest precisely around the critical layer.




The magnitude of $\omega_{\theta}$ increases as the critical layer
shifts radially outwards (see
figure\,\ref{fig_vor_vel_field}). According to Rayleigh's stability
condition, the closer to the outer cylinder one looks, the more
centrifugally stable the flow is locally, and a stronger local wave
amplitude is therefore required to sustain the streak
field. Rayleigh's condition is based on the {\sc ccf} profile, so that
weak Taylor vortices may still arise due to nonlinearly-driven
mean-flow-field distortions in the vicinity of the inner
cylinder. This is the case of solution {\sc c} and provides a physical
explanation as to why the connections between {\sc (d)tvf} and {\sc
  drw} occur in the way discussed in \S\ref{sec_subcritrow}.


\subsection{Stability of {\sc drw} and the onset of chaotic dynamics}
\label{ssec:stability}

Let us now turn our attention to the stability of {\sc drw}
  solutions and their role in engendering a chaotic set.  The count
and type of eigenvalues, as computed through linear stability
analysis, are indicated in brackets for a number of sampled solutions
along the various {\sc drw} branches in
figure~\ref{stab_analysis_subrw4p5}. From saddle-node {\sc sn}$_1$
onwards, the higher-torque branch is stable while the lower-torque
solutions have a single real unstable eigenvalue. This result,
which is a must for one-dimensional dynamical systems exhibiting a
saddle-node bifurcation, is nevertheless rarely encountered in
high-dimensional systems such as Taylor-Couette flow.  In fact, the
same (or closely related) {\sc drw} solutions present a higher number of
unstable eigenvalues when considered in the usual orthogonal domain
\citep{DeMeMe14} or in
parallelograms of size $k_2$ other than 4.5. Our choice of domain is
therefore expressly convenient for the simplest possible
exploration of the onset of chaotic dynamics.

\begin{figure}                                                                 
 \begin{center}
  \begin{tabular}{cc}
   (a) & (b) \\[0.5em]
   \includegraphics[width=.52\linewidth]{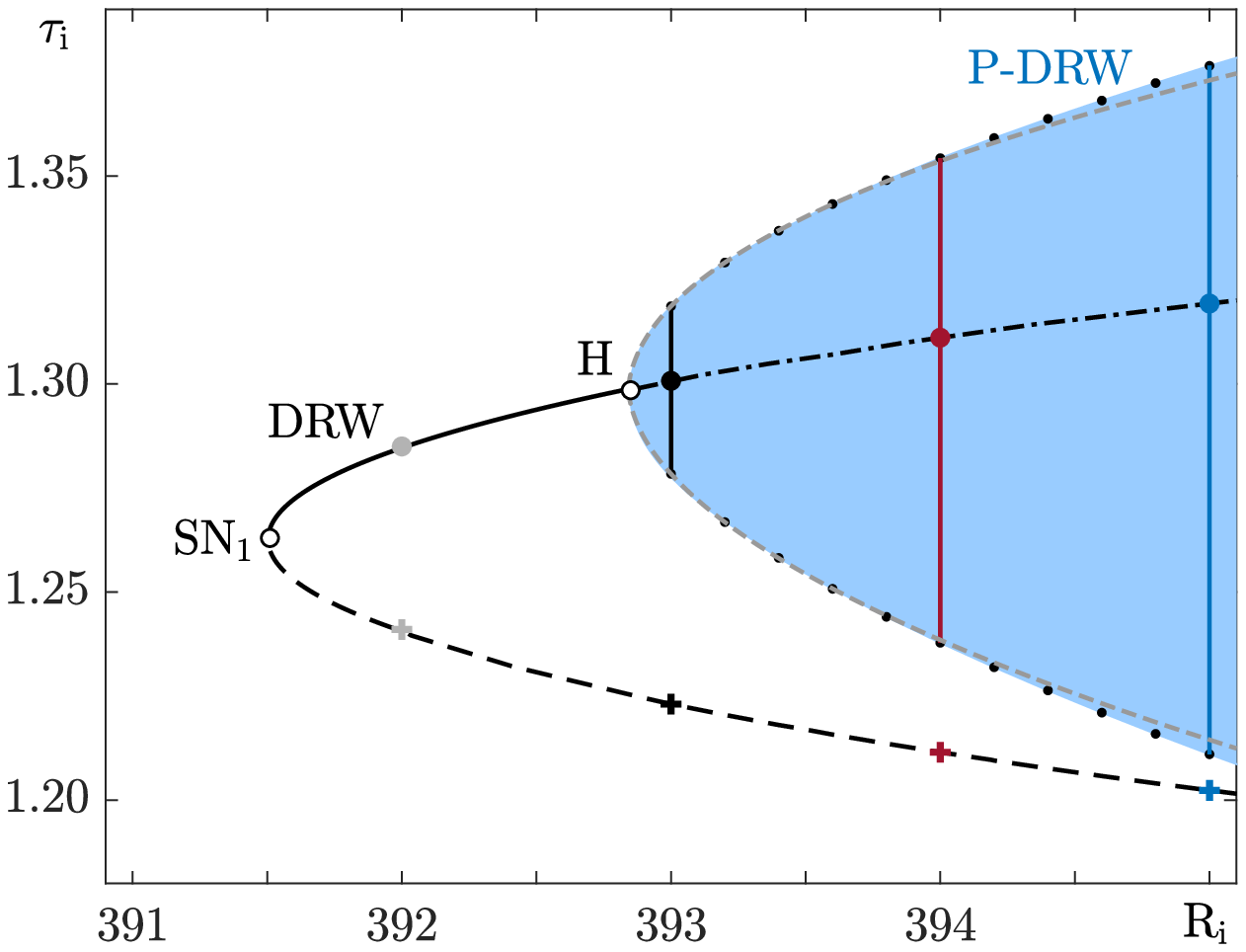} &
    \raisebox{0.05cm}{\includegraphics[width=.37\linewidth]{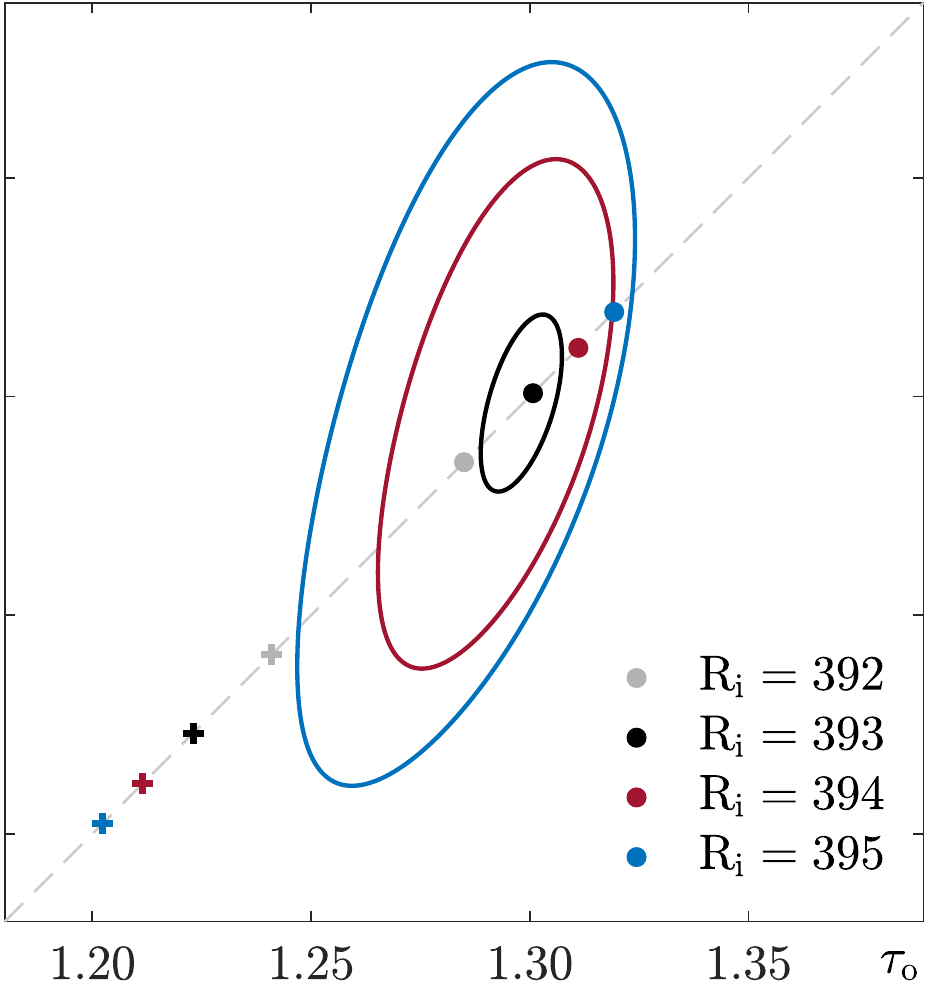}}
   \end{tabular}
  \end{center}  
 \caption{Onset of {\sc p-drw} solutions. (a) {The bifurcation diagram
     near the saddle-node {\sc sn}${_1}$ shown in figure 3. Line
     styles denote {\sc drw}} branches with different stability
   properties: stable (solid), one unstable real eigenvalue (dashed)
   and one unstable complex pair (dash-dotted).  {The small black
     circles indicate the maximum and minimum $\tau_{\rm{i}}$ attained
     by the relative periodic orbit {\sc p-drw}, whose oscillation
     amplitude is delimited by the shaded blue region.  The {\sc
       p-drw} emerges supercritically at the Hopf bifurcation point
     {\sc h}, such that the amplitudes obey locally a square-root fit
     (gray dashed curve). (b) The
     $(\tau_{\rm{o}},\tau_{\rm{i}})$-phase map projections of upper
     and lower branches of {\sc drw} (bullets and crosses,
     respectively) and the {\sc p-drw} limit cycles (solid curves) for
     different values of $\Ri$.}  }
\label{MRW_k4p5_Hopf}              
\end{figure}

The stability of the higher-torque {\sc drw} branch emerged from {\sc
  sn}$_1$ is lost in a Hopf bifurcation at $\Ri^{\rm H} = 392.85$
(point {\sc h} in figure\,\ref{stab_analysis_subrw4p5}), as
implied by the crossing of a pair of complex eigenvalues into the
positive-real-half of the complex plane.
The resulting relative periodic orbit has a structure
  similar to {\sc drw} but is subject to time-periodic amplitude oscillations, hence the acronym {{\sc p-drw}}.  The
Poincar\'e-Newton-Krylov method described in \S\,\ref{sec_formulation}
has been deployed to perform natural continuation of the {{\sc p-drw}} solution branch.
Figure\,\ref{MRW_k4p5_Hopf}a signifies the oscillation amplitude of
the {{\sc p-drw}} as the area (shaded blue region) bounded by the maximum and
minimum inner torque $\tau_{\rm i}$ along a full cycle (black dots).
The solution amplitude obeys a square-root law of the form
$A_{\tau_{\rm i}}=\tau_{\rm i}^{\rm max}-\tau_{\rm i}^{\rm min}\sim(\Ri-\Ri^{\rm
  H})^{1/2}$, as revealed by the least-squares fit (gray dashed
line) to a few of the closest points to the Hopf bifurcation {\sc
  h}. This attests to the supercritical nature of the Hopf
bifurcation. As a result, {\sc p-drw} solutions are stable at onset
and could therefore have been computed by mere time-stepping. The
evolution of {{\sc p-drw}} with $\Ri$ is more clearly illustrated by
the outer-vs-inner torque $(\tau_{\rm{o}},\tau_{\rm{i}})$ phase-map
projections of figure\,\ref{MRW_k4p5_Hopf}b. 
With each increase in $\Ri$, the lower-torque (crosses) and higher-torque (circles) {\sc drw} solutions become further apart in phase space.
A small limit cycle ({\sc p-drw})
clearly orbits the higher-toque {\sc drw} at $\Ri=393$, and its amplitude
grows as $\Ri$ is further increased.

\begin{figure}                                                                 
 \begin{center}
  \begin{tabular}{cc}
   (a)& (b)\\[0.5em]
   \includegraphics[width=.378\linewidth]{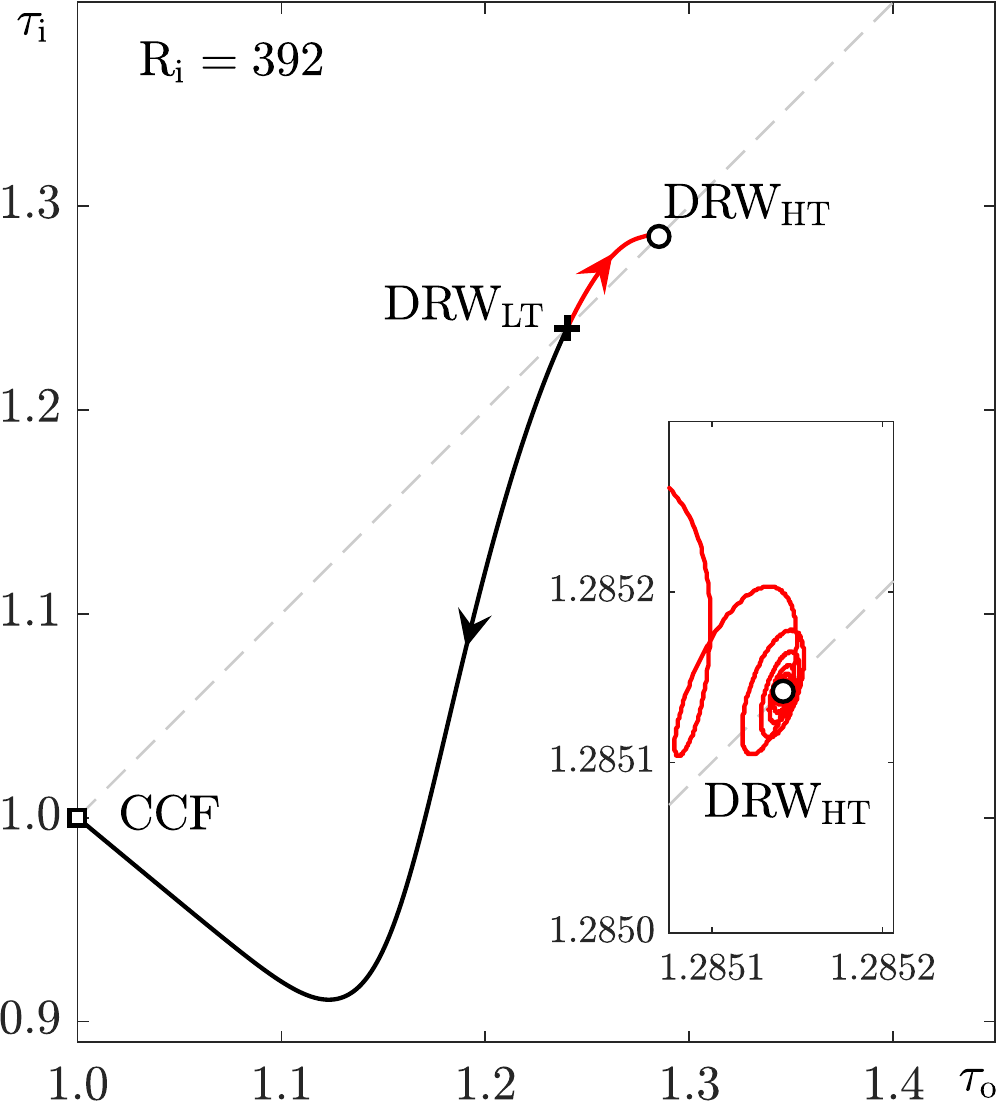} &
   \includegraphics[width=.36\linewidth]{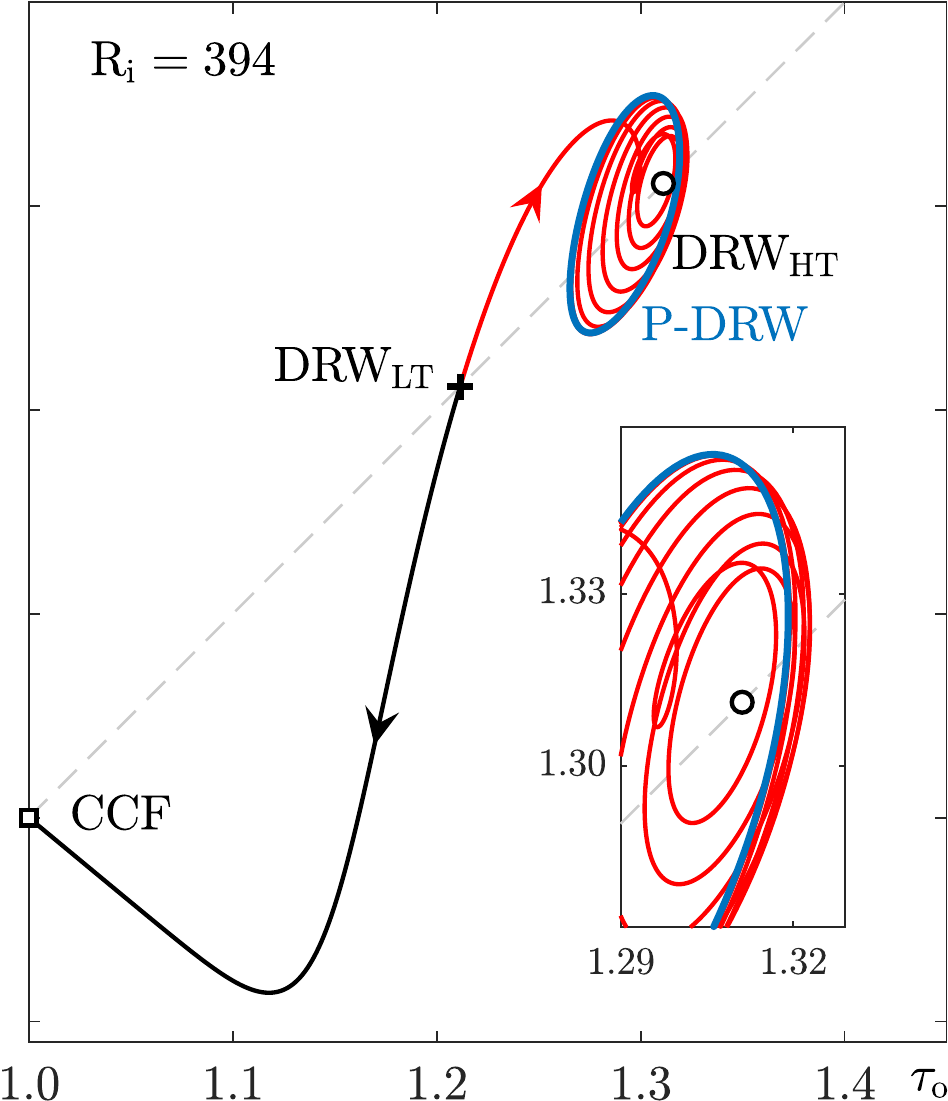}
   \end{tabular}
  \end{center}                                                                 
\caption{Phase-map projections of trajectories issued from the
  unstable lower-torque {\sc drw}$_{\rm LT}$ solution on the
  $(\tau_{\rm o},\tau_{\rm i})$-plane at inner Reynolds number (a)
  $\Ri=392$ and (b) $\Ri=394$. The represented solutions are {\sc ccf}
  (square), lower-torque ({\sc drw}$_{\rm LT}$, cross) and
  higher-torque ({\sc drw}$_{\rm HT}$, circle) {\sc drw} solutions,
  and {\sc p-drw} (blue line). The red/black lines depict {\sc dns}
  trajectories starting from initial conditions taken by scaling {\sc
    drw}$_{\rm LT}$ by $(1+\gamma)$, with $\gamma=\pm10^{-4}$,
  respectively.  }
  \label{phasemap_Ri392_Ri394}   
\end{figure}

In order to establish the dynamical connections among the various
solutions, {\sc dns} from small perturbations to the lower-torque {\sc
  drw} solution have been run. The initial condition has been set by 
scaling the exact solution following $(1+\gamma) \mathbf{a}^{\rm DRW}$, with $|\gamma|\ll1$.
Figure\,\,\ref{phasemap_Ri392_Ri394}a shows the
$(\tau_{\rm{o}},\tau_{\rm{i}})$-phase map projections of a couple
runs at $\Ri=392$.
As we have already noted, the lower-torque {\sc drw}$_{\rm LT}$
solution has only a single real unstable eigenvalue, so that the
diagram reflects but a close approximation to its one-dimensional
unstable manifold. For $\gamma =-10^{-4}$, the flow uneventfully
departs towards {\sc ccf} (black line), while for $\gamma=10^{-4}$ the
flow is captured by the linearly stable higher-torque {\sc drw}$_{\rm
  HT}$ solution (red line). The lower-torque {\sc drw} solution is
therefore acting as an edge state \citep{ItTo01,SkYoEck06}, separating
the basins of attraction of {\sc ccf} on one side and the
higher-torque {\sc drw}$_{\rm HT}$ solution on the other.  The
dynamics remain qualitatively the same as illustrated here within the
parameter range $391.52<\Ri<392.85$, while the upper branch solution
preserves its linear stability.

Slightly above the Hopf bifurcation ($\Ri>\Ri^{\rm H}=392.85$), the
higher-torque {\sc drw}$_{\rm HT}$ solution has become unstable and
the stable {\sc p-drw} has popped up into existence, as illustrated at
$\Ri=394$ by figure\,\ref{phasemap_Ri392_Ri394}b. The dynamics are
qualitatively unaltered as regards {\sc ccf} or {\sc drw}$_{\rm LT}$,
which remains an edge state, but the phase-map trajectory that
previously led to {\sc drw}$_{\rm HT}$ is now only able to transiently
approach it to some extent before being repelled towards {\sc p-drw}
(blue line) in a spiralling fashion. The simulation eventually
converges onto {\sc p-drw}, now the local attractor on this side of
phase-space. The insets clarify the nature of {\sc drw}$_{\rm HT}$,
which remains a focus across the Hopf bifurcation, but switches from
stable to unstable. Sufficiently close to {\sc sn}$_1$ the solution is
instead a node, stable within a neighbourhood of $k_2=4.5$, but
unstable beyond a certain threshold. These facts put together suggest
that the saddle-node and Hopf bifurcations are in fact constituent
pieces of a codimension-2 Takens-Bogdanov bifurcation \citep{Kuz04}.

\begin{figure}                                                                 
 \begin{center}
   \begin{tabular}{cc}
      (a)& (b)\\[0.5em]
            \includegraphics[width=.46\linewidth]{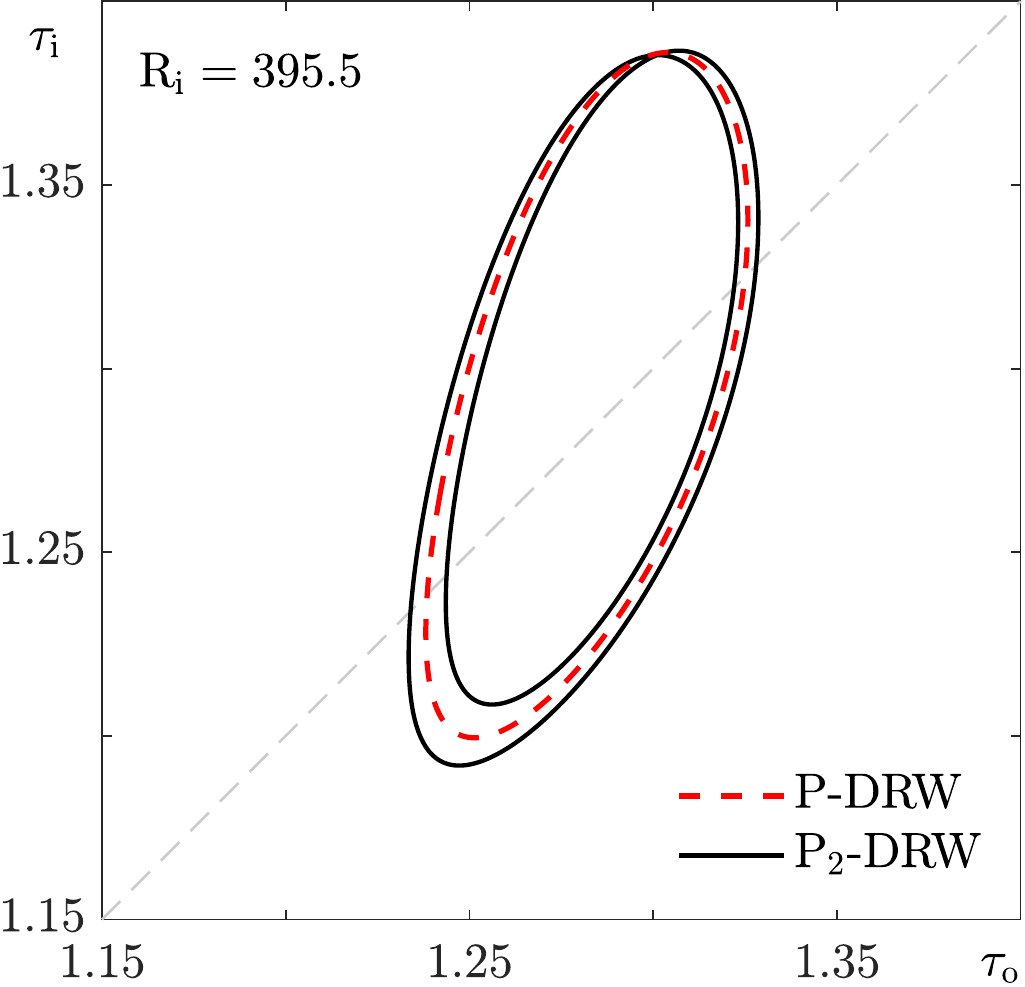} & 
           \includegraphics[width=.433\linewidth]{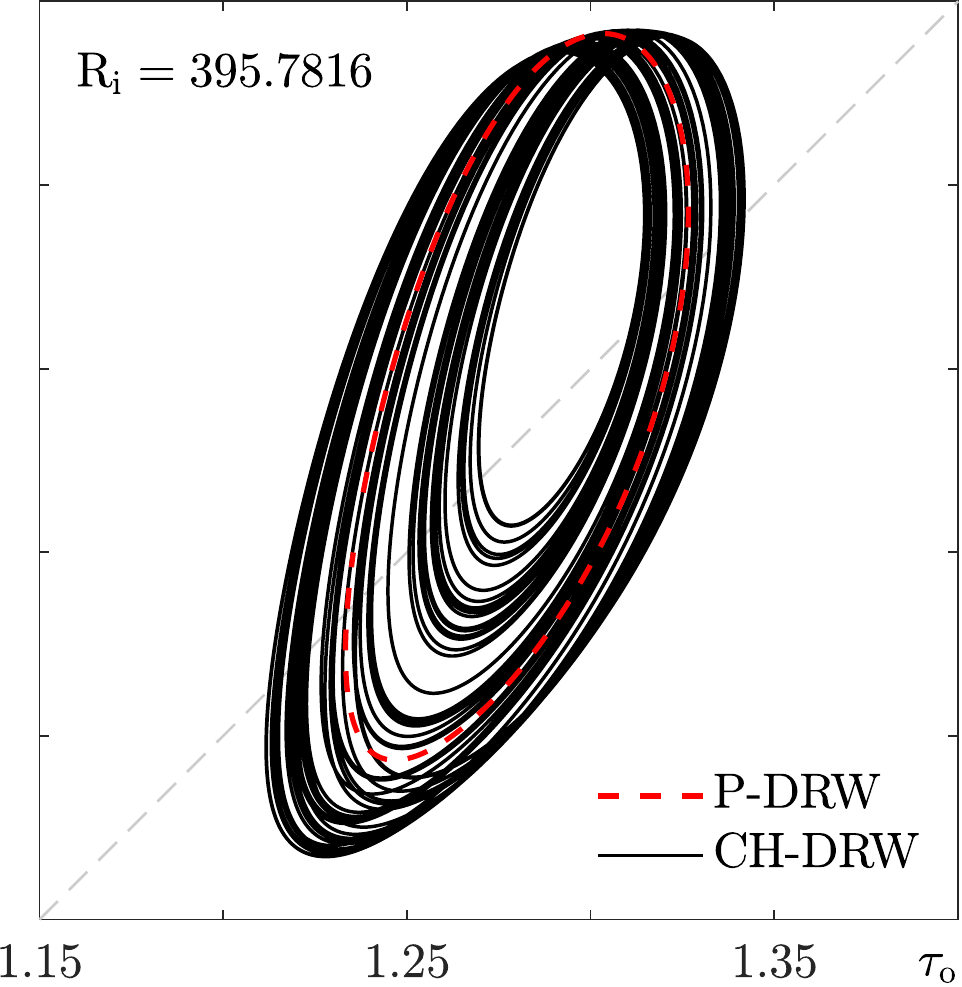} \\
   \end{tabular}
 \end{center}                                                                 
 \caption{Two stages of the period-doubling cascade of {\sc p-drw} as
   illustrated by phase-map projections on the $(\tau_{\rm
     o},\tau_{\rm i})$ plane at (a) $\Ri=395.5$ and (b)
   $\Ri=395.7816$. Shown are the stable attractor (black line) along
   with the unstable {\sc p-drw} solution (red dashed line).  }
  \label{fig_phasemap}   
\end{figure}

At even higher Reynolds number $\Ri = 395.5$, {\sc p-drw} has become
unstable and the new stable limit cycle makes two similar but not
identical revolutions in phase space before closing on itself. The
resulting period-doubled {\sc p}$_2${\sc -drw} solution (black line)
is shown, along with the unstable {\sc p-drw} limit cycle, at the same
$\Ri=395.5$ (dashed red) in figure\,\ref{fig_phasemap}a.
The period doubling that occurs just short of $\Ri=395.5$ is followed
by a number of ensuing period-doublings upon further increasing $\Ri$
that eventually lead to the chaotically-modulated {\sc drw} ({\sc
  ch-drw}) of figure\,\ref{fig_phasemap}b at $\Ri=395.7816$. The
transition route to chaos observed here is suggestive of a
period-doubling cascade scenario, as has been reported for several
other parallel shear flows \citep{KrEc2012,LuKaVeShiKo19}. The stable
state at $\Ri=395.7816$ is only mildly chaotic and the unstable {\sc
  p-drw} solution provides a fair approximation of the attractor
properties. Several complex global bifurcations can be identified
along the period-doubling cascade that are determinant to the
dynamics, but their nature will be discussed elsewhere on account of
the associated intricacies.

The stability analysis reflected in
figure\,\ref{stab_analysis_subrw4p5} reveals at least two more Hopf
bifurcations of the already unstable {\sc drw} as $\Ri$ is increased
beyond the value for the onset of {\sc ch-drw}. Solution branches
issued from these Hopf points and subsequent bifurcation cascades like
the one we report may probably contribute, along with {\sc ch-drw}, through global bifurcations involving crises and mergers,
to the formation of the {\it strange} set that sustains spatiotemporally chaotic dynamics at higher $\Ri$ \citep{KPG21}.

\subsection{Turbulent dynamics at $\Ri=600$}

\begin{figure}                                               
\begin{center}
 \begin{tabular}{cc}
    (a)& (b)\\
\includegraphics[width=0.335\linewidth]{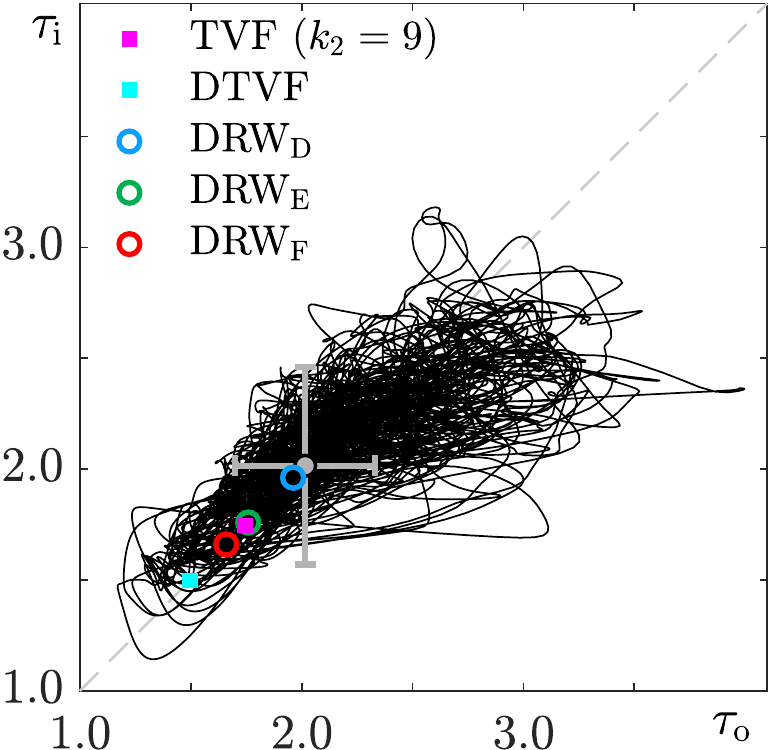} &
\includegraphics[width=0.595\linewidth]{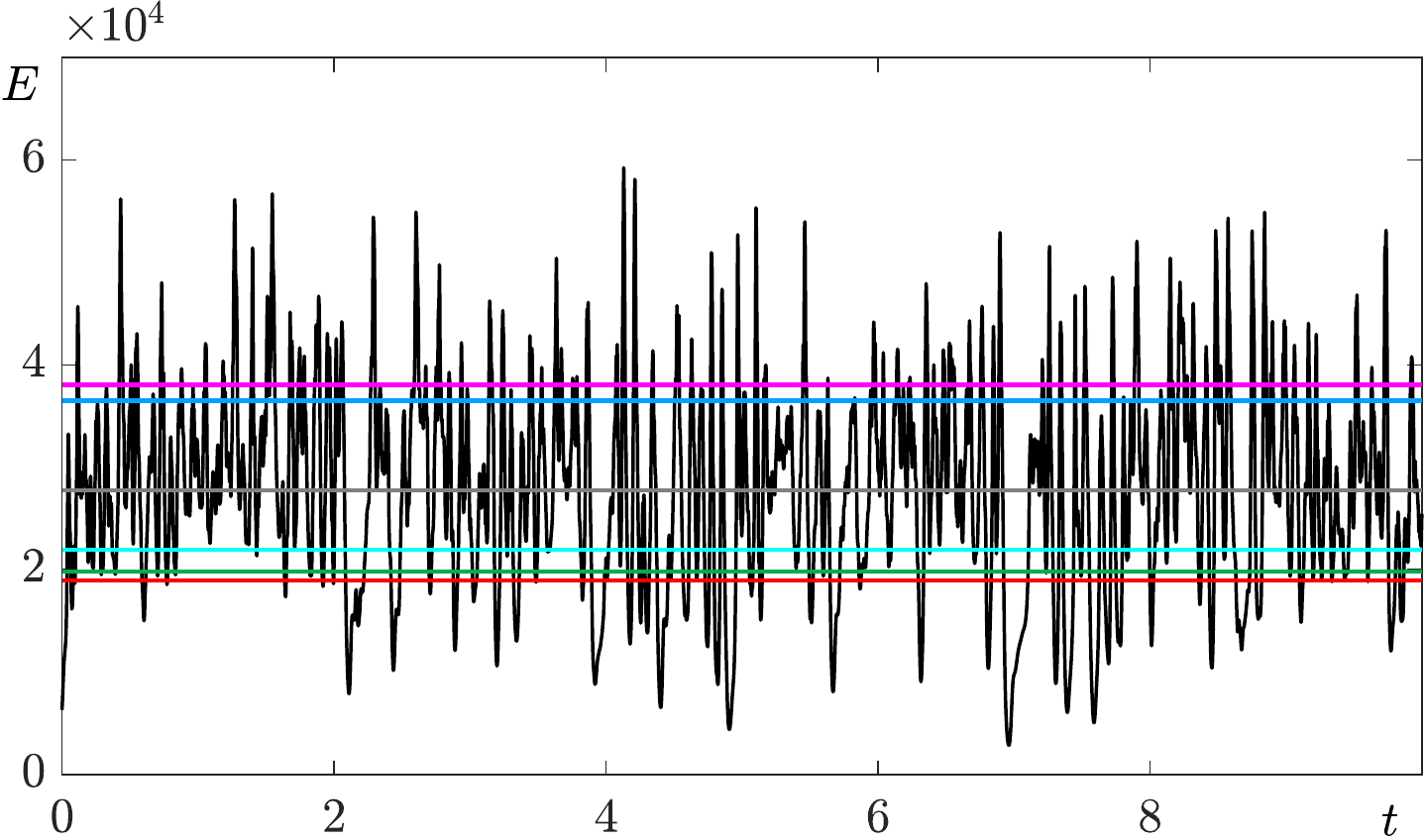} \\
\end{tabular}
\end{center}
\caption{ Spatiotemporally chaotic dynamics at $\Ri=600$ in the short parallelogram
  domain of figure \ref{figdomain}c
  $(n_1,k_1,n_2,k_2)=(10,2,0,4.5)$. (a) Phase-map projections on the
  $(\tau_{\rm o},\tau_{\rm i})$ plane. Gray bullet and error bars
  denote mean and root-mean-square of the fluctuations, respectively. (b) Time series of
  kinetic energy $E$. The gray line indicates mean kinetic energy. The
  colour circles/squares and lines indicate {\sc drw}, {\sc tvf} and {\sc dtvf} solutions labeled in
  figure\,\ref{stab_analysis_subrw4p5}.
}
\label{fig_smallbox_turbulence}
\end{figure}

As shown in figure \ref{fig_smallbox_turbulence} at $\Ri=600$, the
dynamics in the short parallelogram domain exhibits wild
fluctuations alternating rather low-torque/low-kinetic-energy, ostensibly dormant, {\it laminar} phases with actively {\it turbulent} stages.
If these turbulent transients are governed by the coalescence of a
  number of temporally chaotic sets such as {\sc ch-drw}, it is to be
  expected that the periodic orbits at their origin should contribute
  their part to the dynamics.  Unfortunately, current computational
resources have not allowed continuation of the {\sc p-drw} branch this
far up in Reynolds number.  Nonetheless, existing {\sc drw} solutions
at $\Ri=600$ (points {\sc d, e} and {\sc f} indicated in figure
\ref{stab_analysis_subrw4p5}), all of them unstable, seem to still
play a role in scaffolding the transient turbulent state. The
only solution that remains of those issued from {\sc sn}$_1$ ({\sc
  drw}$_{\rm D}$, blue) has a perturbation kinetic energy $E$ somewhat
larger than the mean for the statistically steady turbulent state.
The other two solutions, resulting from {\sc sn}$_{3}$ (higher-torque
solution {\sc drw}$_{\rm E}$ in green and lower-torque solution {\sc
  drw}$_{\rm F}$ in red), have smaller perturbation energy and
torque. We will argue that, in spite of their misleadingly high
  torque and kinetic energy values, the laminar stages of the dynamics
  are indirectly linked to the {\sc tvf} (magenta) and {\sc dtvf}
  (cyan) solutions.

\begin{figure}                                               
\begin{center}
 \begin{tabular}{cc}
    (a)& (b)\\
\raisebox{-2cm}{\includegraphics[width=0.35\linewidth]{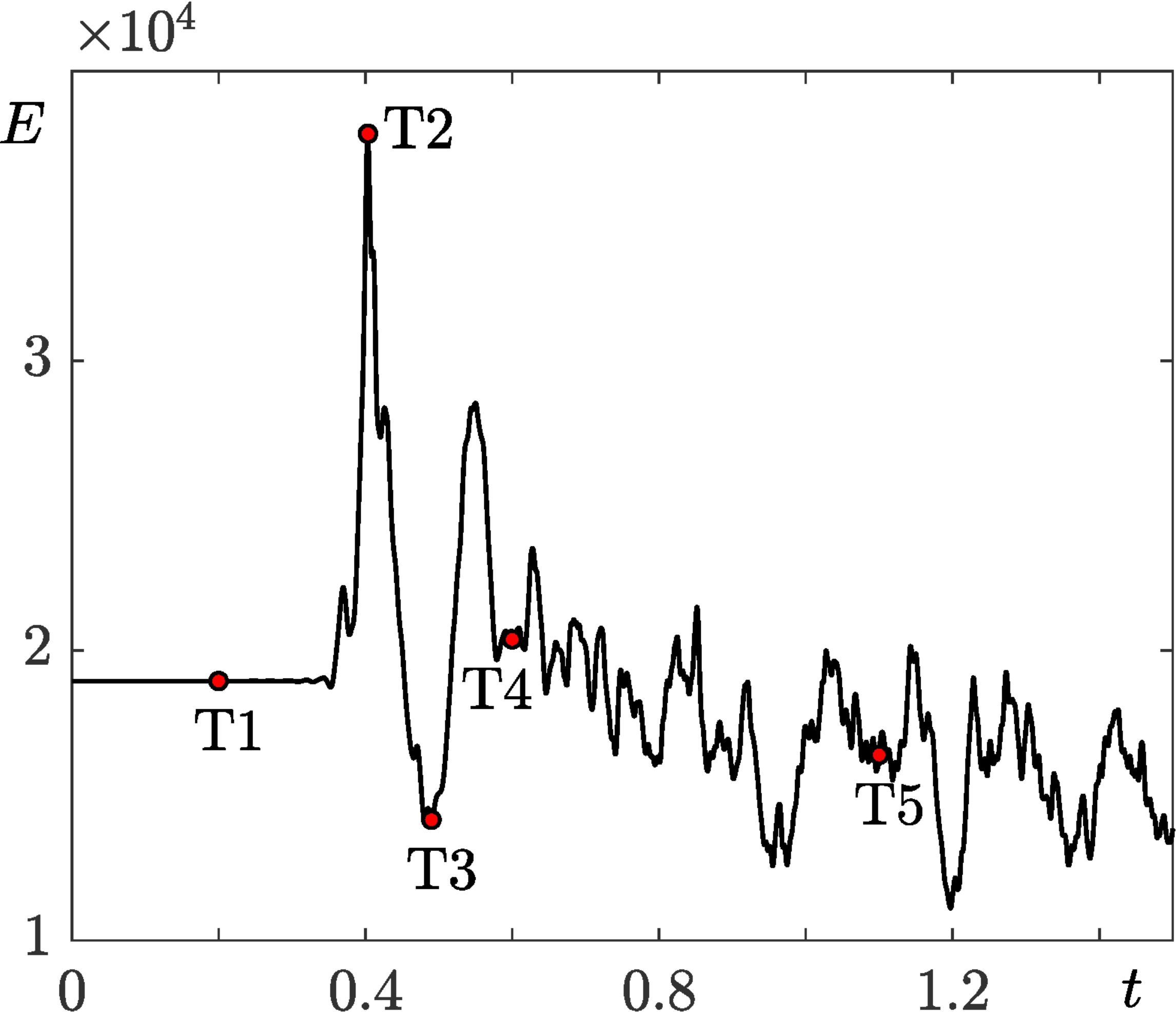} }&
\begin{tabular}{c}
\includegraphics[width=0.6\linewidth]{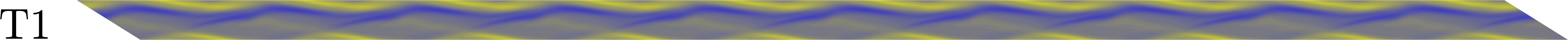}  \\ 
\includegraphics[width=0.6\linewidth]{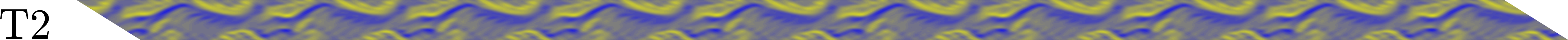}  \\ 
\includegraphics[width=0.6\linewidth]{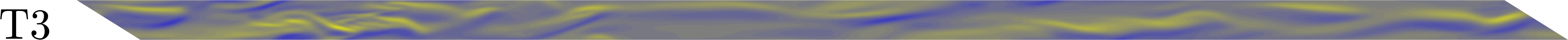}  \\ 
\includegraphics[width=0.6\linewidth]{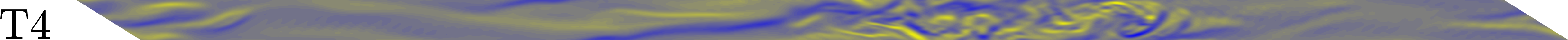}  \\ 
\includegraphics[width=0.6\linewidth]{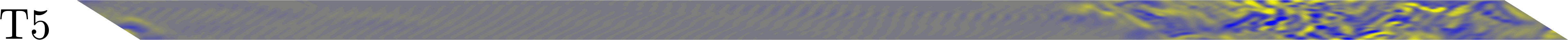}  \\ 
\includegraphics[width=0.6\linewidth]{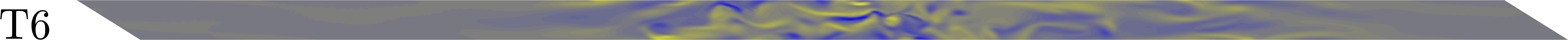}  \\ 
\end{tabular} 
\end{tabular}
\end{center}
\caption{Transient stages of the development of {\sc spt} in the long
  parallelogram domain of figure\,\ref{figdomain}b
  $(n_1,k_1,n_2,k_2)=(10,2,0,4.5)$ at $\Ri=600$, starting from a
  10-fold azimuthal replication of {\sc drw}$_{\rm F}$ (red bullet in
  figure\,\ref{fig_smallbox_turbulence}a). (a) Time series of kinetic
  energy $E$. (b) Selected instantaneous snapshots of radial vorticity
  $\omega_r\in[-4000,4000]$ fields at the mid gap $r_m$. The panels
  T1--T5 correspond to the points indicated in (a), while T6 is taken
  beyond all transients for $t\gg1$. See online movie 4, presented as supplementary material.}
\label{fig_R10th}
\end{figure}

Interesting new phenomena occur when the computational domain is
extended in the azimuthal direction to cover the whole circumference
of the annulus. Figure\,\ref{fig_R10th} shows {\sc dns} results in the
long computational domain of figure\,\ref{figdomain}b characterised by
$(n_1,k_1,n_2,k_2)=(1,2,0,4.5)$.
The flow field has been initialised with a ten-fold azimuthal
replication of the {\sc drw}$_{\rm F}$ solution and thence left to
evolve freely. The simulation deviates very slowly from {\sc
  drw}$_{\rm F}$ in the beginning, such that snapshot T1, taken at
$t=0.2$, is still indistinguishable from {\sc drw}$_{\rm F}$. Soon
after, instability leads to a sudden surge of kinetic energy. Snapshot
T2 is taken at the peak and, as clear from panel b, still markedly
preserves the wavelength of the original pattern. The instability
driving the burst is therefore superharmonic. However, as the vortex
collapses and the energy of the disturbance drops, the wavelength
becomes strongly modulated as illustrated by snapshot T3 at the valley
of the decay. From this point on, the azimuthal inhomogeneity of the
flow field becomes increasingly pronounced as time evolves. Snapshots
T4 and T5 show how this inhomogeneity gradually develops into strong
localisation. After all remnants of transient dynamics, the
statistically turbulent state exhibits azimuthal localisation as
exemplified by the instantaneous snapshot T6.

\begin{figure}                                               
  \begin{center}
    \begin{tabular}{c}
      (a)\\[0.5em]
      \includegraphics[width=0.95\linewidth]{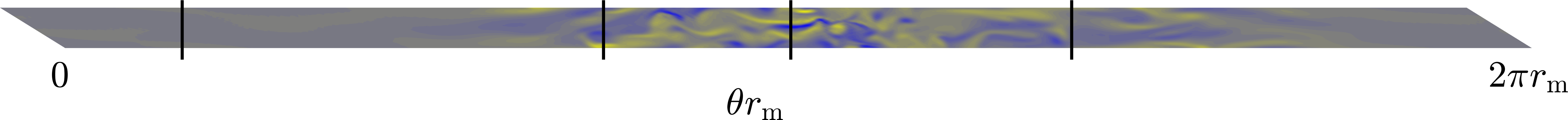} \\
      \begin{tabular}{cccc}
        (b) & (c) & (d) & (e) \\
        \raisebox{-.42cm}{\includegraphics[width=0.235\linewidth]{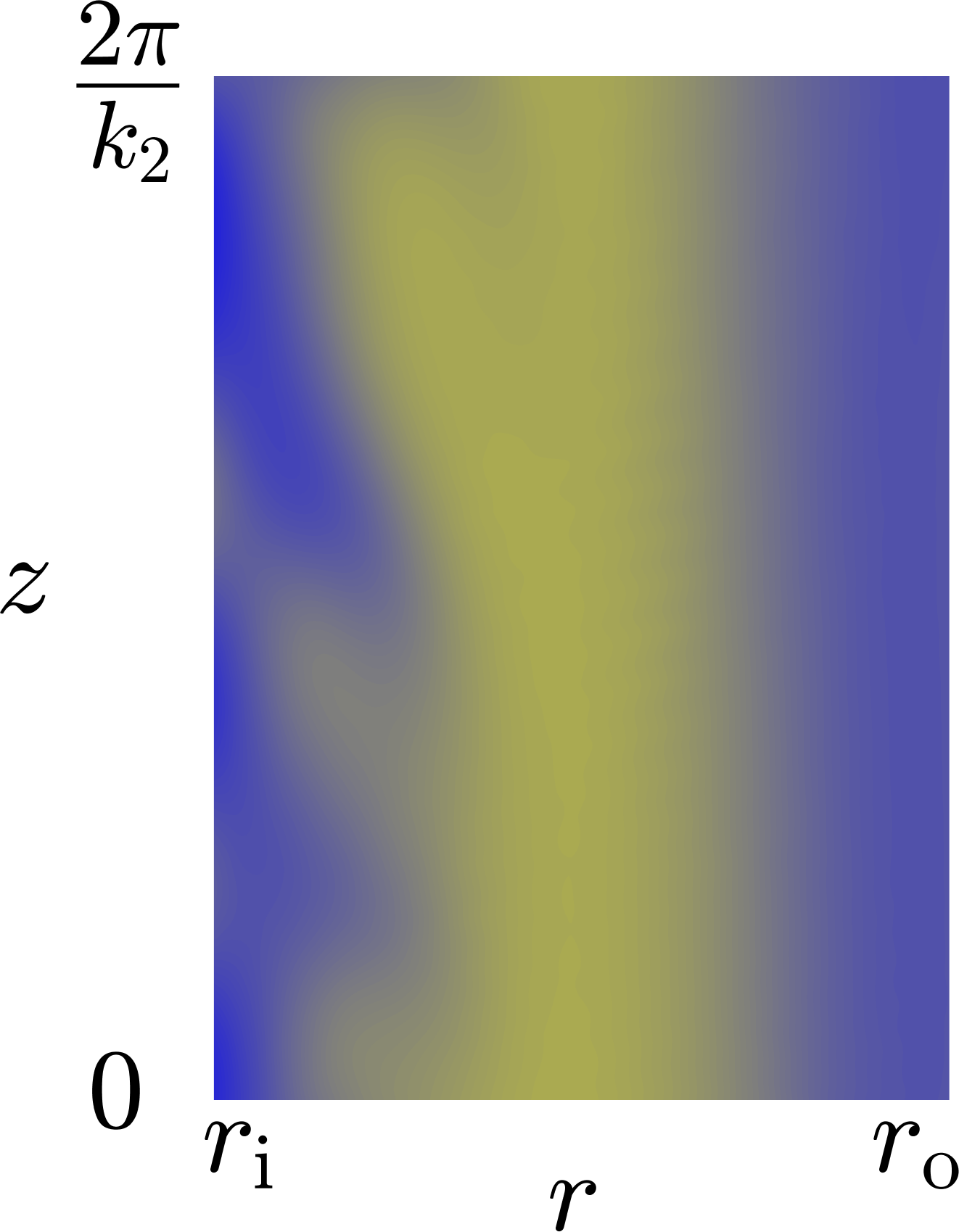}} &
        \includegraphics[width=0.18\linewidth]{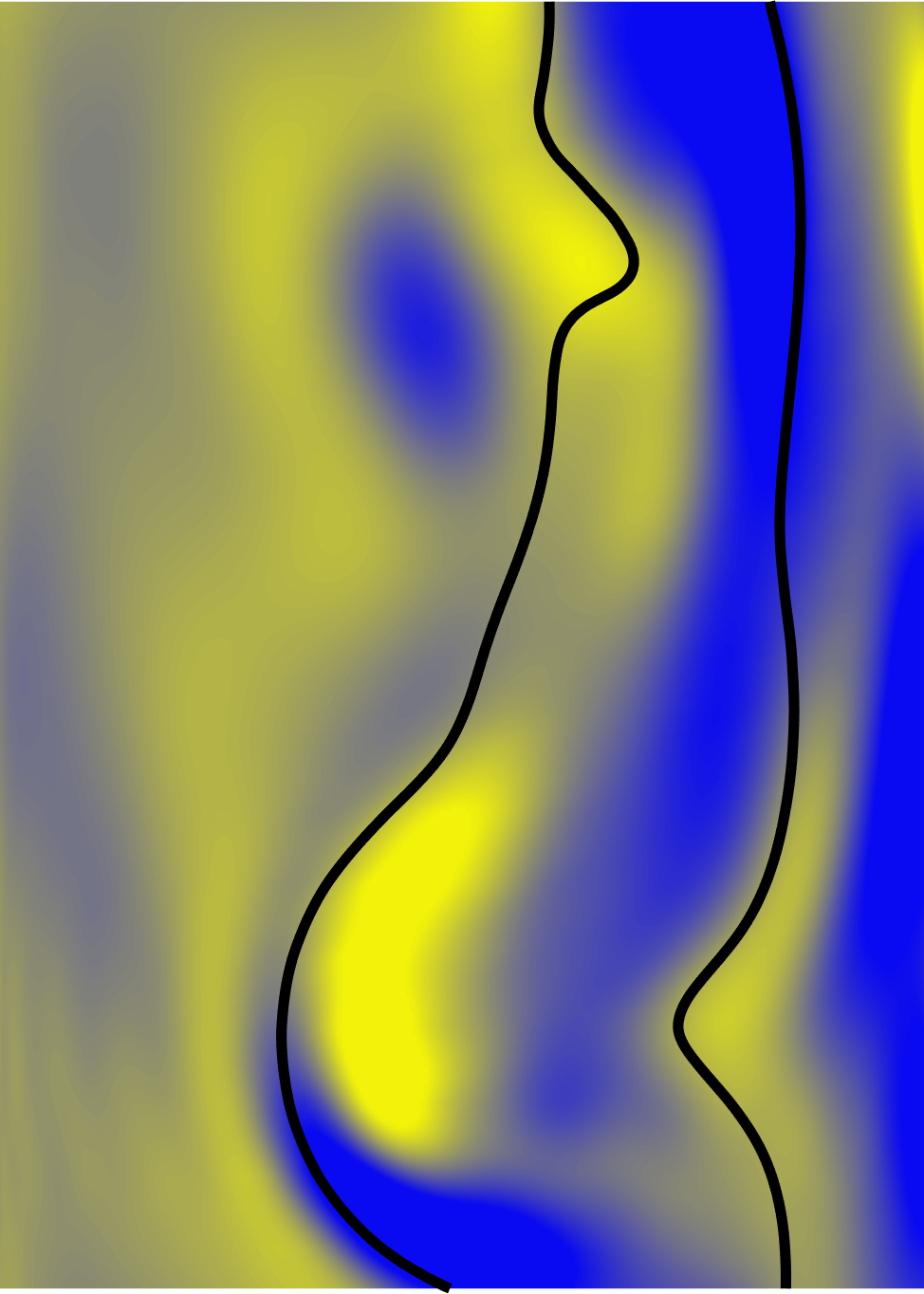} &
        \includegraphics[width=0.18\linewidth]{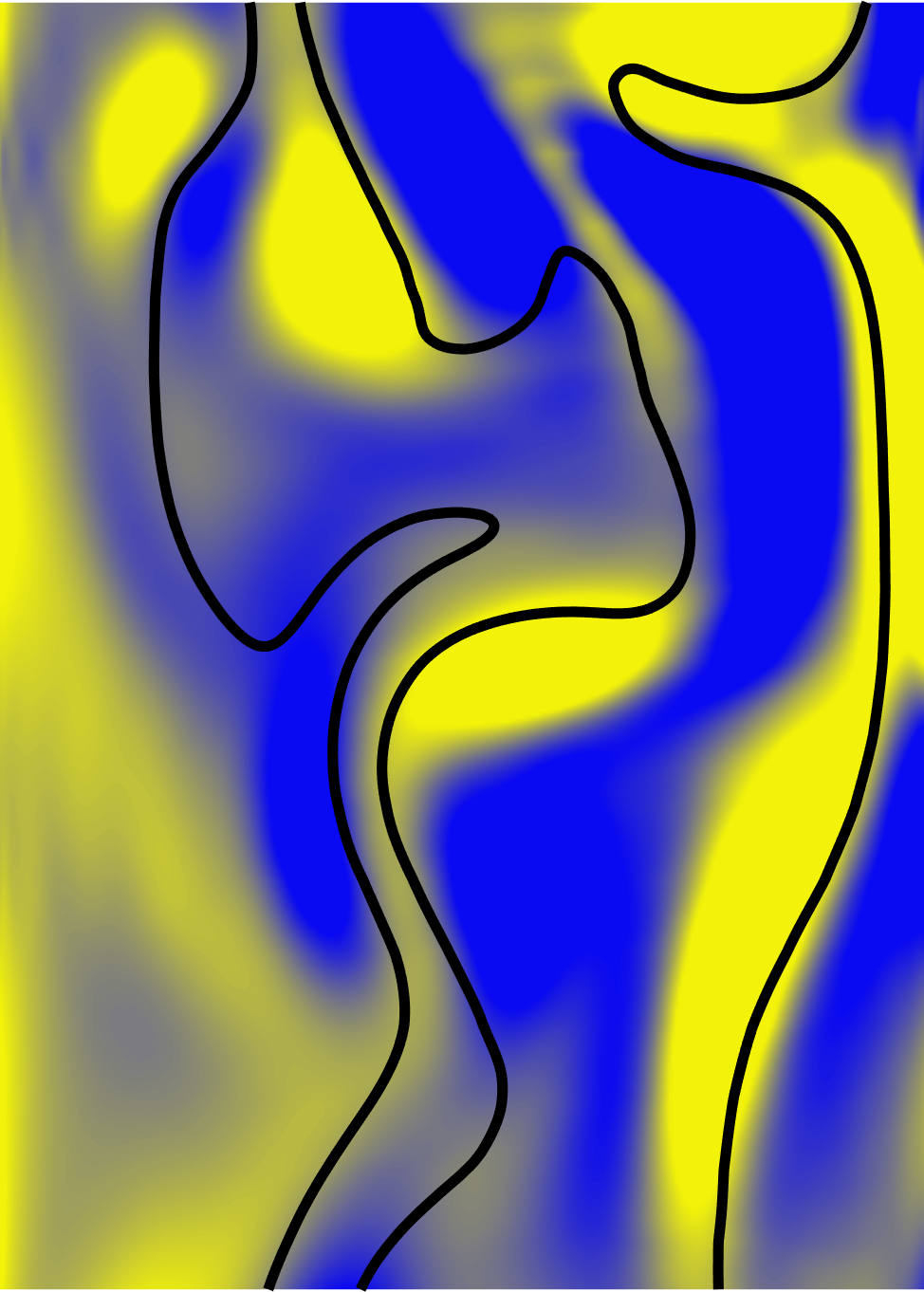} &
        \includegraphics[width=0.18\linewidth]{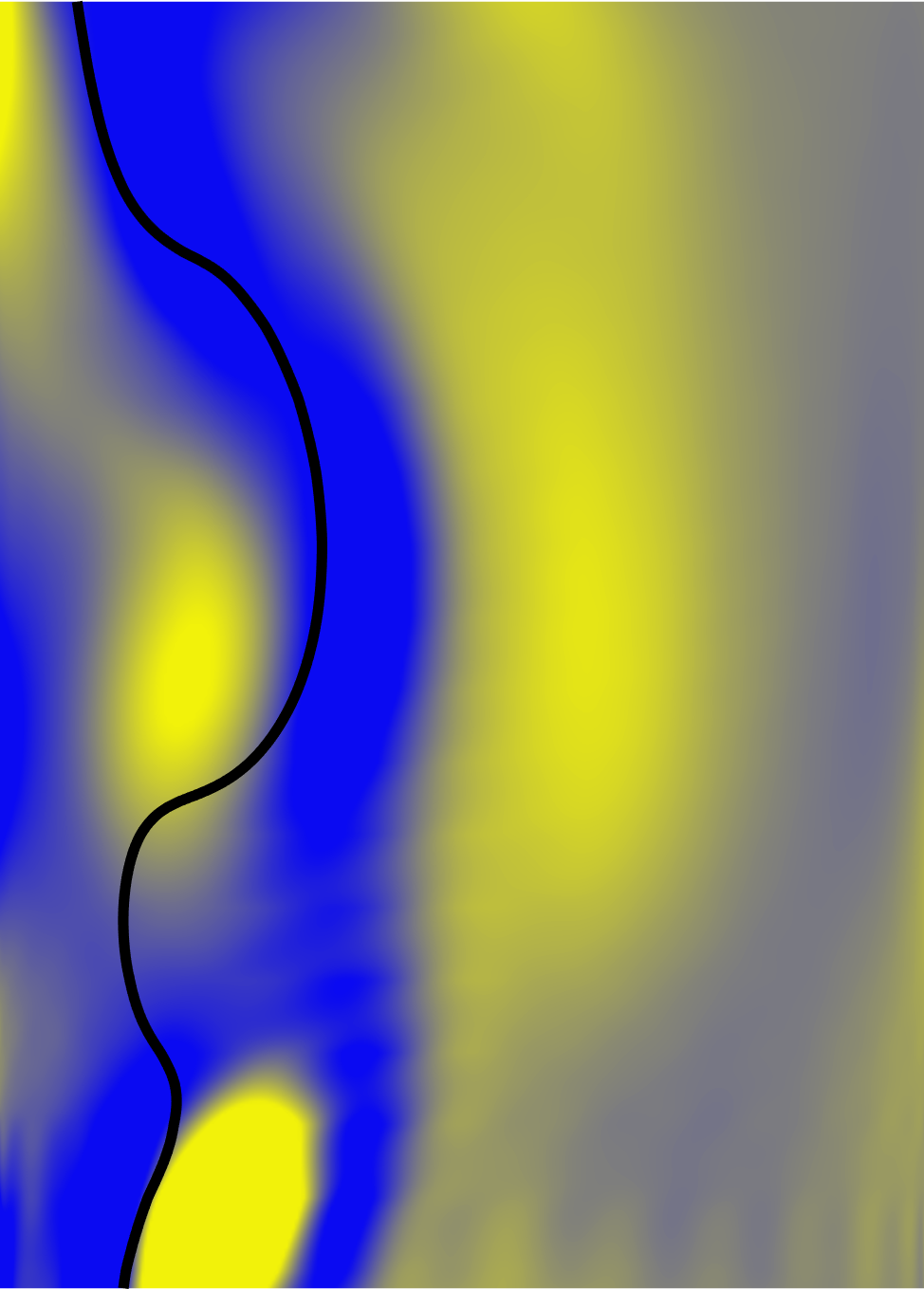}\\
      \end{tabular} 
    \end{tabular}
  \end{center}
\caption{An instantaneous snapshot of the statistically steady {\sc
    spt} state in the long domain at $\Ri=600$. (a) A close up of
  snapshot {\sc f} of figure \ref{fig_R10th}b.  (b)-(e) Azimuthal
  vorticity $\omega_{\theta}$ distribution in the $r$--$z$ plane at
  azimuthal cross-sections $\theta_1 = 0.5$, $\theta_2 = 2.3$,
  $\theta_3 = 3.1$, and $\theta_4 = 4.3$ (black vertical lines in
  panel a). Colour ranges between
  $\omega_{\theta}\in[-1500,1500]$. From left to right, the black
  curves in (c)-(e) correspond to selected isocontours with (c) $V/r =
  -20, -80$, (d) $V/r = 0,-30,-100$, (e) $V/r = 0$. See online movie 5, presented as supplementary material, showing a cross-sectional sweep of the instantaneous flow field along the full perimeter.}
\label{fig_thinbox}
\end{figure}

The flow structure of the instantaneous snapshot of the statistically
steady state T6 is examined in greater detail in
figure\,\ref{fig_thinbox}.
The basic structure within the turbulent region locally displays
sinuous wavy vortices resembling {\sc drw} but the wavelength exhibits
azimuthal modulation. The turbulent band moves from right to left with
mean azimuthal phase speed $\langle c_{\theta}\rangle_t=-35.4$ and
root-mean-square of the fluctuations
${\sqrt{\langle\left(c_{\theta}-\langle
  c_{\theta}\rangle_t\right)^2\rangle_t}=15.2}$, while the local wave
propagation speed is strongly location-dependent. This behaviour of
the flow field is typical of localised solutions in shear flows
\citep[see for example][]{AMRH13,BraGib14,ZamEck2014b,MeMe2015}, and
may be mathematically justified by a Wentzel-Kramers-Brillouin-type
approach (see \citet{BeOs99}, for example).

The local appearance of {\sc ssp}-related flow structures in streamwise-extended domains has been reported by
  \cite{DeHa14} in plane Couette flow using {\sc ecs}.
The observation of the local {\sc ssp} must in this case be based on
the strong correlation existing between the vortex sheet and a
critical layer defined by the local (rather than the
streamwise-averaged) streamwise velocity field.  A similar phenomenon
is observed in figure \ref{fig_thinbox} despite the
flow being turbulent.  The azimuthal vorticity colour map on a
$r$--$z$ azimuthal cross-section near the trailing edge of the
turbulent stripe (figure\,\ref{fig_thinbox}e) reveals a vortex sheet
in the vicinity of the inner cylinder.  Its spatial
  arrangement is closely outlined by an appropriate choice of $V/r$
  contour (black line), which is also true for figures
  \ref{fig_thinbox}c and d.
In the core of the turbulent band, several strong vortex sheets are
visible (figure \ref{fig_thinbox}d), while only those close to the
outer cylinder seem to persist toward the leading edge (figure
\ref{fig_thinbox}c). The presence of vortex sheets at various radial
locations is consistent with the multiplicity of {\sc drw} solutions
existing at this value of $\Ri$, which suggests that each one of them
might participate locally in the deployment of the {\sc ssp}.

\begin{figure}
  \begin{center}
  \begin{tabular}{cc}
  (a) & (c) \\
    \raisebox{-1.8cm}{\includegraphics[width=0.51\textwidth]{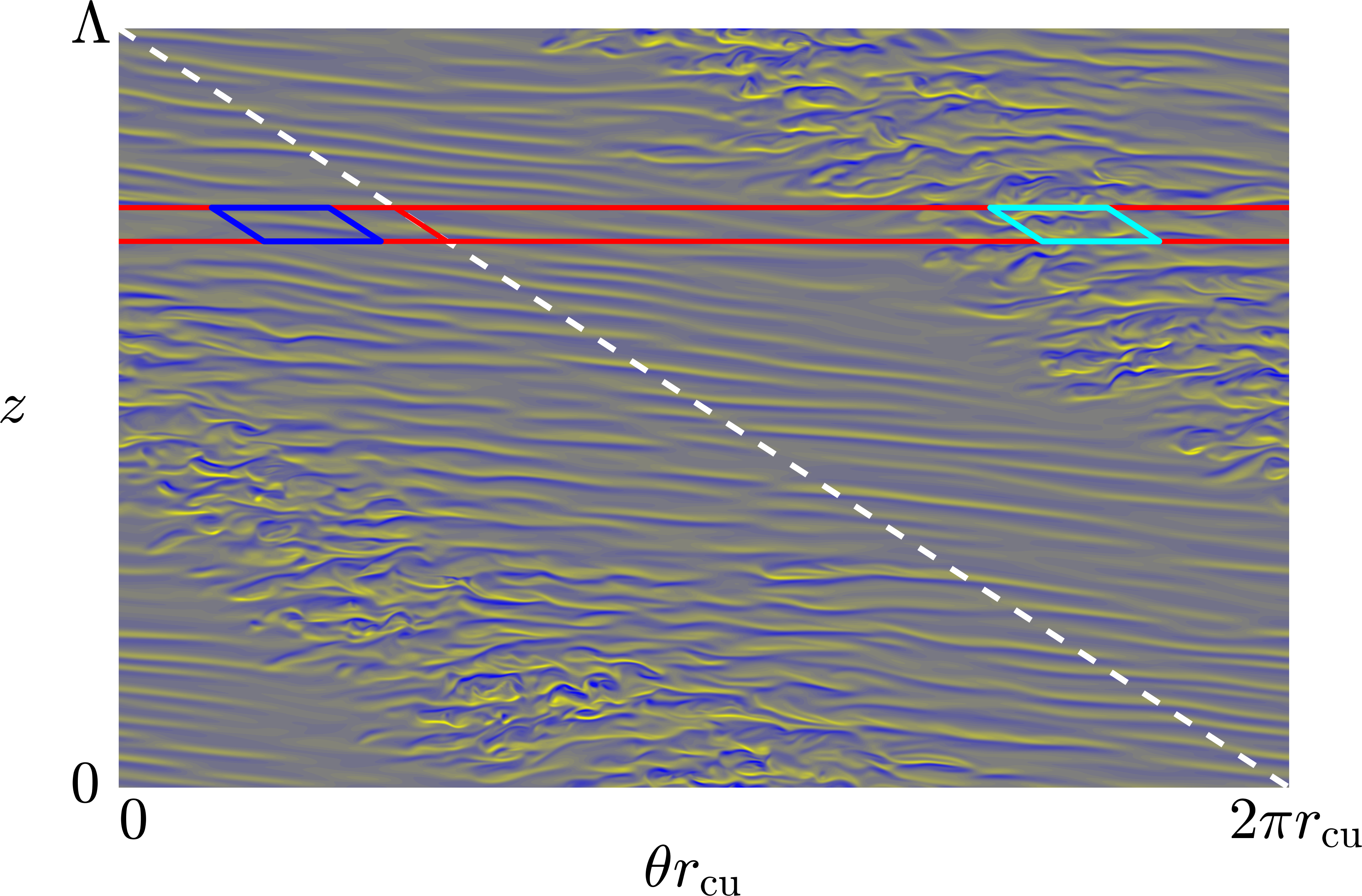}} &
    \begin{tabular}{l}
      \raisebox{0.4cm}{\includegraphics[width=0.465\textwidth]{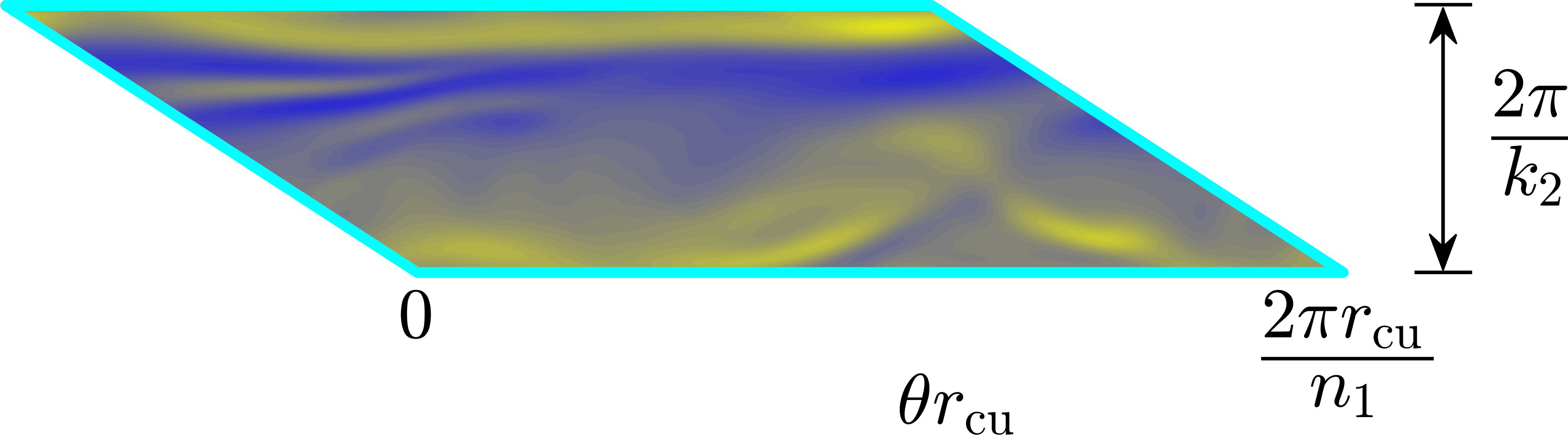}} \\
      \raisebox{0.5cm}{\includegraphics[width=0.465\textwidth]{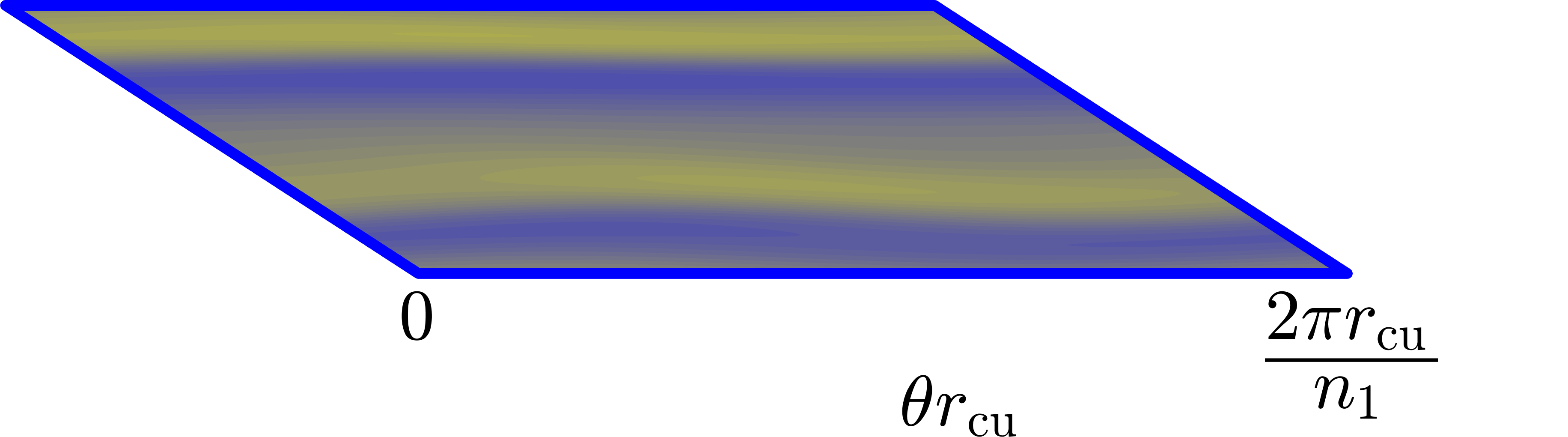}}
    \end{tabular}
  \end{tabular}\\
  (b)\\[0.5em]
 \includegraphics[width=0.9\textwidth]{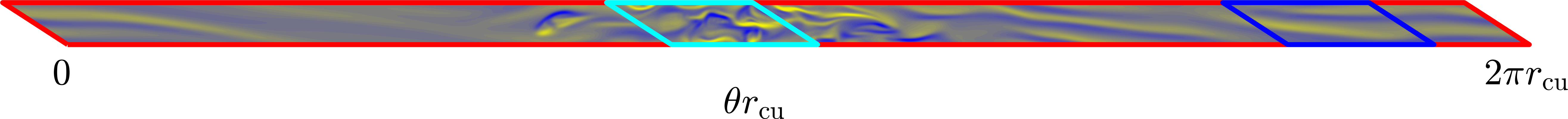} \\
  \end{center}
   \caption{The morphology of {\sc spt} within the centrifugally
     unstable region of the annulus gap. Same as
     figure~\ref{figdomain}, but at $r=r_{\rm cu}=(r_{\rm i}+r_{\rm
       n})/2=7.705<r_{\rm n}$. Colour ranges between
     $\omega_{r}\in[-4000,4000]$. See online supplementary materials: (a) movie 1, (b) movie 2 and (c) movie 3.
    }
\label{fig_SPT_SPI} 
\end{figure}

\begin{figure}
  \begin{center}
    \begin{tabular}{cc}
    (a) & (c) \\
      \raisebox{-1.8cm}{\includegraphics[width=0.51\textwidth]{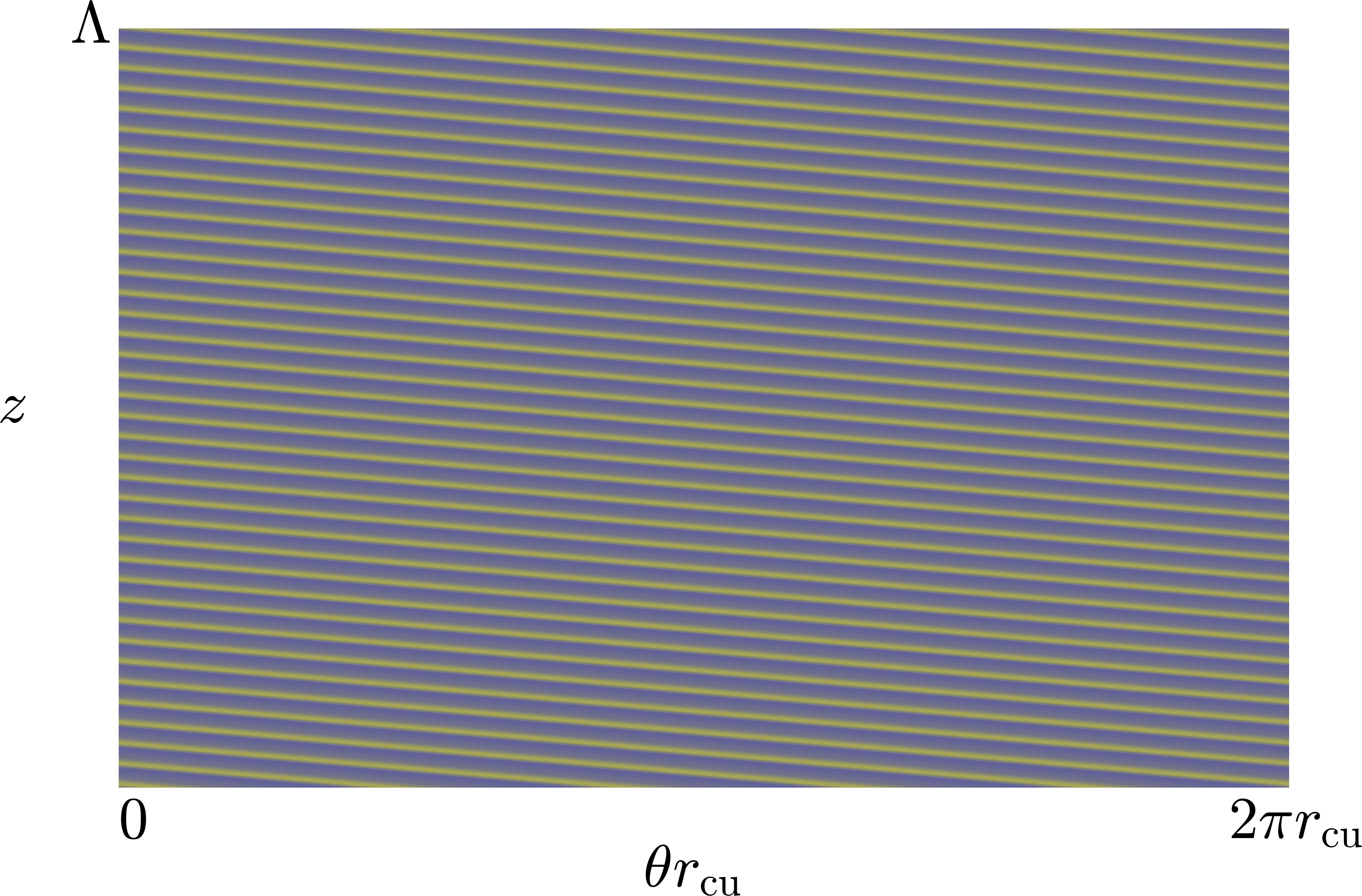} }&
      \begin{tabular}{l}
      \raisebox{0.4cm}{\includegraphics[width=0.465\textwidth]{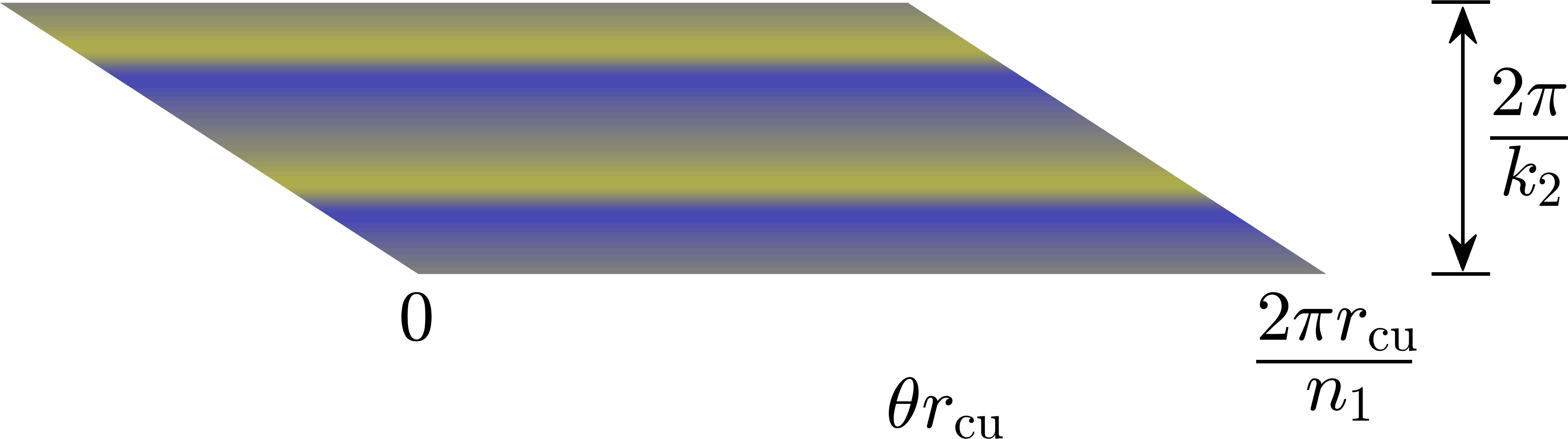}} \\
      \raisebox{0cm}{\includegraphics[width=0.465\textwidth]{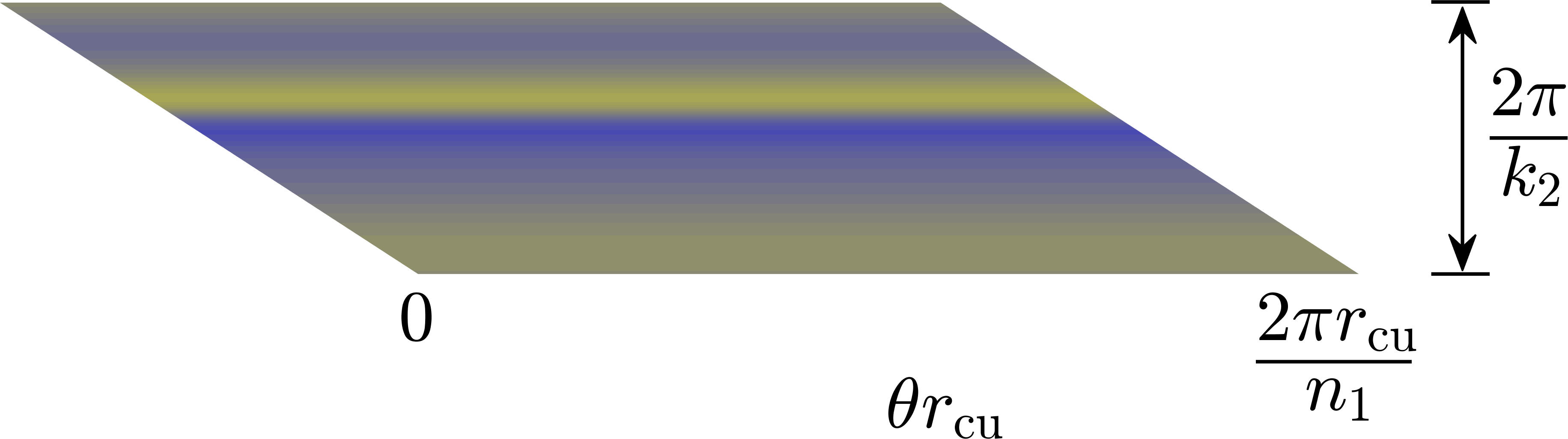}}
    \end{tabular}
  \end{tabular}\\
  (b)\\[0.5em]
 \includegraphics[width=0.9\textwidth]{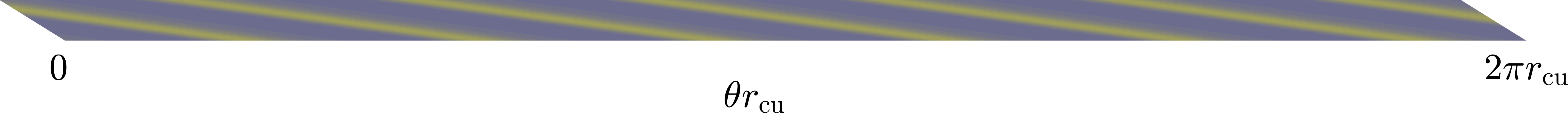} \\
  \end{center}
  \caption{Examples of centrifugally-driven exact solutions in the three domains for $(\Ro,\Ri)=(-1200,600)$, visualised through $\omega_{r}\in[-4000,4000]$ colourmaps at $r=r_{\rm cu}=(r_{\rm i}+r_{\rm n})/2=7.705<r_{\rm n}$. (a) $(n,k)=(5,7.4)$ {\sc spi}. (b) $(n,k)=(6,5.7)$ {\sc spi}. (c) $k=9$ {\sc tvf} (top) and $k=4.5$ {\sc dtvf} (bottom).
    }
\label{fig_SPI_TVF} 
\end{figure}

The basic flow is linearly unstable at $\Ri=600$, which would in
principle be at odds with the observation of a laminar flow region in
figures \ref{fig_thinbox}a and b. Note, however, that the $\theta-z$
cross-section corresponds to mid gap, while the centrifugal
instability is known to develop in the close neighbourhood of the
inner cylinder.
In fact, examination of the cross-section of \ref{fig_thinbox}b reveals the
presence of weak vortices there. 
The structure of these nonlinear vortices is best understood by
examining an unwrapped $\theta-z$ section of the full domain
simulation at $r=r_{\rm cu}=(r_{\rm i}+r_{\rm n})/2=7.705$, in the
midst of the centrifugally unstable region
(figure~\ref{fig_SPT_SPI}a). A spiral-like flow topology clearly
arises away from the turbulent stripe.
Despite the dislocations and irregularities, the vortical structures
in the laminar region compare favourably with exact spiral solutions
\citep{MeMeAvMa09_A} of the right axial-azimuthal wave-numbers (and
therefore tilt). The base {\sc ccf} is unstable to a wide range
  of axial and azimuthal wavenumbers for
  $(\eta,\Ri,\Ro)=(0.883,600,-1200)$, including axisymmetric
  perturbations with $k\in[2.44,14.26]$ and spirals modes with
  ${n\leq8.84}$ ($<9$ if the pattern is to fit exactly an integer number of times around the apparatus). Figure~\ref{fig_SPI_TVF} shows a selection of centrifugally-driven states that fit exactly in either of the three domains considered.
In particular, figure~\ref{fig_SPI_TVF}a shows one such solution
for $(n,k)=(5,7.4)$, evincing that the tilted vortical structures in
the laminar region of {\sc spt} could well be the result of the
interaction of a continuum of spiral solutions of varying wave numbers
around those of the state shown here. The same holds upon inspection
of the same $\theta-z$ section for the narrow long domain in
figure~\ref{fig_SPT_SPI}b, for which dislocated spiral patterns
of wave numbers compatible with those of the exact {\sc spi} of
  figure~\ref{fig_SPI_TVF}b, corresponding to $(n,k)=(6,5.7)$, also
show up away from the turbulent spiral. The azimuthally-extended
  nature of both domains, combined with the spatially-adjusting role of the
  turbulent spiral and, to a lesser degree, of defects and
  dislocations, allows for the formation of spirals not necessarily
  restricted to $n\in\mathbb{Z}$. As a result, one must not expect that the
  spirals observed in the laminar region of {\sc spt} conform to a
  superposition of exact solutions, each one strictly fitting in the
  domain.

In the small domain (figure~\ref{fig_SPT_SPI}c), however, quasi-axisymmetric {\sc tvf}-like structures arise during the laminar stages of the time evolution (bottom) instead of {\sc spi} patterns.
While {\sc spi} solutions of tilt and wavenumbers compatible with those observed in the time evolution exist in the full orthogonal (figure~\ref{fig_SPI_TVF}a) and narrow but long parallelogram (figure~\ref{fig_SPI_TVF}b) domains, such solutions do not exist in the short and narrow parallelogram domain (figure~\ref{fig_SPI_TVF}c). Unstable modes of {\sc ccf} are exclusively axisymmetric in the small parallelogram, which explains why the laminar stages of {\sc dns} display {\sc tvf}- and {\sc dtvf}-like structures in the centrifugally unstable region and not spirals. The laminar phases of the time-evolution, however, reach well below the torque and kinetic energy levels that are characteristic of {\sc tvf} or even {\sc dtvf}. This suggests that it is not the exact states themselves that are being approached but the unstable manifolds of {\sc ccf} that lead toward them. Axisymmetric modes with $k=4.5$, 9 and 13.5 are the three only linear instabilities of {\sc ccf} in the small parallelogram domain at $(\Ro,\Ri)=(-1200,600)$. The one fitting twice in the domain ($k=9$) is by far the most unstable, followed by that fitting just once ($k=4.5$). Mode $k=13.5$, which fits three times in the domain, is only marginally unstable and therefore difficult to observe. As the flow decays from turbulent towards {\sc ccf}, it is repelled along the direction of one of these modes and the phase-map trajectory aligns with the corresponding manifold. As this occurs, the flow fields adopt the shape of one of the unstable modes of {\sc ccf} and undergo a quasi-modal growth that eventually deforms into the nonlinear manifolds connecting to {\sc tvf} or {\sc dtvf}. This is what is actually observed during the laminar period of the time evolution. If the centrifugally-driven solutions were stable (or only marginally unstable), the approach would be consummate (or at close quarters), but since they are strongly unstable, the trajectory is again repelled well before the actual {\sc tvf} or {\sc dtvf} solution can be reached in any of their unstable directions. Online movie 3, presented as supplementary material, clearly shows how the energy-growing quasi-axisymmetric flow fields of the laminar stages of time evolution indeed undergo a wavy destabilisation before eventually triggering the turbulent burst.

Cyclic turbulent bursts similar to those observed in our small parallelogram have been reported by \citet{CoMa96} in Taylor-Couette flow at similar values of the parameters and slightly smaller radius ratio using an orthogonal computational domain that, though extended in the azimuthal direction, was incapable of sustaining {\sc spt} due to an insufficient axial height. In their computations, a laminar pattern of dislocated/defective spirals forms in the centrifugally-unstable region following the linear instability of {\sc ccf}. These spirals are in turn unstable to a modulational instability that reaches beyond $r_{\textrm{n}}$ into the centrifugally stable region, which, at these values of the parameters, happens to behave as a subcritically unstable shear flow. When the modulational instability exceeds a certain threshold, a gap-filling turbulent burst is triggered. During the laminar part of the cycle a fair amount of energy has been stored in the flow's differential rotation but, as the turbulent burst develops, the energy is quickly transferred to the smaller scales and dissipated by viscosity. Once the mean flow energy has been depleted, the turbulent state has no energy source to rely upon and decays. As the flow relaminarises, the system tries to restore the basic {\sc ccf} and clutches onto the unstable manifold that leads towards {\sc spi}, thereby recommencing the cycle. The same process seems to apply to the dynamics in the narrow and short parallelogram domain presented here, but the spiral instability of {\sc ccf} is in our case replaced by an axisymmetric instability connecting to {\sc tvf} due to our restricted azimuthal size.



The flow structure associated with the centrifugally unstable modes is
manifestly different from that of a vortex sheet resulting from the
{\sc ssp}, and both together constitute the fundamental pieces of the
{\sc spt} pattern in supercritical counter-rotating Taylor-Couette
flow. This is of course analogous to what happens for subcritical
parallel shear flows, where alternating bands of the laminar and
turbulent states conform the stripe pattern. However, there is one
essential difference between the two types of banded structures. For
parallel shear flows, both the laminar and turbulent flows are stable
solutions of the equations when considered in minimal flow units
(small periodic domains). The realisation of the banded pattern can
therefore be explained by the local bistability of the system, a fact
that has been recently used as the basis for a statistical approach
employing directed percolation theory \citep{LSAJAH16}. One might thus
expect, by analogy, that both a permanent (or at least long-lived) shear-driven turbulent regime and a stable spiral-like state of centrifugal
origin must exist in Taylor-Couette flow when considering a suitably
small parallelogram domain. However, detection of bistability at
$\Ri=600$ turns out to be utterly elusive. Instead, the laminar and the turbulent states, the former in the shape of axisymmetric {\sc tvf}-like structures instead of {\sc spi} due to modal selection in the minimal flow unit, seem to be embedded in some sort of heteroclinic cyclic loop that drives the dynamics in a chaotic cycle approaching, yet never fully relaxing onto, either state. It would therefore appear
that the formation of {\sc spt} in supercritical flows might be the
result of some sort of large scale interaction, something that would
then be at odds with the current understanding of pattern formation in
parallel shear flows. One might speculate that the heteroclinic cycle the dynamics shadow in time within the small domain, is instead deployed in space for azimuthally long domains. As observed by \citet{CoMa96}, the centrifugally-driven laminar state feeds a subcritical shear instability that triggers a turbulent burst. The resulting turbulent state relies on the pre-existence of a laminar flow field from which energy can be extracted for its long-term sustainment while, at the same time, gap filling-turbulence suppresses the centrifugal instability that drives the laminar state. As a result, the turbulent state is short-lived and cannot self-sustain. In the small domain this can only happen by alternating the laminar and turbulent states in time, while {\sc spt} patterns can arise in azimuthally-long domains as turbulence continuously decays and forms at either one of the laminar-turbulent interfaces. The azimuthally-long domain of \citet{CoMa96} should then, in principle, be capable of sustaining laminar-turbulent patterns. It might be the case that the tilt of the laminar-turbulent interfaces, which is not compatible with their narrow axially-periodic orthogonal domain, is essential to equilibrating the rates of turbulence production and decay. If the characteristic lifetime of the turbulent state is long in comparison with the growth rate of the laminar-state instability that triggers the bursts, no spatio-temporal coexistence can be expected.

In view of the reported observations so far, the narrow long domain is
capable of qualitatively reproducing the {\sc spt} regime of the full
domain, but nothing has been claimed so far as to quantitative
accuracy.  The large length-scale associated to the spiral tilt has
been straightforwardly addressed by the use of the
parallelogram-shaped domain, and the short length-scales of the
underlying {\sc drw} solutions and the even shorter associated to
turbulence are captured by a right choice of $k_2$ and sufficient
resolution, respectively. There might however be intermediate
length-scales related to a modulational instability of {\sc drw} along
the helical direction.  If such intermediate scales play an important
role in {\sc spt}, the elongated domain must be wide enough for the
associated coherent structures to decorrelate, as would happen in an
infinitely (or sufficiently) long apparatus.

The degree to which the statistics of turbulent signals converge as
the domain and/or resolution are modified provides a means of
quantifying the inaccuracies associated to domain shape, size and
discretisation. Here we have used the kinetic energy ($E$), inner
($\tau_{\rm i}$) and outer ($\tau_{\rm o}$) torque, and azimuthal
propagation speed ($c_{\theta}$) signals in the comparison of full
domain and narrow-long domain results. The simulations were run for 20
viscous time units past all foreseeable transients to reach the
statistically steady turbulent state and then run for an additional 30
and 100 viscous time units, for the full and narrow-long domain
respectively, to collect statistics. A similar resolution density was
kept from one domain to the other. All signals were positively checked
for stationarity via the augmented Dickey-Fuller test
\citep{DiFu79,DiFu81} and the mean and root-mean-square of the
fluctuation component (standard deviation) computed. Using stationary
bootstrapping \citep{PoRo94}, 95\% confidence intervals for the two
statistics estimators have been determined with automatic block size
optimisation \citep{PoWhi04,PaPoWhi09}. Since fluctuation amplitude
depends on domain size (note that a system of infinite aspect ratio
would present no fluctuations on account of the averaging properties
of normalised aggregate quantities as we are monitoring here), the
signal variance needs to be scaled accordingly. This is the same as
assuming that the signal fully decorrelates over the height of the
domain, and that the variance is therefore the composition of the
variances of as many independent, randomly distributed, identical
signals as times the narrow domain fits in the full
domain. Fluctuation amplitudes are therefore dependent on domain size,
and must be scaled. Results are summarised in table~\ref{tab:stats}.
\begin{table}
  \centering
  \begin{tabular}{c|cc|cc}
    case & \multicolumn{2}{c|}{full domain} & \multicolumn{2}{c}{long narrow domain} \\
    $(n_1,k_1,n_2,k_2)$ & \multicolumn{2}{c|}{$(1,0,0,0.2)$} & \multicolumn{2}{c}{$(1,0.2,0,4.5)$} \\
    $(L,N,M)$ & \multicolumn{2}{c|}{$(322,322,42)$} & \multicolumn{2}{c}{$(18,322,42)$} \\
    estimator & $\langle\bullet\rangle$ & $\sigma_{\bullet}=\sqrt{\langle\left(\bullet-\langle\bullet\rangle\right)^2\rangle}$ & $\langle\bullet\rangle$ & $\sigma_{\bullet}=\sqrt{\langle\left(\bullet-\langle\bullet\rangle\right)^2\rangle}$\\
    \hline
    $E$ & $16390\pm40$ & $570\pm 30$ & $16960\pm90$ & $520\pm20$ \\
    $\tau_{\rm i}$ & $0.658\pm0.001$ & $0.0195\pm0.0007$ & $0.689\pm0.003$ & $0.0182\pm0.0005$ \\
    $\tau_{\rm o}$ & $0.658\pm0.002$ & $0.0232\pm0.0009$ & $0.690\pm0.004$ & $0.0209\pm0.0006$ \\
    $c_{\theta}$ & $-34.18\pm0.08$ & $3.47\pm0.07$ & $-35.4\pm0.2$ & $3.2\pm0.2$ \\
  \end{tabular}
  \caption{Statistics of several time signals (mean $\langle\bullet\rangle$ and root-mean-square of the fluctuation component $\sigma_{\bullet}=\sqrt{\langle\left(\bullet-\langle\bullet\rangle\right)^2\rangle}$, this latter corrected for domain size) of the statistically-steady turbulent state as computed in the full orthogonal domain and in the azimuthally-long but axially-narrow parallelogram domain. The uncertainty margins correspond to 95\% confidence intervals obtained through stationary bootstrapping. }\label{tab:stats}
\end{table}
The runs are long enough for the mean to be converged to within $1\%$
for all time series. The uncertainty in the root-mean-square is
larger, ranging from $3\%$ to $9\%$. In any case, it is sufficient to
claim, with $95\%$ certainty, that the narrow and long parallelogram
domain produces results that are quantitatively different from the
full domain. Taking the mean of the statistic (mean or
root-mean-square) as the best estimate, however, bounds the inaccuracy
to within $5\%$ for the mean and $10\%$ for the root-mean-square,
which is fair but not extremely precise. While the narrow long domain of figure~\ref{figdomain}b provides reasonable quantitative agreement, wider parallelogram domains would be required to further enhance accuracy, particularly so regarding second-order statistics.

\section{Conclusions}
\label{sec_conclusions}

We have identified numerically three-dimensional nonlinear-wave
solution branches that spread over a wide
  parameter-space region of both subcritical and supercritical counter-rotating
  Taylor-Couette flow. The states have been computed in suitable
minimal parallelogram-shaped domains incorporating the oblique
pseudo-invariance of the spiral turbulence {\sc(spt)} regime, as
observed both experimentally \citep{ColAtt67} and numerically
\citep{MeMeAvMa09_A,Dong2009,DoZhe2011}. We have undertaken here to
ascertain their dynamical relevance as precursors of spiral turbulence
beyond reasonable doubt.


The waves, which mainly rotate in the azimuthal direction but possess
also a mild axial drift, spring from saddle-node bifurcations that
considerably anticipate the primary instability of the base laminar
flow. The lower-torque-waves branch is indirectly connected to
circular Couette flow through an intricate succession of intervening
secondary solutions progressively breaking the symmetries of the base
flow in a complex bifurcation scenario. The sequence involves the
subharmonic bifurcation of slightly subcritical Taylor vortices, an
axial symmetry breaking that introduces the axial drift, and a rupture
of the azimuthal invariance that sets the wave into rotation. The
actual arrangement of solution branches and the way they connect with
one another is highly dependent on the width of the parallelogram
domain, as also are the stability properties of the solutions
themselves.


The strong subcriticality of these waves implies that some mechanism,
other than centrifugal instability, must be accountable for their
self-sustainment. Axisymmetric solutions bifurcated directly from the
base flow cannot be continued far into the subcritical region, which
suggests that the three-dimensional streamwise-dependent component of
the drifting-rotating waves holds the key to self-sustainment. The
origin of the three-dimensional wave might be explained by the
inviscid instability of the streak, understood as the axisymmetric
component of the azimuthal velocity field. Compelling evidence that
this is the underlying mechanism at play is provided by the strong
wave-like vortex sheet that concentrates around the critical layer,
where the inviscid problem becomes singular. It is precisely this
vortex sheet that allegedly drives the axisymmetric roll-streak field
through the action of the Reynolds stresses, thereby closing the
interaction feed-back loop between the axisymmetric and
three-dimensional components. This is reminiscent of the
roll-streak-wave cycle that is characteristic of parallel shear flows
\citep{Wa97,WangGiWa07,HaSh2010}, except that the streak and the roll
mutually interact through both the lift-up and Coriolis effects, and
not only the former. The latter coupling term might probably affect
the formal asymptotic structure, a detailed analysis of which is
beyond the scope of this paper.


Arnoldi stability analysis and direct numerical time integration of
the Navier-Stokes equations reveal that the lower-torque branch of
drifting-rotating waves acts as an edge state separating the basin of
attraction of the base circular Couette flow from that of nontrivial
nonlinear states. Moreover, in a neighbourhood of the saddle-node and
for suitably chosen streamwise width of the parallelogram domain, the
higher-torque branch happens to be initially linearly stable. This
situation is very convenient in that it can be fruitfully exploited to
one's advantage for the detailed analysis of the onset of chaotic
motion as the governing parameter is increased further
\citep{MeEck2012,KrEc2012,LuKaVeShiKo19}.

In the case under scrutiny, the route to chaos commences with a
supercritical Hopf bifurcation of the drifting-rotating wave that
issues a branch of stable time-periodic solutions. We have found
evidence that this relative periodic orbit undergoes a period-doubling
cascade that eventually engenders a chaotic set. As the supercritical
regime is approached, the drifting-rotating wave undergoes a
succession of additional linear instabilities whence a number of
bifurcation cascades akin to the one just mentioned are expected to
ensue. It is the concurrent action of several chaotic sets thus
generated that would ultimately beget turbulent dynamics. Notably,
drifting-rotating waves, albeit unstable, can be continued all the way
up to the supercritical region of parameter space where {\sc spt} is
ubiquitous. A comparison of these solutions with turbulence as
computed in the small parallelogram domain for the same values of the
parameters suggests that the former are fairly descriptive of the
self-sustained process ({\sc ssp}) that drives the complex dynamics of
the latter.

Simulations within a narrow parallelogram domain, extended in the
azimuthal direction to wrap completely around the apparatus gap, have
been deployed to explore the relationship between the localised
turbulent stripe and the {\sc ssp}. In the elongated domain,
streamwise inhomogeneity naturally arises in the form of banded
localisation of turbulence. The final statistically-steady state
captures surprisingly well the features of {\sc spt} as computed in
the full orthogonal domain, both qualitatively and, to a large extent,
also quantitatively. The active part of inhomogeneous turbulence
within {\sc spt} is characterised by the {\sc ssp}, where streaks and
vortex layers, analogous to those we report in small periodic domains,
are also observed. Meanwhile, in the apparently quiescent flow region,
spiral vortices arise in the close proximity of the inner cylinder
following the centrifugal instability to which this region is subject.

  


Unlike what happens for subcritical parallel shear flows featuring
laminar-turbulent patterns, {\sc spt} is observed in supercritical
counter-rotating Taylor-Couette flow despite the instability of the
base laminar flow. In the case of parallel flows, the
laminar-turbulent banded pattern alternates patches of the laminar and
the turbulent states, both {\it stable} in suitably chosen small
computational domains. Present results show that this seems to not be
the case for mildly supercritical counter-rotating
Taylor-Couette flow. Neither the laminar nor the turbulent
  states can be considered permanent, and therefore {\it stable}, in
  minimal flow units. In particular, no stable state has been found
  that may account for the spiral-like waves that appear in the
  laminar region of spiral turbulence due to the centrifugal
  instability. Instead, the dynamics in small domains are invariably
  chaotic, and alternate in time short relaxation periods onto some
  sort of centrifugally-driven axisymmetric laminar state resembling
  Taylor vortex flow, and sudden bursts into short-lived shear-driven
  turbulence. To some extent, this behaviour is similar to that
  identified by \citet{CoMa96} in axially narrow orthogonal
  domains for a slightly wider gap, except that the laminar transients consisted of a dislocated spirals pattern instead of axisymmetric {\sc
    tvf}-like vortices. This might require reconsidering the mechanisms that
underly laminar-turbulent pattern formation when centrifugal effects
are at stake. To give but one example, directed percolation theory,
which has been lately employed in explaining subcritical turbulence in
parallel flows \citep{LSAJAH16,SaTa2016,ChaTuBar2017}, does not apply
to the supercritical regime. It might therefore be worthwhile
examining how the theory is affected by the centrifugal instability.

The exact spiral solutions found by \citet{MeMeAvMa09_B} are highly
unstable within the small periodic domain, so that {\sc tvf}-type
  axisymmetric modes take the lead in governing the nearly-laminar
  phases of the chaotic dynamics. Nevertheless, it might still be the
case that spiral-like structures govern the dynamics within the
laminar regions in azimuthally-extended domains, and that the
patterns observed therein, including wave dislocations, 
arise from the competition of a continuum of such spirals of
varying wavelength and tilt. The key point is that, since
neither full-fledged turbulence nor stable
  centrifugally-driven exact solutions are to be observed in the
supercritical regime within the small periodic domain, some
large-scale effect must be responsible for the permanent
  sustainment of turbulence in localised bands and for the
stabilisation of laminar spiral patterns within the laminar patches of {\sc spt}. Large-scale
pattern formation such as that observed in these laminar regions is
sometimes explained by amplitude equations derived from weakly
nonlinear theory, together with some external noise term
\citep{PriGreChaDauSaa02,BeDiZhuVerStVanLo2020}. Whether this
artificial noise term can somehow be replaced by the highly nonlinear
and autonomous {\sc ssp} structures observed here remains unclear.

Another question that arises naturally is whether the long but narrow domain
can accomodate exact solutions that may be capable of explaining the
actual topology of laminar-turbulent patterns. Exact localised
solutions resulting from subharmonic bifurcation of periodic
wave-train solutions sharing with the laminar-turbulent stripes their
main large-scale features, have been found for several parallel shear
flows \citep{ReKrSc19,PaDuHo20}. These solutions typically develop
from modulational instability of some subcritical wave to
long-wavelength perturbations \citep{ChaWiKe2014,MeMe2015}, and their
nonlinear evolution leads to the formation of localised wave fronts
that connect a patch of the nontrivial state with the base flow at
either side. Besides direct stability analysis and branch
continuation, edge-tracking has emerged as a powerful method to
compute boundary states. Unfortunately, this approach requires stable
laminar and turbulent states, something that is missing in
supercritical Taylor-Couette flow as investigated here.


Whether one employs a deterministic or a statistical approach,
  the parallelogram-shaped domain we have developed here opens the
  path to a more detailed yet affordable analysis of a wide variety of
  problems in Taylor-Couette flow that could hitherto only be
  addressed for parallel shear flows and, therefore, in the absence of
  centrifugal effects.

\backsection[Supplementary data]{\label{SupMat}Supplementary material and movies are available at \\https://doi.org/10.1017/jfm.2019...}


\backsection[Funding]{\label{acknow}
KD's research was supported by Australian Research Council Discovery
Early Career Researcher Award (DE170100171). BW, RA, FM and AM research
was supported by the Spanish Ministerio de Econom\'ia y Competitivdad (grant numbers FIS2016-77849-R and FIS2017-85794-P) and Ministerio de Ciencia e Innovaci\'on
(grant number PID2020-114043GB-I00),
and the Generalitat de Catalunya (grant 2017-SGR-785). BW's research was
also supported by the Chinese Scholarship Council (grant CSC No. 201806440152).
  Where no specific funding has been provided for research, please provide the following statement: "This research received no specific grant from any funding agency, commercial or not-for-profit sectors."
}

\backsection[Declaration of interests]{
  The authors report no conflict of interest.
}

\backsection[Data availability statement]{
  The data that support the findings of this study may be obtained from the corresponding author upon reasonable request.
}



\bibliography{local}

\begin{thebibliography}{75}
\expandafter\ifx\csname natexlab\endcsname\relax\def\natexlab#1{#1}\fi
\def\au#1{#1} \def\ed#1{#1} \def\yr#1{#1}\def\at#1{#1}\def\jt#1{\textit{#1}}
  \def\bt#1{#1}\def\bvol#1{\textbf{#1}} \def\vol#1{#1} \def\pg#1{#1}
  \def\publ#1{#1}\def\arxiv#1{#1}\def\org#1{#1}\def\st#1{\textit{#1}}

\bibitem[Andereck {\em et~al.\/}(1986)Andereck, Liu \& Swinney]{ALS86}
{\sc \au{Andereck, C.~D.}, \au{Liu, S.~S.} \& \au{Swinney, H.~L.}} \yr{1986}
  \at{Flow regimes in a circular {Couette} system with independently rotating
  cylinders}.  \jt{J.\,Fluid Mech.}  \bvol{164},  \pg{155--}.

\bibitem[Avila {\em et~al.\/}(2013)Avila, Mellibovsky, Roland \& Hof]{AMRH13}
{\sc \au{Avila, M.}, \au{Mellibovsky, F.}, \au{Roland, N.} \& \au{Hof, B.}}
  \yr{2013}  \at{Streamwise-localized solutions at the onset of turbulence in
  pipe flow}.  \jt{Phys.\ Rev.\ Lett.}  \bvol{110},  \pg{224502}.

\bibitem[Ayats {\em et~al.\/}(2020{\natexlab{{\em a\/}}})Ayats, Deguchi,
  Mellibovsky \& Meseguer]{AyDeMeMe20}
{\sc \au{Ayats, R.}, \au{Deguchi, K.}, \au{Mellibovsky, F.} \& \au{Meseguer,
  A.}} \yr{2020{\natexlab{{\em a\/}}}}  \at{Fully nonlinear mode competition in
  magnetised {Taylor-Couette} flow}.  \jt{J.\,Fluid Mech.}  \bvol{897},
  \pg{A14}.

\bibitem[Ayats {\em et~al.\/}(2020{\natexlab{{\em b\/}}})Ayats, Meseguer \&
  Mellibovsky]{AyMeMe20}
{\sc \au{Ayats, R.}, \au{Meseguer, A.} \& \au{Mellibovsky, F.}}
  \yr{2020{\natexlab{{\em b\/}}}}  \at{Symmetry-breaking waves and space-time
  modulation mechanisms in two-dimensional plane {Poiseuille} flow}.
  \jt{Phys.\ Rev.\ Fluids}  \bvol{5},  \pg{094401}.

\bibitem[Barkley \& Tuckerman(2005)]{BARTUC05}
{\sc \au{Barkley, D.} \& \au{Tuckerman, L.~S.}} \yr{2005}  \at{Computational
  study of turbulent laminar patterns in {Couette} flow}.  \jt{Phys.\ Rev.\
  Lett.}  \bvol{94},  \pg{014502}.

\bibitem[Bender \& Orszag(1999)]{BeOs99}
{\sc \au{Bender, C.~M.} \& \au{Orszag, S.~A.}} \yr{1999} {\em Advanced
  Mathematical Methods for Scientists and Engineers I: Asymptotic Methods and
  Perturbation Theory\/}, 1st edn.  \publ{Springer}.

\bibitem[Berghout {\em et~al.\/}(2020)Berghout, Dingemans, Zhu, Verzicco,
  Stevens, van Saarloos \& Lohse]{BeDiZhuVerStVanLo2020}
{\sc \au{Berghout, P.}, \au{Dingemans, R.~J.}, \au{Zhu, X.}, \au{Verzicco, R.},
  \au{Stevens, R. J. A.~M.}, \au{van Saarloos, W.} \& \au{Lohse, D.}} \yr{2020}
   \at{Direct numerical simulations of spiral taylor–couette turbulence}.
  \jt{J.\,Fluid Mech.}  \bvol{887},  \pg{A18}.

\bibitem[Boberg \& Brosa(1988)]{BOBRO88}
{\sc \au{Boberg, L.} \& \au{Brosa, U.}} \yr{1988}  \at{Onset of turbulence in a
  pipe}.  \jt{Z. Naturforsch.}  \bvol{43a},  \pg{697--726}.

\bibitem[Brand \& Gibson(2014)]{BraGib14}
{\sc \au{Brand, E.} \& \au{Gibson, J.~F.}} \yr{2014}  \at{A doubly localized
  equilibrium solution of plane {Couette} flow}.  \jt{J.\,Fluid Mech.}
  \bvol{750}.

\bibitem[Canuto {\em et~al.\/}(2007)Canuto, Hussaini, Quarteroni \&
  Zang]{CHQZ2007}
{\sc \au{Canuto, C.~G.}, \au{Hussaini, M.~Y.}, \au{Quarteroni, A.} \& \au{Zang,
  T.~A.}} \yr{2007} {\em Spectral methods: evolution to complex geometries and
  applications to fluid dynamics\/}.  \publ{Springer}.

\bibitem[Canuto {\em et~al.\/}(2010)Canuto, Hussaini, Quarteroni \&
  Zang]{CHQZ2010}
{\sc \au{Canuto, C.~G.}, \au{Hussaini, M.~Y.}, \au{Quarteroni, A.} \& \au{Zang,
  T.~A.}} \yr{2010} {\em Spectral methods: fundamentals in single domains\/},
  2nd edn.  \publ{Springer}.

\bibitem[Chantry {\em et~al.\/}(2017)Chantry, Tuckerman \&
  Barkley]{ChaTuBar2017}
{\sc \au{Chantry, M.}, \au{Tuckerman, L.~S.} \& \au{Barkley, D.}} \yr{2017}
  \at{Universal continuous transition to turbulence in a planar shear flow}.
  \jt{J.\,Fluid Mech.}  \bvol{824}.

\bibitem[Chantry {\em et~al.\/}(2014)Chantry, Willis \& Kerswell]{ChaWiKe2014}
{\sc \au{Chantry, M.}, \au{Willis, A.~P.} \& \au{Kerswell, R.~R.}} \yr{2014}
  \at{Genesis of streamwise-localized solutions from globally periodic
  traveling waves in pipe flow}.  \jt{Phys.\ Rev.\ Lett.}  \bvol{112},
  \pg{164501}.

\bibitem[Chossat \& Iooss(1994)]{ChoIoo94}
{\sc \au{Chossat, P.} \& \au{Iooss, G.}} \yr{1994} {\em The {Couette-Taylor}
  Problem\/}.  \publ{Springer-Verlag}.

\bibitem[Clever \& Busse(1992)]{CleBu92}
{\sc \au{Clever, R.~M.} \& \au{Busse, F.~H.}} \yr{1992}  \at{Three-dimensional
  convection in a horizontal fluid layer subjected to a constant shear}.
  \jt{Journal of Fluid Mechanics}  \bvol{234},  \pg{511}.

\bibitem[Clever \& Busse(1997)]{CleBu97}
{\sc \au{Clever, R.~M.} \& \au{Busse, F.~H.}} \yr{1997}  \at{Tertiary and
  quaternary solutions for plane couette flow}.  \jt{J.\,Fluid Mech.}
  \bvol{344},  \pg{137--153}.

\bibitem[Coles \& Van~Atta(1967)]{ColAtt67}
{\sc \au{Coles, D.} \& \au{Van~Atta, C.~W.}} \yr{1967}  \at{Digital experiment
  in spiral turbulence}.  \jt{Phys.\ Fluids\,Suppl.}  \bvol{121},  \pg{S120--}.

\bibitem[Coughlin \& Marcus(1996)]{CoMa96}
{\sc \au{Coughlin, K.} \& \au{Marcus, P.S.}} \yr{1996}  \at{Turbulent bursts in
  couette-taylor flow}.  \jt{Phys.\ Rev.\ Lett.}  \bvol{77(11)},
  \pg{2214--2217}.

\bibitem[Deguchi(2016)]{Deguchi_JFM17}
{\sc \au{Deguchi, K.}} \yr{2016}  \at{The rapid-rotation limit of the neutral
  curve for {Taylor-Couette} flow}.  \jt{J.\,Fluid Mech.}  \bvol{808},
  \pg{R2}.

\bibitem[Deguchi \& Altmeyer(2013)]{DeAlt2013}
{\sc \au{Deguchi, K.} \& \au{Altmeyer, S.}} \yr{2013}  \at{Fully nonlinear mode
  competitions of nearly bicritical spiral or {Taylor} vortices in
  {Taylor-Couette} flow}.  \jt{Phys. Rev. E}  \bvol{87},  \pg{043017}.

\bibitem[Deguchi \& Hall(2014)]{DeHa14}
{\sc \au{Deguchi, K.} \& \au{Hall, P.}} \yr{2014}  \at{The high-reynolds-number
  asymptotic development of nonlinear equilibrium states in plane {Couette}
  flow}.  \jt{J.\,Fluid Mech.}  \bvol{750},  \pg{99--112}.

\bibitem[Deguchi \& Hall(2015)]{DeguchiHall2015}
{\sc \au{Deguchi, K} \& \au{Hall, P.}} \yr{2015}  \at{Asymptotic descriptions
  of oblique coherent structures in shear flows}.  \jt{J.\,Fluid Mech.}
  \bvol{782},  \pg{356--367}.

\bibitem[Deguchi {\em et~al.\/}(2014)Deguchi, Meseguer \&
  Mellibovsky]{DeMeMe14}
{\sc \au{Deguchi, K.}, \au{Meseguer, A.} \& \au{Mellibovsky, F.}} \yr{2014}
  \at{Subcritical equilibria in {Taylor-Couette} flow}.  \jt{Phys.\ Rev.\
  Lett.}  \bvol{112},  \pg{184502}.

\bibitem[Deguchi \& Nagata(2011)]{DeNa2011}
{\sc \au{Deguchi, K.} \& \au{Nagata, M.}} \yr{2011}  \at{Bifurcations and
  instabilities in sliding {Couette} flow}.  \jt{J.\,Fluid Mech.}  \bvol{678},
  \pg{156--178}.

\bibitem[Dickey \& Fuller(1979)]{DiFu79}
{\sc \au{Dickey, D.~A.} \& \au{Fuller, W.~A.}} \yr{1979}  \at{Distribution of
  the estimators for autoregressive time-series with a unit root}.  \jt{J. Am.
  Stat. Assoc.}  \bvol{74}~(366),  \pg{427--431}.

\bibitem[Dickey \& Fuller(1981)]{DiFu81}
{\sc \au{Dickey, D.~A.} \& \au{Fuller, W.~A.}} \yr{1981}  \at{Likelihood ratio
  statistics for autoregressive time-series with a unit-root}.
  \jt{Econometrica}  \bvol{49}~(4),  \pg{1057--1072}.

\bibitem[Dong(2009)]{Dong2009}
{\sc \au{Dong, S.}} \yr{2009}  \at{Evidence for internal structures of spiral
  turbulence}.  \jt{Phys. Rev. E}  \bvol{80},  \pg{067301}.

\bibitem[Dong \& Zheng(2011)]{DoZhe2011}
{\sc \au{Dong, S.} \& \au{Zheng, X.}} \yr{2011}  \at{Direct numerical
  simulation of spiral turbulence}.  \jt{J.\,Fluid Mech.}  \bvol{668},
  \pg{150--}.

\bibitem[Drazin \& Reid(1981)]{DraRe81}
{\sc \au{Drazin, P.~G.} \& \au{Reid, W.~H.}} \yr{1981} {\em Hydrodynamic
  Stability\/}.  \publ{Cambridge University Press}.

\bibitem[Duguet {\em et~al.\/}(2010)Duguet, Schlatter \&
  Henningson]{DuSchHen2010}
{\sc \au{Duguet, Y.}, \au{Schlatter, P.} \& \au{Henningson, D.\, S.}} \yr{2010}
   \at{Formation of turbulent patterns near the onset of transition in plane
  {Couette} flow}.  \jt{J.\,Fluid Mech.}  \bvol{650},  \pg{119}.

\bibitem[Eckhardt {\em et~al.\/}(2007)Eckhardt, Schneider, Hof \&
  Westerweel]{EcScHoWe07}
{\sc \au{Eckhardt, B.}, \au{Schneider, T.~M.}, \au{Hof, B.} \& \au{Westerweel,
  J.}} \yr{2007}  \at{Turbulence transition in pipe flow}.  \jt{Ann.\ Rev.\
  Fluid\ Mech.}  \bvol{39},  \pg{447--468}.

\bibitem[Esser \& Grossmann(1996)]{EsGr96}
{\sc \au{Esser, A.} \& \au{Grossmann, S.}} \yr{1996}  \at{Analytic expression
  for {Taylor-Couette} stability boundary}.  \jt{Phys.\ Fluids}  \bvol{8},
  \pg{1814--1819}.

\bibitem[Faisst \& Eckhardt(2003)]{FAIECK03}
{\sc \au{Faisst, H.} \& \au{Eckhardt, B.}} \yr{2003}  \at{Travelling waves in
  pipe flow}.  \jt{Phys.\ Rev.\ Lett.}  \bvol{91},  \pg{224502}.

\bibitem[Feynman(1964)]{Fey64}
{\sc \au{Feynman, R.~P.}} \yr{1964} {\em Lecture Notes in Physics\/}, ,
  \vol{vol.~2}.  \publ{Reading, MA: Addison-Wesley}.

\bibitem[Graham \& Floryan(2021)]{GraFlo21}
{\sc \au{Graham, M.~D} \& \au{Floryan, D.}} \yr{2021}  \at{Exact coherent
  states and the nonlinear dynamics of wall-bounded turbulent flows}.
  \jt{Ann.\ Rev.\ Fluid\ Mech.}  \bvol{53},  \pg{227--253}.

\bibitem[Grossmann(2000)]{Grossmann2000}
{\sc \au{Grossmann, S.}} \yr{2000}  \at{The onset of shear flow turbulence}.
  \jt{Rev. Mod. Phys.}  \bvol{72},  \pg{603--618}.

\bibitem[Hall \& Sherwin(2010)]{HaSh2010}
{\sc \au{Hall, P.} \& \au{Sherwin, S.}} \yr{2010}  \at{Streamwise vortices in
  shear flows: harbingers of transition and the skeleton of coherent
  structures}.  \jt{J.\,Fluid Mech.}  \bvol{661},  \pg{178--205}.

\bibitem[Hamilton {\em et~al.\/}(1995)Hamilton, Kim \& Waleffe]{HamKimWa95}
{\sc \au{Hamilton, J.~M.}, \au{Kim, J.} \& \au{Waleffe, F.}} \yr{1995}
  \at{Regeneration mechanisms of near-wall turbulence}.  \jt{J.\,Fluid Mech.}
  \bvol{287},  \pg{317--348}.

\bibitem[Itano \& Toh(2001)]{ItTo01}
{\sc \au{Itano, T.} \& \au{Toh, S.}} \yr{2001}  \at{The dynamics of bursting
  process in wall turbulence}.  \jt{J. Phys. Soc. Japan}  \bvol{70}~(3),
  \pg{703--716}.

\bibitem[Kawahara {\em et~al.\/}(2012)Kawahara, Uhlmann \& Van~Veen]{KaUhVe12}
{\sc \au{Kawahara, G.}, \au{Uhlmann, M.} \& \au{Van~Veen, L.}} \yr{2012}
  \at{The significance of simple invariant solutions in turbulent flows}.
  \jt{Ann.\ Rev.\ Fluid\ Mech.}  \bvol{44},  \pg{203--225}.

\bibitem[Kelley(2003)]{Kel03}
{\sc \au{Kelley, C.T.}} \yr{2003} {\em Solving nonlinear equations with
  Newton's method\/}.  \publ{Philadelphia: SIAM}.

\bibitem[Kerswell(2005)]{Kerswell2005}
{\sc \au{Kerswell, R.~R.}} \yr{2005}  \at{Recent progress in understanding the
  transition to turbulence in a pipe}.  \jt{Nonlinearity}  \bvol{18},
  \pg{R17--R44}.

\bibitem[Kreilos \& Eckhardt(2012)]{KrEc2012}
{\sc \au{Kreilos, T.} \& \au{Eckhardt, B.}} \yr{2012}  \at{Periodic orbits near
  onset of chaos in plane {Couette} flow}.  \jt{CHAOS}  \bvol{22},
  \pg{047505}.

\bibitem[Krygier {\em et~al.\/}(2021)Krygier, Pughe-Sanford \&
  Grigoriev]{KPG21}
{\sc \au{Krygier, M.~C.}, \au{Pughe-Sanford, J.~L.} \& \au{Grigoriev, R.~O.}}
  \yr{2021}  \at{Exact coherent structures and shadowing in turbulent
  taylor–couette flow}.  \jt{J.\,Fluid Mech.}  \bvol{923},  \pg{A7}.

\bibitem[Kuznetsov(2004)]{Kuz04}
{\sc \au{Kuznetsov, Y.~A.}} \yr{2004} {\em Elements of Applied Bifurcation
  Theory, 3rd Ed.\/}.  \publ{Springer-Verlag}.

\bibitem[Lemoult {\em et~al.\/}(2016)Lemoult, Shi, Avila, Jalikop, Avila \&
  Hof]{LSAJAH16}
{\sc \au{Lemoult, G.}, \au{Shi, L.}, \au{Avila, K.}, \au{Jalikop, S.~V.},
  \au{Avila, M.} \& \au{Hof, B.}} \yr{2016}  \at{Directed percolation phase
  transition to sustained turbulence in couette flow}.  \jt{Nature Physics}
  \bvol{12},  \pg{254--258}.

\bibitem[Lin(1955)]{Lin55}
{\sc \au{Lin, C.}} \yr{1955} {\em The theory of hydrodynamic stability\/}.
  \publ{Cambridge University Press}.

\bibitem[Lustro {\em et~al.\/}(2019)Lustro, Kawahara, van Veen, Shimizu \&
  Kokubu]{LuKaVeShiKo19}
{\sc \au{Lustro, J.\!R.\!T.}, \au{Kawahara, G.}, \au{van Veen, L.},
  \au{Shimizu, M.} \& \au{Kokubu, H.}} \yr{2019}  \at{The onset of transient
  turbulence in minimal plane {Couette} flow}.  \jt{J.\,Fluid Mech.}
  \bvol{862},  \pg{R2}.

\bibitem[Mellibovsky \& Eckhardt(2011)]{MeEck2011}
{\sc \au{Mellibovsky, F.} \& \au{Eckhardt, B.}} \yr{2011}
  \at{{Takens–Bogdanov} bifurcation of travelling-wave solutions in pipe
  flow}.  \jt{J.\,Fluid Mech.}  \bvol{670},  \pg{96--129}.

\bibitem[Mellibovsky \& Eckhardt(2012)]{MeEck2012}
{\sc \au{Mellibovsky, F.} \& \au{Eckhardt, B.}} \yr{2012}  \at{From travelling
  waves to mild chaos: a supercritical bifurcation cascade in pipe flow}.
  \jt{J.\,Fluid Mech.}  \bvol{709},  \pg{149--190}.

\bibitem[Mellibovsky \& Meseguer(2015)]{MeMe2015}
{\sc \au{Mellibovsky, F.} \& \au{Meseguer, A.}} \yr{2015}  \at{A mechanism for
  streamwise localisation of nonlinear waves in shear flows}.  \jt{J.\,Fluid
  Mech.}  \bvol{779},  \pg{R1}.

\bibitem[Meseguer(2020)]{MeBookFNM20}
{\sc \au{Meseguer, A.}} \yr{2020} {\em Fundamentals of Numerical Mathematics
  for Physicists and Engineers\/}.  \publ{Hoboken, N.J.: John Wiley \& Sons}.

\bibitem[Meseguer {\em et~al.\/}(2007)Meseguer, Avila, Mellibovsky \&
  Marques]{EPJSTMAMM07}
{\sc \au{Meseguer, A.}, \au{Avila, M.}, \au{Mellibovsky, F.} \& \au{Marques,
  F.}} \yr{2007}  \at{Solenoidal spectral formulations for the computation of
  secondary flows in cylindrical and annular geometries}.  \jt{Eur.\ Phys.\
  J.\,Special\,Topics}  \bvol{146},  \pg{249--259}.

\bibitem[Meseguer {\em et~al.\/}(2009{\natexlab{{\em a\/}}})Meseguer,
  Mellibovsky, Avila \& Marques]{MeMeAvMa09_B}
{\sc \au{Meseguer, A.}, \au{Mellibovsky, F.}, \au{Avila, M.} \& \au{Marques,
  F.}} \yr{2009{\natexlab{{\em a\/}}}}  \at{Families of subcritical spirals in
  highly counter-rotating {Taylor-Couette} flow}.  \jt{Phys.\ Rev.\ E}
  \bvol{79},  \pg{036309}.

\bibitem[Meseguer {\em et~al.\/}(2009{\natexlab{{\em b\/}}})Meseguer,
  Mellibovsky, Avila \& Marques]{MeMeAvMa09_A}
{\sc \au{Meseguer, A.}, \au{Mellibovsky, F.}, \au{Avila, M.} \& \au{Marques,
  F.}} \yr{2009{\natexlab{{\em b\/}}}}  \at{Instability mechanisms and
  transition scenarios of spiral turbulence in {Taylor-Couette} flow}.
  \jt{Phys.\ Rev.\ E}  \bvol{80},  \pg{046315}.

\bibitem[Moser {\em et~al.\/}(1983)Moser, Moin \& Leonard]{MoMoLe83}
{\sc \au{Moser, R.~D.}, \au{Moin, P.} \& \au{Leonard, A.}} \yr{1983}  \at{A
  spectral numerical method for the {Navier-Stokes} equations with applications
  to {Taylor-Couette} flow}.  \jt{J. Comp. Phys.}  \bvol{52},  \pg{524--544}.

\bibitem[Nagata(1990)]{Nag90}
{\sc \au{Nagata, M.}} \yr{1990}  \at{Three-dimensional finite-amplitude
  solutions in plane {Couette} flow: bifurcation from infinity}.  \jt{J.\,Fluid
  Mech.}  \bvol{217},  \pg{519--527}.

\bibitem[Paranjape {\em et~al.\/}(2020)Paranjape, Duguet \& Hof]{PaDuHo20}
{\sc \au{Paranjape, C.~S.}, \au{Duguet, Y.} \& \au{Hof, B.}} \yr{2020}
  \at{Oblique stripe solutions of channel flow}.  \jt{J.\,Fluid Mech.}
  \bvol{897},  \pg{A7}.

\bibitem[Patton {\em et~al.\/}(2009)Patton, Politis \& White]{PaPoWhi09}
{\sc \au{Patton, A.}, \au{Politis, D.~N.} \& \au{White, H.}} \yr{2009}
  \at{Correction to “automatic block-length selection for the dependent
  bootstrap” by d. politis and h. white}.  \jt{Econom. Rev.}  \bvol{28}~(4),
  \pg{372--375}.

\bibitem[Politis \& Romano(1994)]{PoRo94}
{\sc \au{Politis, D.~N.} \& \au{Romano, J.~P.}} \yr{1994}  \at{The stationary
  bootstrap}.  \jt{J. Am. Stat. Assoc.}  \bvol{89}~(428),  \pg{1303--1313}.

\bibitem[Politis \& White(2004)]{PoWhi04}
{\sc \au{Politis, D.~N.} \& \au{White, H.}} \yr{2004}  \at{Automatic
  block-length selection for the dependent bootstrap}.  \jt{Econom. Rev.}
  \bvol{23}~(1),  \pg{53--70}.

\bibitem[Prigent {\em et~al.\/}(2002)Prigent, Gr\'egoire, Chat\'e, Dauchot \&
  van Saarloos]{PriGreChaDauSaa02}
{\sc \au{Prigent, A.}, \au{Gr\'egoire, G.}, \au{Chat\'e, H.}, \au{Dauchot, O.}
  \& \au{van Saarloos, W.}} \yr{2002}  \at{Finite-wavelength modulation within
  turbulent shear flows}.  \jt{Phys.\ Rev.\ Lett.}  \bvol{89},  \pg{014501}.

\bibitem[Reetz {\em et~al.\/}(2019)Reetz, Kreilos \& Schneider]{ReKrSc19}
{\sc \au{Reetz, F.}, \au{Kreilos, T.} \& \au{Schneider, T.~M.}} \yr{2019}
  \at{Exact invariant solution reveals the origin of self-organized oblique
  turbulent-laminar stripes}.  \jt{{Nat. Commun.}}  \bvol{10},  \pg{2277}.

\bibitem[Sano \& Tamai(2016)]{SaTa2016}
{\sc \au{Sano, M.} \& \au{Tamai, K.}} \yr{2016}  \at{A universal transition to
  turbulence in channel flow}.  \jt{Nat. Phys.}  \bvol{12(3)},  \pg{249--253}.

\bibitem[Schneider {\em et~al.\/}(2008)Schneider, Gibson, Lagha, De~Lillo \&
  Eckhardt]{SchGiLaDeLiEck08}
{\sc \au{Schneider, T.~M.}, \au{Gibson, J.~F.}, \au{Lagha, M.}, \au{De~Lillo,
  F.} \& \au{Eckhardt, B.}} \yr{2008}  \at{Laminar-turbulent boundary in plane
  {Couette} flow}.  \jt{Phys.\ Rev.\ E}  \bvol{78},  \pg{037301}.

\bibitem[Shi {\em et~al.\/}(2013)Shi, Avila \& Hof]{ShAvHo2013}
{\sc \au{Shi, L.}, \au{Avila, M.} \& \au{Hof, B.}} \yr{2013}  \at{Scale
  invariance at the onset of turbulence in {Couette} flow}.  \jt{Phys.\ Rev.\
  Lett.}  \bvol{110},  \pg{204502}.

\bibitem[Skufca {\em et~al.\/}(2006)Skufca, Yorke \& Eckhardt]{SkYoEck06}
{\sc \au{Skufca, J.~D.}, \au{Yorke, J.~A.} \& \au{Eckhardt, B.}} \yr{2006}
  \at{Edge of chaos in a parallel shear flow}.  \jt{Phys.\ Rev.\ Lett.}
  \bvol{96},  \pg{174101}.

\bibitem[Trefethen \& Bau(1997)]{Trebooknla}
{\sc \au{Trefethen, L.~N.} \& \au{Bau, D.}} \yr{1997} {\em Numerical Linear
  Algebra\/}.  \publ{Philadelphia: SIAM}.

\bibitem[Tuckerman {\em et~al.\/}(2020)Tuckerman, Chantry \&
  Barkley]{TuChaBar2020}
{\sc \au{Tuckerman, L.~S.}, \au{Chantry, M.} \& \au{Barkley, D.}} \yr{2020}
  \at{Patterns in wall-bounded shear flows}.  \jt{Ann. Rev. Fluid Mech.}
  \bvol{52},  \pg{010719--060221}.

\bibitem[Tuckerman {\em et~al.\/}(2014)Tuckerman, Kreilos, Schrobsdorff,
  Schneider \& Gibson]{Tucketal2014}
{\sc \au{Tuckerman, L.~S.}, \au{Kreilos, T.}, \au{Schrobsdorff, H.},
  \au{Schneider, T.~M.} \& \au{Gibson, J.~F.}} \yr{2014}  \at{Turbulent-laminar
  patterns in plane {Poiseuille} flow}.  \jt{Phys.\ Fluids}  \bvol{26},
  \pg{114103}.

\bibitem[Waleffe(1997)]{Wa97}
{\sc \au{Waleffe, F.}} \yr{1997}  \at{On a self-sustaining process in shear
  flows}.  \jt{Phys.\ Fluids}  \bvol{9},  \pg{883--900}.

\bibitem[Waleffe(2003)]{Wa03}
{\sc \au{Waleffe, F.}} \yr{2003}  \at{Homotopy of exact coherent structures in
  plane shear flows}.  \jt{Phys.\ Fluids}  \bvol{15(6)},  \pg{1517--1534}.

\bibitem[Wang {\em et~al.\/}(2007)Wang, Gibson \& Waleffe]{WangGiWa07}
{\sc \au{Wang, J.}, \au{Gibson, J.} \& \au{Waleffe, F.}} \yr{2007}  \at{Lower
  branch coherent states in shear flows: Transition and control}.  \jt{Phys.\
  Rev.\ Lett.}  \bvol{98},  \pg{204501}.

\bibitem[Wedin \& Kerswell(2004)]{WEDKER04}
{\sc \au{Wedin, H.} \& \au{Kerswell, R.~R.}} \yr{2004}  \at{Exact coherent
  structures in pipe flow: travelling wave solutions}.  \jt{J.\,Fluid Mech.}
  \bvol{508},  \pg{333--371}.

\bibitem[Zammert \& Eckhardt(2014)]{ZamEck2014b}
{\sc \au{Zammert, S.} \& \au{Eckhardt, B.}} \yr{2014}  \at{Streamwise and
  doubly-localised periodic orbits in plane {Poiseuille} flow}.  \jt{J.\,Fluid
  Mech.}  \bvol{761},  \pg{348--359}.

\end{thebibliography}
\bibliographystyle{jfm}

\end{document}